\documentclass[10pt,a4paper]{article}
\usepackage[left=2cm, right=2cm, top=2cm]{geometry}
\usepackage{url}
\usepackage{float} 
\usepackage{graphicx}
\usepackage{amsmath}
\usepackage{mathtools}
\usepackage{amssymb}
\usepackage{bm}
\usepackage{framed}
\usepackage{subfig}
\usepackage{hyperref}
\usepackage{nomencl}
\usepackage{color}
\usepackage{textcomp}
\usepackage{graphics}
\usepackage{epsfig}
\usepackage{epstopdf}
\usepackage{amsthm}
\usepackage{algorithm}
\usepackage{algorithmicx}
\usepackage{algpseudocode}
\usepackage{ulem}
\usepackage{lineno}
\usepackage{latexsym,amsfonts}
\usepackage{url}
\usepackage{longtable}
\usepackage[figuresright]{rotating}
\usepackage{listings}
\usepackage{footnote}
\usepackage{xcolor}
\usepackage{soul}
\usepackage{authblk} 
\usepackage{etoolbox}
\usepackage[bb=dsserif]{mathalpha}
\makeatletter
\newtheorem{theorem}{Theorem}

\newtheorem{proposition}[theorem]{Proposition}

\newcommand{\address}[1]{\gdef\@address{#1}}
\newcommand{\email}[1]{\gdef\@email{\url{#1}}}
\newcommand{\@endstuff}{\par\vspace{\baselineskip}\noindent\small
\begin{tabular}{@{}l}\scshape\@address\\\textit{E-mail address:} \@email\end{tabular}}
\AtEndDocument{\@endstuff}
\makeatother
\newcommand\myeq{\mathrel{\stackrel{\makebox[0pt]{\mbox{\normalfont\tiny def}}}{=}}}
\newcommand{\argmax}{\mathop{\mathrm{arg\,max}}}

\newcommand{\D}{\bf{D}}
\newcommand{\Dkl}{D_{\tiny{\mbox{KL}}}}
\newcommand{\prior}{\pi_{\tiny{\mbox{pr}}}}
\newcommand{\post}{\pi_{\tiny{\mbox{pos}}}}
\newcommand{\xs}{\bf{x}^{\tiny{\mbox{src}}}}
\newcommand{\xss}{{\bf{x}}_{r}^{\tiny{\mbox{src}}}}
\newcommand{\nbr}{\small{\mbox{nbr:}}}
\newcommand{\mprior}{\bm{\mu}_{\tiny{\mbox{pr}}}}
\newcommand{\mpost}{\bm{\mu}_{\tiny{\mbox{pos}}}}
\newcommand{\Sigprior}{\bm{\Sigma}_{\tiny{\mbox{pr}}}}
\newcommand{\Sigpost}{\bm{\Sigma}_{\tiny{\mbox{pos}}}}
\newcommand{\trace}{\hbox{Tr}}
\newcommand{\risk}{\hbox{Risk}}

\DeclareMathOperator{\Ig}{\mathcal{I}}
\DeclareMathOperator{\Ibar}{\bar{\mathcal{I}}}
\definecolor{tocorrect}{rgb}{0.97, 0.04, 0.56}
\newcounter{lnote}

\title{Greedy selection of optimal location of sensors for uncertainty reduction in seismic moment tensor inversion}
\author{Ben Mansour Dia$^{1,2,}$\thanks{corresponding author}, Michael Fehler$^{3}$, SanLinn I. Kaka$^{4}$, Andrea Scarinci$^{5}$, Umair bin Waheed$^{4}$, and Chen Gu$^{6}$}
\address{Authors affiliations\\
\\
$^{1}$Center for Integrative Petroleum Research, College of Petroleum Engineering and Geosciences, King Fahd University of \\Petroleum and Minerals, Dhahran 31261, Kingdom of Saudi Arabia\\
\\
$^{2}$Department of Mathematics, College of Computing and Mathematics, King Fahd University of \\Petroleum and Minerals, Dhahran 31261, Kingdom of Saudi Arabia\\
\email{ben.dia@kfupm.edu.sa}
\\
$^{3}$Department of Earth, Atmospheric, and Planetary Sciences, Massachusetts Institute of Technology, Cambridge, MA 02139 USA\\
\\
$^{4}$Department of Geosciences, College of Petroleum Engineering and Geosciences, King Fahd  University of Petroleum and \\Minerals, Dhahran 31261, Kingdom of Saudi Arabia\\
\\
$^{5}$Department of Aeronautics and Astronautics, Massachusetts Institute of Technology, Cambridge, MA 02139 USA\\
\\
$^{6}$Department of Civil Engineering, Tsinghua University, Beijing, 100084, China
\\}

\date{}

\begin{document}
\maketitle


\begin{abstract}
We address an optimal sensor placement problem through Bayesian experimental design for seismic full waveform inversion for the recovery of the associated moment tensor. The objective is that of optimally choosing the location of the sensors (stations) from which to collect the observed data. The Shannon expected information gain is used as the objective function to search for the optimal network of sensors. A closed form for such objective is available due to the linear structure of the forward problem, as well as the Gaussian modeling of the observational errors and prior distribution. 
The resulting problem being inherently combinatorial, a greedy algorithm is deployed to sequentially select the sensor locations that form the best network for learning the moment tensor. Numerical results  are presented and analyzed under several instances of the problem, including: use of full three-dimensional velocity-models, cases in which the earthquake-source location is unknown, as well as moment tensor inversion under model misspecification.
\end{abstract} 



\tableofcontents

\section{Introduction}
The seismic moment tensor (MT) has a broad range of applications in seismology on all scales including but not limited to providing information about the regional stress field, defining the seismic source in terms of size and type of faulting, basic information about the event magnitude, as well as identifying the orientation of fractures  in a reservoir. The MT can be estimated by linear inversion of observed waveforms assuming that the subsurface structure (particularly the waveform propagation velocities) and the event location are known. 
As such, approaches based on least-squares and least-squares-based methods have been commonplace in seismic inversion \cite{Sipkin1982, Sileny1992, Sileny1996}.  Correlation functions between waveforms have also been adopted \cite{Kong2021}. However, important challenges in MT inversion are associated to uncertainties in the velocity-model, event location,  azimuthal coverage as well as  low signal-to-noise ratio. 
In an effort to quantify the uncertainty in the estimated MT, several approaches have been addressed in a series of papers. In \cite{Weber2018},  a probabilistic method that infers simultaneously the MT, the hypocentral location, and the source time function is presented. Bayesian approaches to MT inversion have also been proposed in \cite{Mustac2015,Chen2017,Steinberg2021}. More recently, a robust approach using Bayesian inference of the MT with the misfit formulated using an optimal transport distance is discussed in \cite{Andrea2022} that helps mitigate the impact of imperfect representation of the subsurface in the velocity-model.

Our focus is on experimental design, that is, choosing the optimal sensor location to maximize the information extraction from the data. This of course also brings reduced costs in terms of the number of sensor that may be needed to achieve equally informative results. A comprehensive review of deterministic optimal experimental design is discussed in \cite{Atkinson1992} and applications to geophysical inversion problems in \cite{Curtis1999, Curtis2004, Thim2017, Xin2021, Guo2023, Sandhu2023}. \\

We focus on applying a Bayesian framework to optimal experimental design, specifically to the problem of moment tensor inversion, where the locations of a limited number of stations serve as the control variable of the amount of uncertainty shown by the MT posterior. More precisely, this consists in finding the optimal setup of stations to maximize the expectation of the Kullback–Leibler divergence \cite{kullback1959, kullback1951}, also referred to as \textit{the Shannon expected information gain} or \textit{the expected information gain} for short. The statement is that the further the posterior gets from the prior the greater the amount of information has been extracted through inversion.

Solving such problem poses a number of computational challenges, which we address in various ways. First, to ease the computation of the posterior (mean and covariance), we frame our inverse problem of learning the MT as linear-inverse Gaussian problem. This is possible given the linear relation between the seismic displacement and the Green's functions (impulse responses to differential operator of the elastic wave equation) and the choice of a (conjugate) Gaussian prior for our setup. Besides the computational challenges of the inverse problem, maximizing the expected information gain (EIG) translates a combinatorial optimization problem,  clearly prohibitive since the number of possible candidate subsets increases exponentially with the number of available stations. For this reason, we adopt a greedy algorithm, see \cite{Ranieri2014, Jiang2019, Mohan2021, Aretz2023, Strutz2023}, that sequentially selects the stations to form the target network. The greedy algorithm is heuristic in the sense that the resulting network of stations is near-optimal, but it provides an ordering structure of the selected subset of stations and it does so in a way that information provided by a station choice at a given step in the algorithm aggregates the marginal information from the previously selected stations. 

Given this setup, our numerical experiments start by assuming that the earthquake-source location is known and that the velocity-model has a known one-dimensional layered structure and we assess in several ways the relevance of using the designed network of sensors for inverting for the MT. We first investigate the evolution of the posterior uncertainty on the estimated MT as a function of the number of stations. We also investigate how the greedy-optimal network changes due the earthquake-source depth. We also address the problem of constructing the greedy-optimal network of stations, refereed to as the consensus greedy-optimal network, when the velocity-model is not known by optimizing the EIG over a set of plausible one-dimensional layered the velocity-models, estimated using representative well-logs from a three-dimensional synthetic velocity-model. We then repeat a similar test by optimizing over a set of earthquake-source locations representing the spread of possible locations expected during an induced seismicity study. Numerical results for our analyses highlight the robustness of our proposed method.
\\

The structure of the paper is as follows.  Section \ref{sec:model} presents the forward model (seismic wave propagation) formulation. In Section \ref{sec:bayesian}, we set up the Bayesian approach for inverting the MT. Section \ref{sec:design} introduces the optimal sensor placement as the combinatorial maximization of the EIG, followed by the greedy approach that provides the near-optimum network of stations referred to as the greedy-optimal network of stations. Numerical experiments in this section mainly aim at testing the setup under nominal circumstances. Section \ref{sec:cons} tests show instead the behavior of the proposed OED design under circumstances of unknown velocity-models, an unknown earthquake-source location, and misspecified data.

\section{Forward modeling}\label{sec:model}
We denote by $\D \subset \mathbb{R}^3$ a heterogeneous Earth medium and $T$ a positive real number  standing for the study duration. We define the space-time seismic wave displacement field induced by an earthquake occurring at time $t_0 \in [0,T]$ at the location $\xs$ in $\D$ as $\bm{u}=(u_1, u_2, u_3)$. We also assume it is governed by the elastic wave equation with the velocity-model $\bm{V} = \{ V_p, V_s\}$
\begin{eqnarray*}
 V_p = \sqrt{\frac{\lambda + 2 \mu}{\rho}} \quad \hbox{and} \quad  V_s = \sqrt{\frac{\mu}{\rho}},
\end{eqnarray*}
where 
$\lambda$ and $\mu$ are the first and second Lam\'e parameters and together constitute a parametrization of the elastic moduli for a homogeneous isotropic medium, see \cite{mavko_mukerji_dvorkin_2009}.
The triplet $(\rho, \lambda, \mu)$ or $(\rho, V_p, V_s)$ defines the Earth's material properties that vary with spatial position in a general velocity-model $\bm{V} = \{ V_p, V_s\}$. 

The inversion of the full MT implies comparing the simulated waveforms to those recorded by a seismograph at a location $\bm{x}_r$, given the geophysical properties of the medium. For $i=1,2,3$,  $u_i$ is the displacement in the $i$-{th} direction, while $G_i$ is the $i$-{th} component of the elastodynamics Green's function   and $\bm{M}$ the MT. We have:  
\begin{eqnarray*}
u_i(t,\bm{x}_r) = \sum_{k,l=1}^3  G_{{i}_{kl}}\bm{M}_{_{kl}} * s(t),
\end{eqnarray*} 
where $*$ is the temporal convolution operator, and $s(t)$ is the source time function. 
The MT  $\bm{M}$  and the Green's function component $G_i$ are $3\times 3$ symmetric matrices given by
\begin{eqnarray*}
\bm{M} = \left[\begin{array}{ccc}
M_{_{11}}   & M_{_{12}}  & M_{_{13}}\\
M_{_{12}}   & M_{_{22}}  & M_{_{23}}\\
M_{_{13}}   & M_{_{23}}  & M_{_{33}}
\end{array}\right] \quad \hbox{and} \quad
G_i = \left[\begin{array}{ccc} 
g_{_{i_1}}   & g_{_{i_4}}  & g_{_{i_5}}\\
g_{_{i_4}}   & g_{_{i_2}}  & g_{_{i_6}}\\
g_{_{i_5}}   & g_{_{i_6}}  & g_{_{i_3}}\\
\end{array}\right].
\end{eqnarray*} 
To simplify the notation, we consider the vectorized form of the upper triangular portion of $G_i$ and $\bm{M}$ and represent it with vectors $\bm{g}_i$ and $\bm{m}$ of six elementary components, respectively,  $\bm{g}_i = \left(g_{i_1}, g_{i_2}, \ldots, g_{i_6}\right)$ and $\bm{m} = \left(m_{_1}, m_{_2}, \ldots, m_{_6}\right)$, where $m_{_1} = M_{_{11}}$, $m_{_2} = M_{_{22}}$, $m_{_3} = M_{_{33}}$, $m_{_4} = M_{_{12}}$, $m_{_5} = M_{_{13}}$, and $m_{_6} = M_{_{23}}$. Therefore, the overall displacement $\bm{u}$ is given by
\begin{eqnarray}
\label{eq:u}
  \bm{u}(t,\bm{x}_r) = \mathcal{G}(t,\bm{x}_r;t_0, {\xs}, \bm{V}) \cdot \bm{m}^\top *s(t), 
\end{eqnarray}
where the matrix $\mathcal{G}$ is obtained from vertical concatenation of $\bm{g}_i$, $i=1,2,3$, $\mathcal{G} = [\bm{g}_1, \bm{g}_2, \bm{g}_3]^\top$.

Let ${n_{_t}}$ be the number of time-points of the signal, the discrete form $U$ of the displacement $\bm{u}$ at the location $\bm{x}_r$ is a time-series concatenating the three unidirectional waveforms, 
\begin{eqnarray}
\label{eq:forward}
   U(\bm{x}_r)  = \bm{G}(\bm{x}_r;t_0, {\xs}, \bm{V})\cdot \bm{m}^\top,
\end{eqnarray}
where the matrix $\bm{G}$ is the discrete form of the Green's function with dimension $(3n_{_t},\; 6)$. Calculating the matrix $\bm{G}$ requires coordinates of the earthquake-source location $\xs$, coordinates of the station $\bm{x}_r$, and characteristics (P-waves velocity $V_p$, S-waves velocity $V_s$, and the density $\rho$) of the medium. 

In this study, we assume the velocity model to be a one-dimensional layered model with a fixed number  of layers, each being homogeneous in terms of P-wave and S-wave velocities as well as density. To compute the forward simulations, we use Axitra, which is based on the  reflectivity method to approximate Green's functions \cite{Bouchon1980, Bouchon1981, Coutant1997}.

\section{Bayesian formulation}\label{sec:bayesian}

\subsection{Prior distribution}
The Bayesian approach of inverting for the seismic MT consists in updating a prior knowledge that is characterized by a probability density function (PDF) to the posterior one. There is more than one way to characterize that prior belief about the seismic MT. For example, an improper prior is set in \cite{Chen2017, Mustac2015} to bypass the problem of choosing any specific prior belief on the parameter space. An advanced setting of the prior is presented in \cite{Stahler2014}, incorporating two auxiliary parameters both following the uniform distribution in the unit ball. This setting encompasses also the source time function and the isotropic component of the MT.
Another common way to characterize that prior belief is to normalize each of its components $m_i$ over the interval $[-1,1]$ and to assign to each component an uniform prior distribution $m_i \sim \mathcal{U}([-1,1])$. That approach signifies that  all the trial values of each component of the MT have equal weight in the interval $[-1,1]$, before considering data.  

While we retain the nomalization of the MT components to the interval $[-1,1]$, in our optimal experimental design context, we choose instead a Gaussian centered prior distribution $\mathcal{N}(0,\sigma^2_{_p})$ for each component of the parameter-of-interest. This allows us to frame the problem as a conjugate-Gaussian inverse problem. The standard deviation $\sigma_{_p}$ is chosen such that realizations from $\mathcal{N}(0,\sigma^2_{_p})$ mostly fall in the interval $[-1,1]$. This choice has to be performed accurately: for a large value of $\sigma_{_p}$, realizations from $\mathcal{N}(0,\sigma^2_{_p})$ would likely  fall outside that interval, while too small values of $\sigma_{_p}$ would lead instead to a prior distribution $\mathcal{N}(0,\sigma^2_{_p})$ artificially shrunk around $0$. In this last scenario, the numbers close to -1 and 1 from the interior of the interval $[-1,1]$ are excluded and the inversion of MT can be underestimated. 

\subsection{Data representation}
When dealing with observed seismograms (waveforms), as with any other data, it  is important to model observational errors associated with the measurement process and instruments. The characterization of such noise is often linked to the choice of misfit function and impacts as well the choice of  likelihood function. Beside the classic $\ell_2$ norm choice deriving from the assumption of observational Gaussian noise, and thus Gaussian likelihood, sever other choices have emerged recently. For its flexibility in comparing waveform-type data such as in seismic inversion, the Wasserstein metric introduced in \cite{Engquist2014, Engquist2016} has become more popular \cite{Chen2018, Dia2019, Motamed2019, Yang2016}. 

In our paper, since the focus is on optimal experimental design we retain the classic additive Gaussian noise formulation with Gaussian likelihood given by:
\begin{equation}
\label{eq:datamodel}
\bm{y}(\bm{x}_r) =  \bm{G}(\bm{x}_r)\cdot \bm{m}_t^\top + \bm{\epsilon},
\end{equation}
where the measurement errors $\bm{\epsilon}$ are independent and identically distributed (i.i.d.) zero-mean Gaussian random variables with covariance matrix $\bm{\Sigma_{\epsilon}}$, that is, $\bm{\epsilon} \sim \mathcal{N}\left(0, \bm{\Sigma_{\epsilon}}\right)$ and $\bm{m}_t$ is the "\textit{true}" value of the MT. We choose the following values $m_{_1}= 0.269$, $m_{_2}= 0.700$, $m_{_3}=-0.969$, $m_{_4}=-0.454$, $m_{_5}=-0.195$, and $m_{_6}= 0.0592$ for the true MT $\bm{m}_t$ for our numerical experiments.

The most conventional choice for the covariance matrix $\bm{\Sigma_{\epsilon}}$ is the scaled identity, which is equivalent to  assuming an uncorrelated noise structure among the time-point values of a waveform. That formulation however does not take into account of the most prominent features of time-series features i.e. the fact the various data-points are correlated in some way.  The analysis in \cite{Yagi2011} shows, in fact, that the conventional diagonal structure of the covariance matrix is not enough to cover uncertainties in the Green's function. Another issue associated with this type of covariance structure is the numerical stability induced by its condition number \cite{Ababou1994}. We therefore choose the Mat\'{e}rn covariance form given by
\begin{eqnarray}
\label{eq:cov}
  (\bm{\Sigma_{\epsilon}})_{_{ij}} = \sigma_{_\varepsilon}^2 \exp\left( - \frac{|t_i - t_j|}{T}\right),
\end{eqnarray}
where $\sigma_{_\varepsilon}$ is the noise deviation factor, and $t_i$ and $t_j$ are the $i$-th and $j$-th time-points. From \eqref{eq:cov}, each time-point value of the waveform is fully correlated to itself and the correlation with any other time-point value of the waveform drops exponentially with the difference in time. The correlation band limit is given by $\sigma_{_\varepsilon}$. An empirical approach proposed in \cite{Mustac2015}  using a number of waveforms obtained in a separate event  to estimate $\sigma_{_\varepsilon}$. Instead, we estimate $\sigma_{_\varepsilon}$ for each specific station $\bm{x}_r$ with a relative error bound, between the synthetic noise-free waveform and the observed waveform, with a 10 percent amount of noise in the $L^2$-norm sense. The noise generation is a random Gaussian stationary process. 
\begin{figure}
  \centering
  \includegraphics[width=1\textwidth]{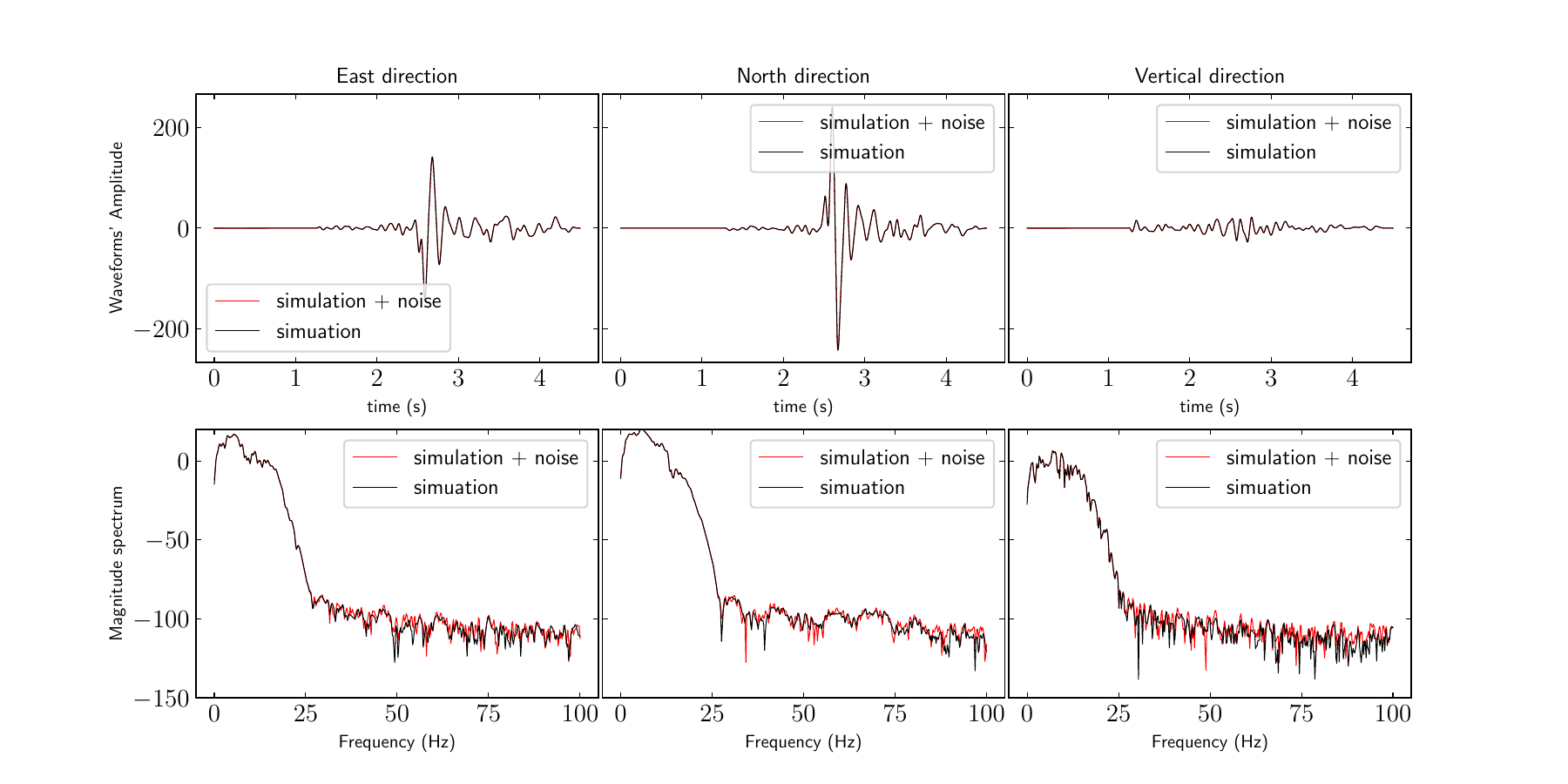}
  \caption{Noise-free and perturbed waveforms in time and frequency domains.}
  \label{fig:waveform}
\end{figure}

Figure \ref{fig:waveform} shows waveforms from simulations and the noise-added versions both in time and frequency domains. One important aspect when adding noise to simulated outputs with additive measurement errors is to make sure that the resulting spectrum of the perturbed signal does not overpass the band limit of the simulated waveforms. The purpose of choosing the standard deviation with majoring the relative error  at a rate of $0.1$ is to ensure that, as illustrated in Figure \ref{fig:waveform}.

\subsection{Conjugate-Gaussian Bayesian inverse problem}
The \textit{true} vector value $\bm{m}_t$, of the seismic MT  is viewed as a random variable denoted by $\bm{m} \in \Theta \subseteq \mathbb{R}^d$, with $d=6$. The prior PDF $\prior(\bm{m})$ of the MT stands for our prior belief in it, and is characterized  by the Gaussian distribution of mean $\mprior$ and covariance matrix $\Sigprior$, that is,  $\prior(\bm{m}) =  \mathcal{N}\left(\mprior,\Sigprior \right)$. According to the data noise representation in \eqref{eq:datamodel}, the likelihood function is in the form
 \begin{eqnarray}
\label{eq:likelihood}
 p(\bm{y}(\bm{x}_r) \vert \bm{m}) =  \det\left(2\pi\bm{\Sigma_\epsilon} \right)^{-\frac{1}{2}} \exp \left( -\frac{1}{2} \left\| \bm{G}(\bm{x}_r)\cdot\bm{m}^\top - \bm{y}(\bm{x}_r) \right\|^2_{\bm{\Sigma_\epsilon}^{-1}} \right),
\end{eqnarray}
where the matrix norm is given by $ \|\bm{X}\|^{2}_{\bm{\Sigma_\epsilon}^{-1}} = \bm{X}^\top \bm{\Sigma_\epsilon}^{-1} \bm{X}$ for a vector $\bm{X}$ and the covariance matrix $\bm{\Sigma_\epsilon}$. The posterior PDF 
$\post(\bm{m} \vert \bm{y}(\bm{x}_r))$ of the MT is given by Bayes' rule as follows
\begin{eqnarray}
\label{eq:bayes}
 \post(\bm{m} \vert \bm{y}(\bm{x}_r)) = \frac{p(\bm{y}(\bm{x}_r) \vert \bm{m}) \prior(\bm{m})}{p(\bm{y}(\bm{x}_r))},
\end{eqnarray}
where the normalizing constant $p(\bm{y}(\bm{x}_r))$ describes the distribution of the data collected at the location $\bm{x}_r$. The linearity of \eqref{eq:forward} with respect to the parameter-of-interest  $\bm{m}$ together with the Gaussian prior distribution and Gaussian centered measurements error give us a conjugate-Gaussian inverse problem. As a result, the posterior distribution $\post(\bm{m} \vert \bm{y}(\bm{x}_r))$ is also Gaussian with mean and covariance denoted by  $\mpost$ and $\Sigpost$, respectively. The closed form the posterior PDF $\post(\bm{m} \vert \bm{y}(\bm{x}_r)) = \mathcal{N}\left(\mpost,\Sigpost \right)$ avoids dealing with computational issues when sampling from it and its first- and second-order moments, according to \cite{Attia2018}, are given by
\begin{eqnarray}
\label{eq:post_moments}
\mpost = \Sigpost \left(\Sigprior^{-1} \mprior + \bm{G}^\top(\bm{x}_r)\bm{\Sigma}^{-1}_{\epsilon}(\bm{x}_r)\bm{y}(\bm{x}_r)\right) \quad \hbox{and} \quad
\Sigpost = \left(\bm{G}^\top(\bm{x}_r)\bm{\Sigma}^{-1}_{\epsilon}(\bm{x}_r)\bm{G}(\bm{x}_r) + \Sigprior^{-1} \right)^{-1}.
\end{eqnarray}
Note that the Green matrix $\bm{G}$, the covariance matrices of the prior $\Sigprior$ and of the data noise $\bm{\Sigma}^{-1}_{\epsilon}$ suffice to obtain the posterior covariance matrix while the posterior mean $\mpost$ is conditioned with the realization $\bm{y}(\bm{x}_r)$ of the data. The more the data $\bm{y}(\bm{x}_r)$ is informative, the less is the uncertainty in the posterior mean. In the next section, we detail our approach on selecting the most informative data. 

\section{Design optimization}\label{sec:design}
In this section, we set the location of a fixed number of sensors as the control variable to optimize the Bayesian inversion for the MT. 

\subsection{Optimal sensor placement problem}
\begin{figure}
  \centering
  \includegraphics[width=.6\textwidth]{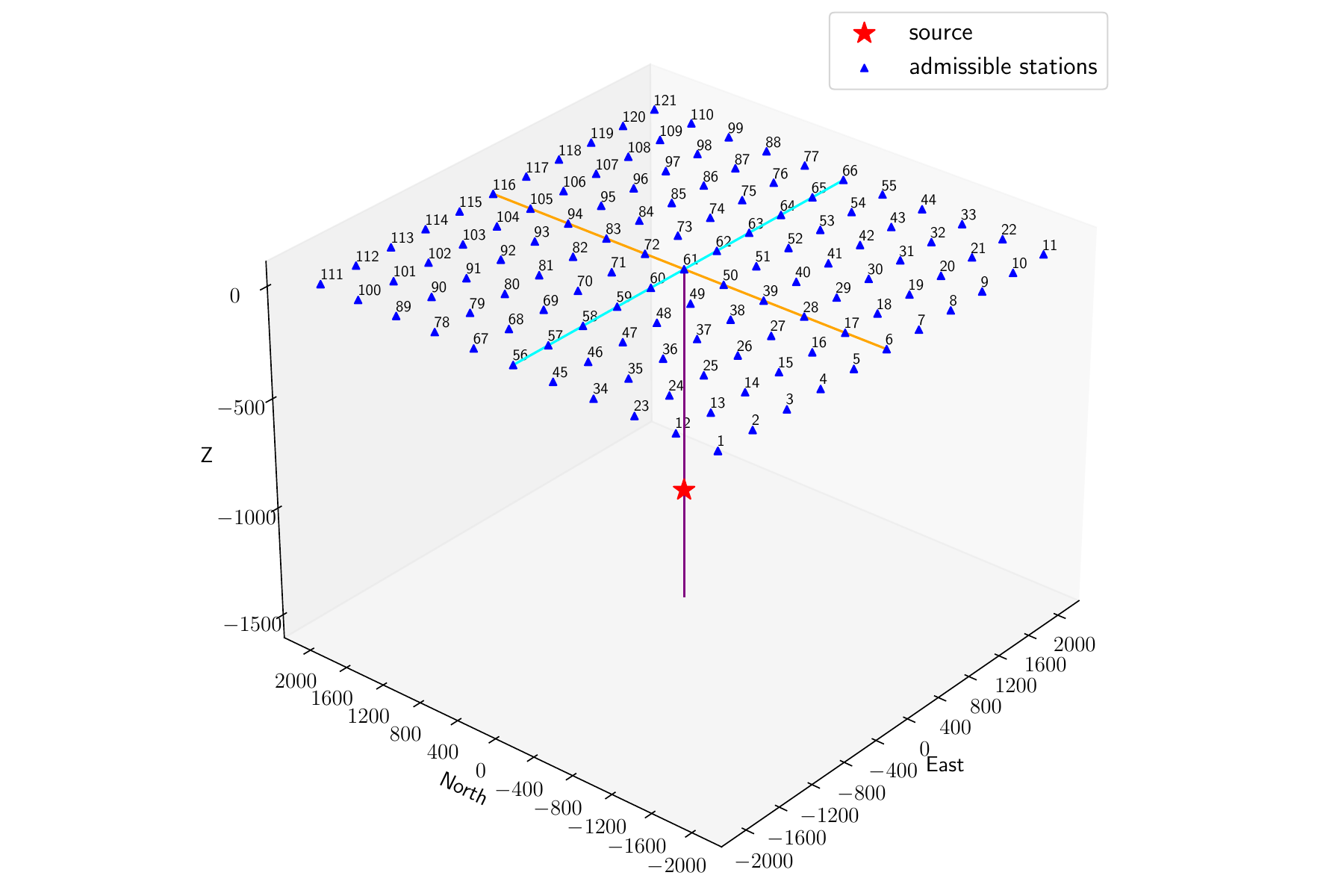}
  \caption{Physical domain $\D$ 4Km$\times$4Km$\times$1.5Km with a grid of $n=11\times 11$ candidate receiver locations and an earthquake-source at 1Km depth.}
  \label{fig:domain}
\end{figure}

Our configuration is such that  $\D$ is a rectangular parallelepiped volume  standing for a region of the Earth, the earthquake-source is located at ${\xs} = (0, 0, x_{3}^{\tiny{\hbox{src}}})$ in a three-dimensional space (East, North, Depth),  and  the propagation velocity field is a one-dimensional layered structure or can be represented by a set of one-dimensional layered velocity-models. The free-surface of the domain $\D$ (representing the Earth's surface) is meshed with  a uniform grid of $n$ possible locations for receivers denoted by  $\bm{x}_{_1}, \bm{x}_{_2}, \ldots, \bm{x}_{_n}$ as illustrated in Figure \ref{fig:domain}.
We intend to use observations from only $k$ stations among $n$. In what follows, we denote by $\mathbb{N}_n$ the set of integers from $1$ to $n$, $\mathbb{N}_n = \{1, 2, \ldots, n\}$, by $\mathcal{P}(\mathbb{N}_n)$ the power set of $\mathbb{N}_n$. A subset of indexes of $k$ stations is denoted by $\mathcal{S}_k$ for $k=1, \ldots, n$. 

The cardinality $k$ (number of stations) of the number of stations chosen can also be part of the optimization problem, together with the location of the stations themselves. That means one would ideally look first for the best subset among those with cardinality $k$ and then look for the best one among those subsets. 
This formulation is computationally prohibitive since it scales as $2^n$. Another disadvantage of this approach is that one needs to beforehand investigate the submodularity (supermodular or submodular) of the objective set function to be maximized in order to figure out the computational complexity of related algorithms \cite{Mohan2021}.  We adopt the constrained cardinality, that is, we require exactly $k$ stations and we refer to \cite{Krause2014, Badanidiyuru2020} for more details in treating $k$ as part of the problem unknowns.
 
We denote by $\mathcal{S}^{\tiny{\hbox{opt}}}_k$ the optimal subset of stations refereed as the best network of stations.  Given the constraints in the cardinality $k$, there are $\frac{n !}{k!(n-k)!}$ candidate subsets. Each candidate subset $\mathcal{S}_k$ comes with a different level of uncertainty in the seismic MT. 
As each network $\mathcal{S}_k$  is associated with an index set  $\{\hbox{i}_1, \hbox{i}_2, \ldots, \hbox{i}_k\}$, we treat the design of $\mathcal{S}^{\tiny{\hbox{opt}}}_k$ as the problem of choosing the indexes $\hbox{i}_l$ of its elements, $l=1, \ldots,k$ among the  $n$ available.  In the next section, we formulate the search of $\mathcal{S}^{\tiny{\hbox{opt}}}_k$ as a Bayesian optimal experimental design problem with an associated objective function.

\subsection{Expected information gain}
We choose the Kullback-Leibler divergence, denoted by $\Dkl$, see \cite{key13, key12}, as the objective function to measure the amount of information provided by the data. It is an entropy function measuring the uncertainty contraction in $\bm{m}$ between the prior $\prior(\bm{m})$ and the posterior $\post(\bm{m} \vert \bm{Y}, \mathcal{S}_k)$ and is given by
\begin{equation}
\label{eq:dkl1}
\Dkl \left(\post(\bm{m}|\bm{Y}, \mathcal{S}_k) || \prior(\bm{m}) \right) = \int_{\Theta}{ \log \left( \frac{\post(\bm{m} \vert \bm{Y}, \mathcal{S}_k)}{\prior(\bm{m})} \right)  \hbox{d} \post(\bm{m} \vert \bm{Y}, \mathcal{S}_k)},
\end{equation}
where $\bm{Y}$ represents the collection of the individual observed waveforms at the $k$ stations of the network $\mathcal{S}_k$ i.e. $\bm{Y} = \left[\bm{y}(\bm{x}_{\hbox{i}_1}), \bm{y}(\bm{x}_{\hbox{i}_2}), \ldots, \bm{y}(\bm{x}_{\hbox{i}_k})\right]$ and $\Theta \subseteq \mathbb{R}^6$ is the set of possible values for the MT. The larger the value of $\Dkl$ is, the more informative the data $\bm{Y}$ is about $\bm{m}$. To generalize this concept to any possible data realization,  the expectation of $\Dkl$ is taken over the sample space $\mathcal{Y}^k$, with $\mathcal{Y} \subseteq \mathbb{R}^{q}$, and referred to as the EIG:
\begin{align}
\Ig (\mathcal{S}_k) = & \int_{\mathcal{Y}}{\int_{\Theta}\Dkl \left(\post(\bm{m}|\bm{Y}, \mathcal{S}_k) || \prior(\bm{m}) \right) \hbox{d}\bm{Y}} \nonumber\\
                              = & \int_{\Theta} \int_{\mathcal{Y}} \log \left( \frac{p(\bm{Y} \vert \bm{m}, \mathcal{S}_k)}{p(\bm{Y}| \mathcal{S}_k)} \right)  p(\bm{Y} \vert \bm{m}, \mathcal{S}_k) \hbox{d} \bm{Y} \prior(\bm{m}) \hbox{d} \bm{m}. \label{eq:expecinfgain}
\end{align}
The latter equality in \eqref{eq:expecinfgain} follows from  Bayes' rule \eqref{eq:bayes}. According to the likelihood function definition in \eqref{eq:likelihood}  using the waveforms from an individual station, we have 
\begin{eqnarray}
 p(\bm{Y} \vert \bm{m}, \mathcal{S}_k) &=& \prod_{l=1}^k  p(\bm{y}(\bm{x}_{\hbox{i}_l}) \vert \bm{m})\nonumber\\
\label{eq:likelihood2} & = &  \left(2\pi\right)^{-\frac{3kq}{2}} \prod_{l=1}^k\det\left(\bm{\Sigma_\epsilon} (\bm{x}_{\hbox{i}_l})\right)^{-\frac{1}{2}} \exp \left( -\frac{1}{2}  \sum_{l=1}^k \left\| \bm{y}(\bm{x}_{\hbox{i}_l}) - \bm{G}(\bm{x}_{\hbox{i}_l})\cdot \bm{m}^\top\right\|^2_{\bm{\Sigma_\epsilon}^{-1}(\bm{x}_{\hbox{i}_l})} \right), \nonumber
\end{eqnarray}
where $\bm{\Sigma_\epsilon}(\bm{x}_{\hbox{i}_l})$ is  the covariance matrix of the measurement noise at the station $\bm{x}_{\hbox{i}_l}$. In general, the evaluation of $\Ig$ is not possible in closed form or it would require a large number of evaluations of the forward model for Monte Carlo sampling: the nested Monte Carlo sampling, known as double loop Monte Carlo method, was introduced in \cite{Ryan2003} and consists in deploying the Monte Carlo approach twice at \eqref{eq:expecinfgain}. Subsequent enhancements have been achieved through polynomial chaos expansions \cite{HuanMarzouk2013}, using Laplace-based importance sampling \cite{Ryan2015, BDELT2018}, and based on different mesh sizes for the physical discretization \cite{Beck2020, Goda2020}.

In our case, however,  the linearity of the forward model together with the Gaussianity of the prior PDF and of the likelihood function  allow us to derive a closed form of the EIG,
\begin{eqnarray}
\label{eq:eig_network}
\Ig (\mathcal{S}_k) &=&  \frac{1}{2} \log \det \left(\bm{G}^\top(\mathcal{S}_k)\bm{\Sigma}^{-1}_{\epsilon}(\mathcal{S}_k)\bm{G}(\mathcal{S}_k)\Sigprior + \bm{I}_d \right),
\end{eqnarray}
where $\bm{I}_d$ is the $d$-dimension identity matrix, the network Green's matrix $\bm{G}(\mathcal{S}_k)$ is obtained by vertical concatenation  of the individual Green's matrices $\bm{G}(\bm{x}_{\hbox{i}_l})$, $l = 1, \ldots,k$ and the network covariance matrix $\bm{\Sigma}_{\epsilon}(\mathcal{S}_k)$ is the diagonal concatenation of the individual covariance matrices $\bm{\Sigma}^{-1}_{\epsilon}(\bm{x}_{\hbox{i}_l})$, $l = 1, \ldots,k$,
\begin{eqnarray}
\label{eq:Green_network}
 \bm{G}(\mathcal{S}_k) = \left[\bm{G}(\bm{x}_{\hbox{i}_1}) \vert \bm{G}(\bm{x}_{\hbox{i}_2}) \vert \ldots \vert \bm{G}(\bm{x}_{\hbox{i}_k})\right]^\top
 \quad \hbox{and} \quad
 \bm{\Sigma}_{\epsilon}(\mathcal{S}_k) = \hbox{diag} \left(\bm{\Sigma}^{-1}_{\epsilon}(\bm{x}_{\hbox{i}_1}), \bm{\Sigma}^{-1}_{\epsilon}(\bm{x}_{\hbox{i}_2}), \ldots, \bm{\Sigma}^{-1}_{\epsilon}(\bm{x}_{\hbox{i}_k})\right). 
\end{eqnarray}
In the next section, we discuss the problem of the direct maximization of $\Ig (\mathcal{S}_k)$.

\subsection{Greedy optimization algorithm}
The objective of our approach is to select the best subset $\mathcal{S}^{\tiny{\hbox{opt}}}_k$ as an optimal Bayesian experimental design problem. The best subset of stations $\mathcal{S}^{\tiny{\hbox{opt}}}_k$  solves the following optimization problem
\begin{eqnarray}
\label{pr:com_op}
  \mathcal{S}^{\tiny{\hbox{opt}}}_k =  \argmax_{\substack{\mathcal{S}_k \in \mathcal{P}(\mathbb{N}_{n}) \\\\ |\mathcal{S}_k| = k}} \Ig (\mathcal{S}_k).
\end{eqnarray} 
Equation \eqref{pr:com_op} is  a combinatorial optimization problem \cite{Blum2001, Korte2012}.  In fact, its solution requires proving that the EIG for the particular subset $\mathcal{S}^{\tiny{\hbox{opt}}}_k$ is higher than the one of any other candidate subset $\mathcal{S}_k$ of cardinality $k$.  Thereupon, we adopt a greedy algorithm that sequentially selects the index element of the optimal subset in a similar fashion to the batch greedy maximization method presented in \cite{Mohan2021} (with batch size equals to one). 
\begin{algorithm}[H]
\caption{Sequential selection of $k$ best stations in a grid of $n$ stations.} \label{algo:seq}
\begin{algorithmic}[1]
  \State Set $\mathcal{S} = \mathbb{N}_n$
  \State Parameterize the data noise through $\sigma_\epsilon$ for all candidate stations
  \State Initialize $\mathcal{S}^{\tiny{\hbox{opt}}}_k = \emptyset$
  \State Set the prior PDF: $\mprior =0$ and $\Sigprior = \sigma_p^2 \bm{I}_d$
  \For{\texttt{$l = 1$ to $k$}}
      \State $\hbox{i}_l = \argmax_{\substack{j \in \mathcal{S}}} \Ig(\{\hat{\bm{x}}_j\})$     \label{setp6}
      \State $\mathcal{S}^{\tiny{\hbox{opt}}}_k \gets \mathcal{S}^{\tiny{\hbox{opt}}}_k \cup \{\hbox{i}_l\}$
      \State $\mathcal{S} \gets \mathcal{S}\backslash \{\hbox{i}_l\}$
      \State Compute the covariance matrix $\Sigpost$ of the posterior PDF according to \eqref{eq:post_moments}
      \State Update the prior moment:  $\Sigprior \gets \Sigpost$ \label{step10}
  \EndFor
  \State \textbf{return} index set $\mathcal{S}^{\tiny{\hbox{opt}}}_k$
  \end{algorithmic}
\end{algorithm}

Because the EIG is expressed only with the covariance matrices of the prior and posterior distributions, from one iteration to another we change the prior by updating only the covariance matrix while we reset the prior mean to $0$, see step \ref{step10} in  Algorithm \ref{algo:seq}.  In fact the posterior mean expression at \eqref{eq:post_moments} includes also the data $\bm{y}$, hence updating the prior mean with the posterior mean  revokes the validity of \eqref{eq:eig_network}, for the following iterations, that is derived with the assumption that the prior mean $\mprior$ does not depend on the data $\bm{y}$.

\subsection{Setting of the forward simulations}\label{sec:setting_forward}
We detail the setup for performing the numerical experiments. We consider a physical domain $\bm{D}$ = [-4000 m, 4000 m]$\times$[-4000 m, 4000 m]$\times$ [0, -4000 m] with an earthquake-source depth at $x_{3}^{\tiny{\hbox{src}}}$ = 2000 m. The Earth's surface is meshed with a total of $n=25, 921$ candidate station locations. The candidate sensor locations form a grid of $161$ stations in each horizontal direction with a spacing of 50 m.
\begin{figure}
  \centering
  \includegraphics[height=7.5cm, width=4cm]{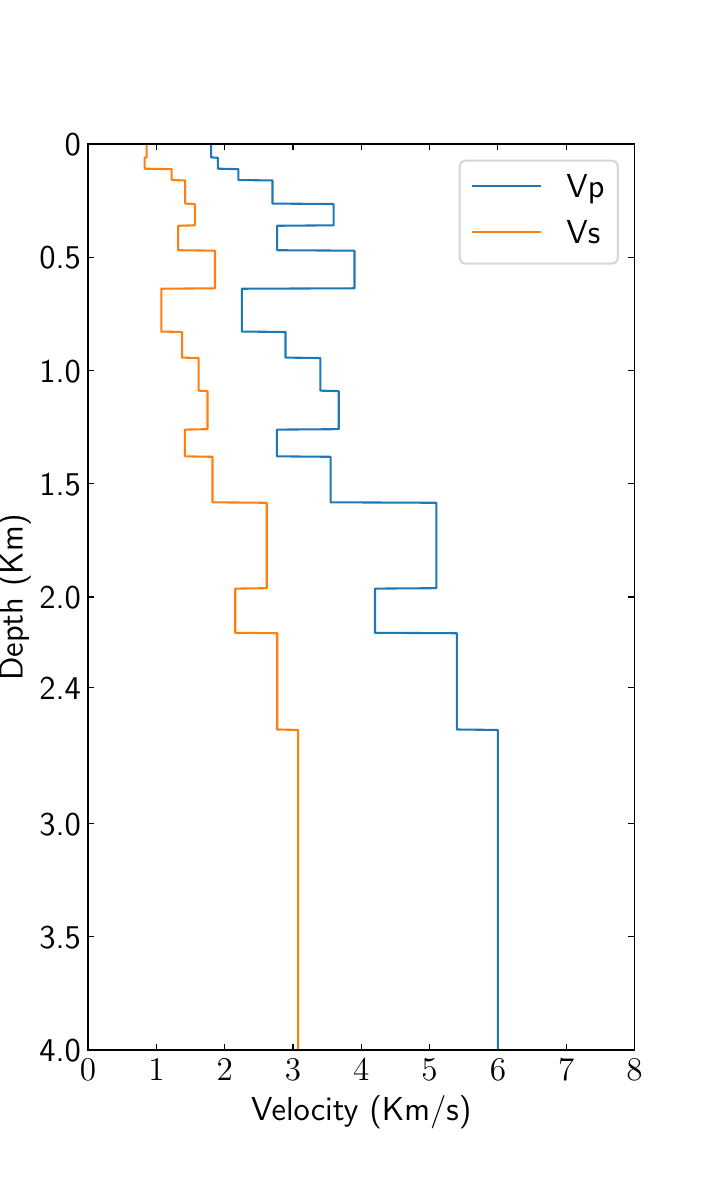}\hspace{-0.5cm}
  \includegraphics[height=7.5cm, width=13cm]{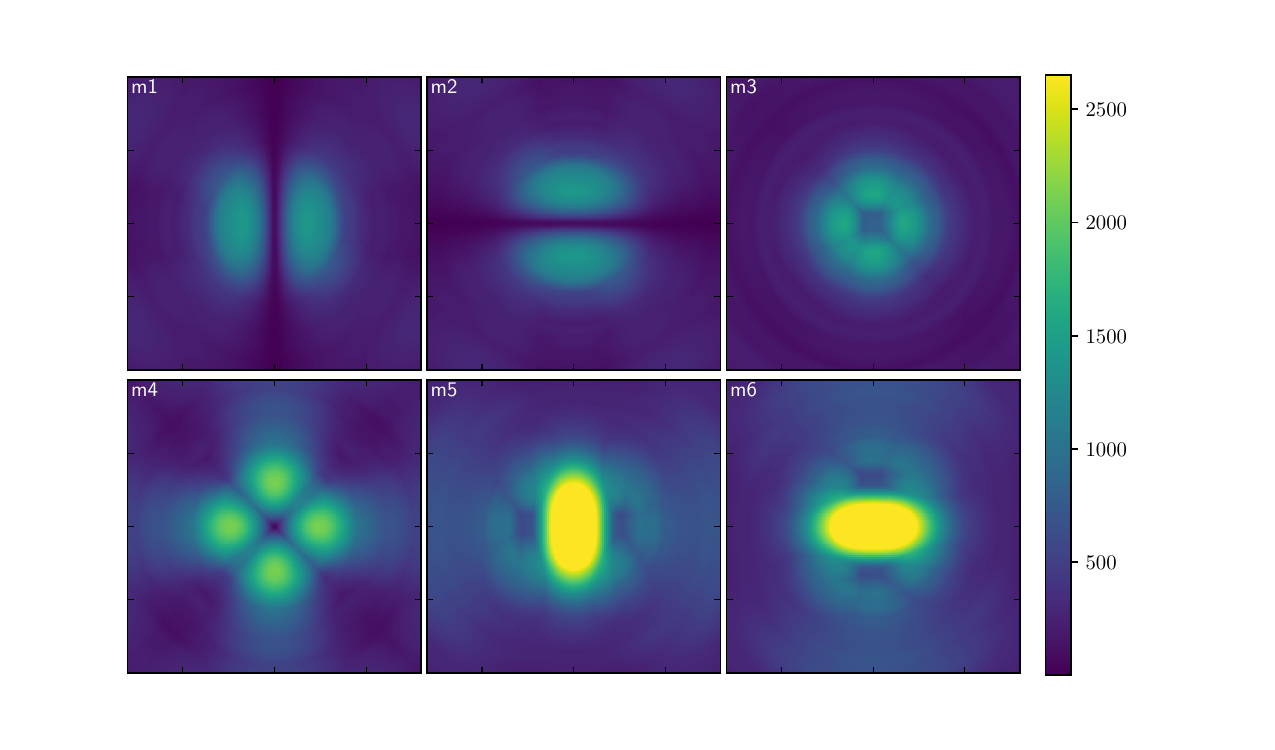}
  \caption{Left: the layered velocity-model from \cite{Chen2017}. Right: the maximum amplitude field of the Green's function at the surface with the source located at $\xs$ = (0, 0, -2000 m), where each panel corresponds to an component of the moment tensor.}
  \label{fig:vel_model_test1}
\end{figure}

Figure \ref{fig:vel_model_test1} shows at the left the one-dimensional layered profile of P-wave ($V_p$) and the S-wave ($V_s$) velocity-models for a three-dimensional layered medium, as treated in \cite{Chen2017}, and at the right the propulsion patterns described the maximum amplitude in the $L^2$-norm of the waveforms from the six components of the MT for an earthquake occurring at ${\xs}$. 

In the simulations, we choose a causal source time history that is band limited to below about 15Hz. While the P- and S-wave velocities vary from layer-to-layer, we use a constant density of 2000 km/m$^3$ and a constant attenuation factor Q of 1000. The high value of Q was chosen have minimal impact on the character of the waveforms within the frequency band of interest. The duration of the simulation is $4.5$ seconds and the time-step for the waveforms is $5.10^{-3}$second. In the simulation, we applied a bandpass filter  to both the Axitra waveform and the source time function before convolution. The filtering leads to a smoother variation of the amplitude with receiver position. 
The simulation output is the waveform $U$. To obtain the Green matrix $\bm{G}$ from \eqref{eq:forward}, we perform six runs, where the MT is set to the single entry vector $e_i$, for $i=1,\ldots,6$: all entries are $0$ except the $i$-th entry, which is equal to $1$. The resulting waveform vector $U$ is the $i$-th column of the Green matrix.

\subsection{Geometry of the greedy-optimal network of stations}
\begin{figure}
  \centering
  \includegraphics[width=0.8\textwidth]{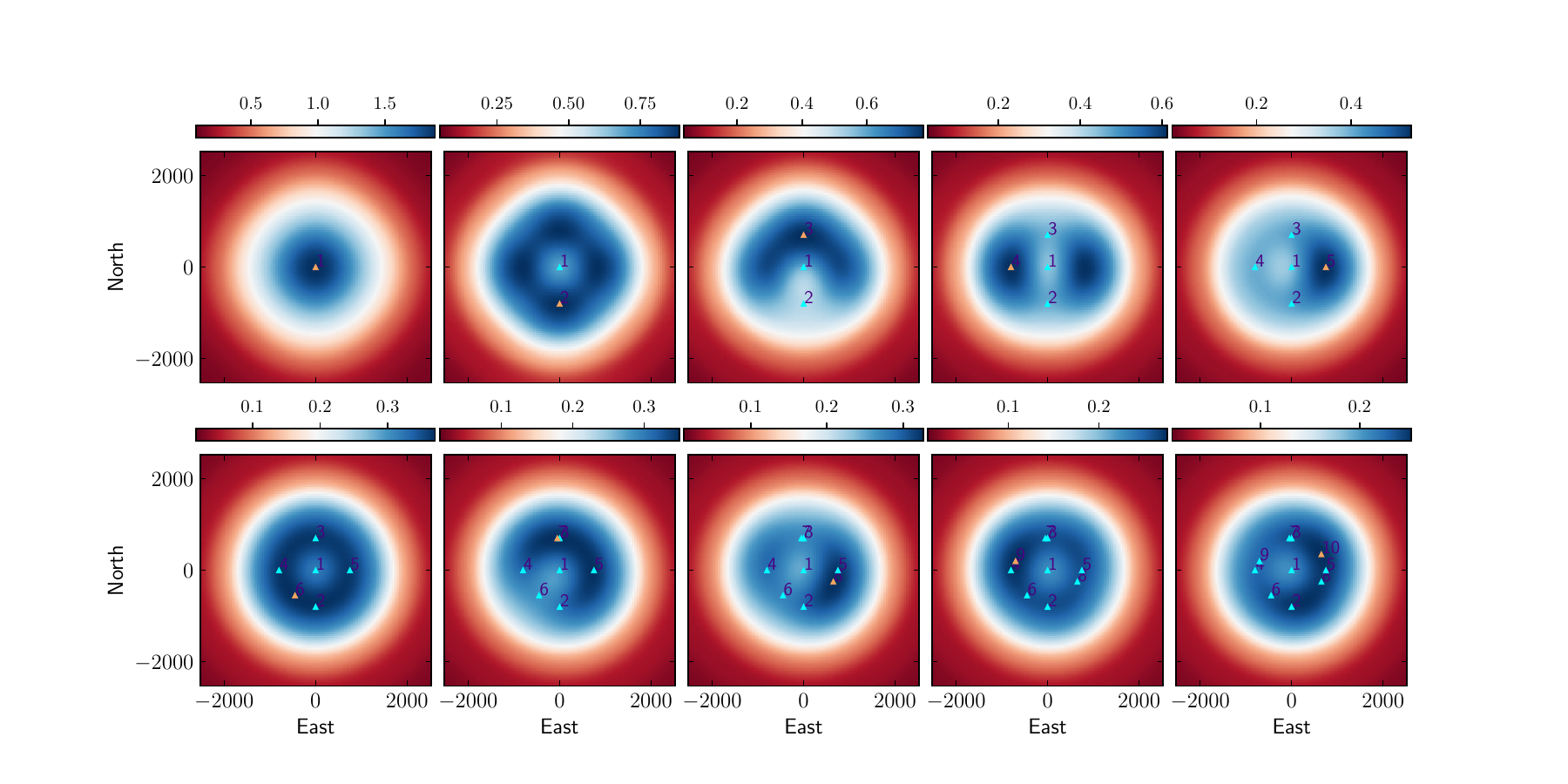}
  \caption{Expected information gain surfaces from selecting the best 10th stations}
  \label{fig:optimal_network_test1}
\end{figure}

We run Algorithm \ref{algo:seq} for the successive selection of $k=10$ stations to build the greedy-optimal network. First we compute the EIG for all the candidate stations and the one with the highest value of EIG is selected as the first best station and is denoted by $\nbr 1$, see the first plot in the top line of Figure \ref{fig:optimal_network_test1}. Then the prior covariance matrix is updated with the posterior covariance matrix obtained with the Green matrix and noise covariance matrix from using station $\nbr1$. Then, we compute the EIG for all the stations except that previously selected one and we repeat the process to select the second best station.

Figure \ref{fig:optimal_network_test1} depicts the selection process of the greedy-optimal network with the best 10 stations for the MT inversion.  The colored field represents the EIG field when picking  the $l$-th best station, which is represented in sandy-brown color, while the previous $l-1$ already selected stations are represented with the indigo color triangles. We observe that for the layered medium, the first station is located above the source and the remaining stations of the greedy-optimal network form a ring centered around the first station. As expected, the increase in information provided by each station decreases as more stations are selected.

\subsection{Posterior distributions}
We now turn to verify how the increase in information gain translates into uncertainty reduction in the posterior distributions for  the MT. For a network $\mathcal{S}_k$, the Gaussian posterior distribution $\post(\bm{m} \vert \bm{Y}) = \mathcal{N}\left(\mpost(\mathcal{S}_k),\Sigpost (\mathcal{S}_k) \right)$ whom moments $\mpost$ and $\Sigpost$ are conditioned with the data collected over the stations of the network $\mathcal{S}_k$, and are given by 
\begin{eqnarray*}
\label{eq:post_momentsY}
\mpost = \Sigpost \left(\Sigprior^{-1} \mprior + \bm{G}^\top(\mathcal{S}_k)\bm{\Sigma}^{-1}_{\epsilon}(\mathcal{S}_k)\bm{Y}(\mathcal{S}_k)\right) \quad \hbox{and} \quad
\Sigpost = \left(\bm{G}^\top(\mathcal{S}_k)\bm{\Sigma}^{-1}_{\epsilon}(\mathcal{S}_k)\bm{G}(\mathcal{S}_k) + \Sigprior^{-1} \right)^{-1},
\end{eqnarray*}
where the network Green's matrix $\bm{G}(\mathcal{S}_k)$  and the network covariance matrix $\bm{\Sigma}_{\epsilon}(\mathcal{S}_k)$ are given in \eqref{eq:Green_network}.
\begin{figure}
  \centering
  \includegraphics[width=.5\textwidth]{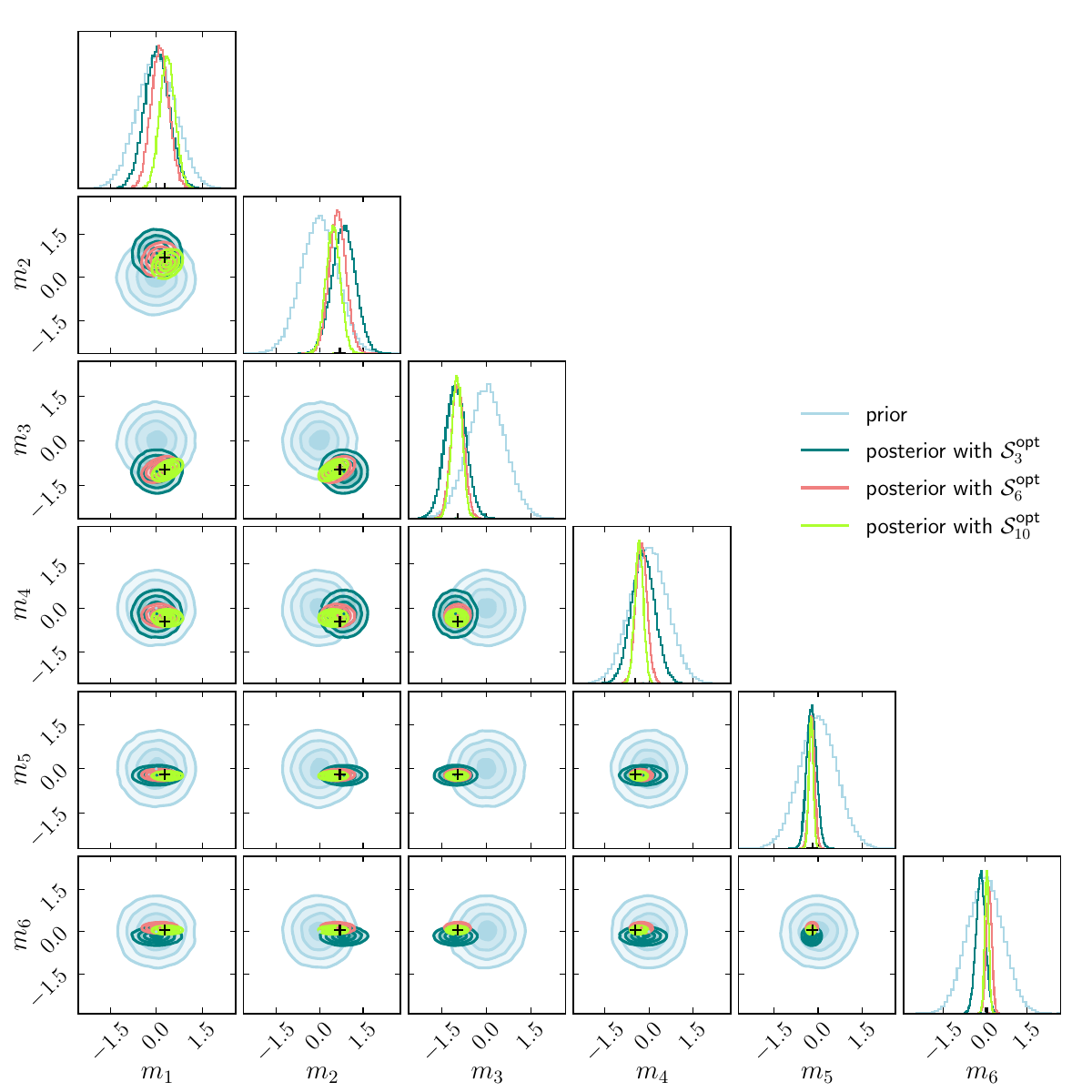}
  \caption{Posterior distributions with different cardinalities, black + shows the true values.}
  \label{fig:posterior_test1}
\end{figure}
 Figure \ref{fig:posterior_test1} displays the prior distribution and the posterior distribution achieved with the observations collected from the stations of the networks $\mathcal{S}^{\tiny{\hbox{opt}}}_3$, $\mathcal{S}^{\tiny{\hbox{opt}}}_{6}$, and $\mathcal{S}^{\tiny{\hbox{opt}}}_{10}$. In addition to the significant reduction of the uncertainty level in the posterior distribution, we observe a bias reduction (true MT marked with black +) for large values $k$.  

\begin{figure}
  \centering
  \includegraphics[width=0.5\textwidth]{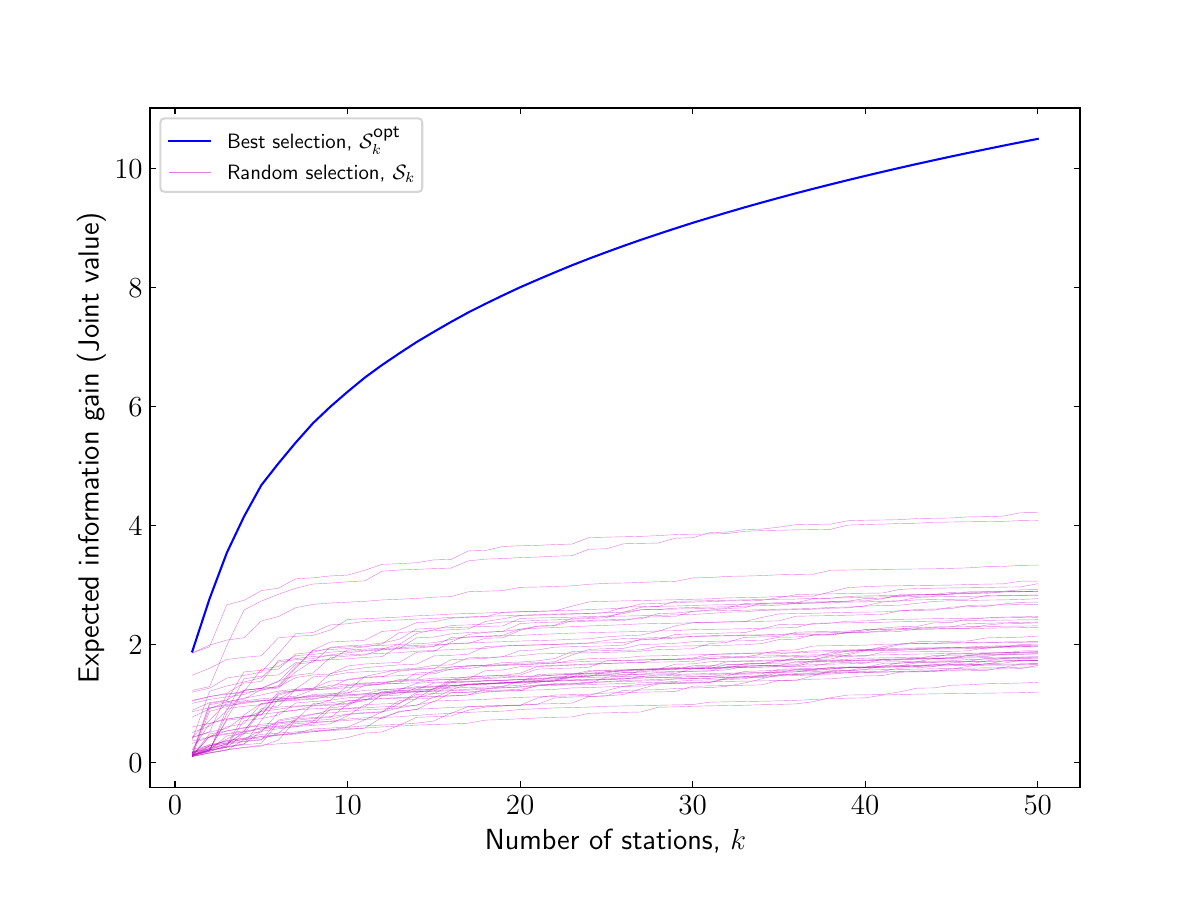}
  \caption{Cumulative information gain for the greedy-optimal network $\mathcal{S}^{\tiny{\hbox{opt}}}_{_{k}}$ and 50 random $\mathcal{S}_{_{k}}$.}
  \label{fig:compare_nets}
\end{figure}

\begin{figure}
  \centering
  \includegraphics[width=0.35\textwidth]{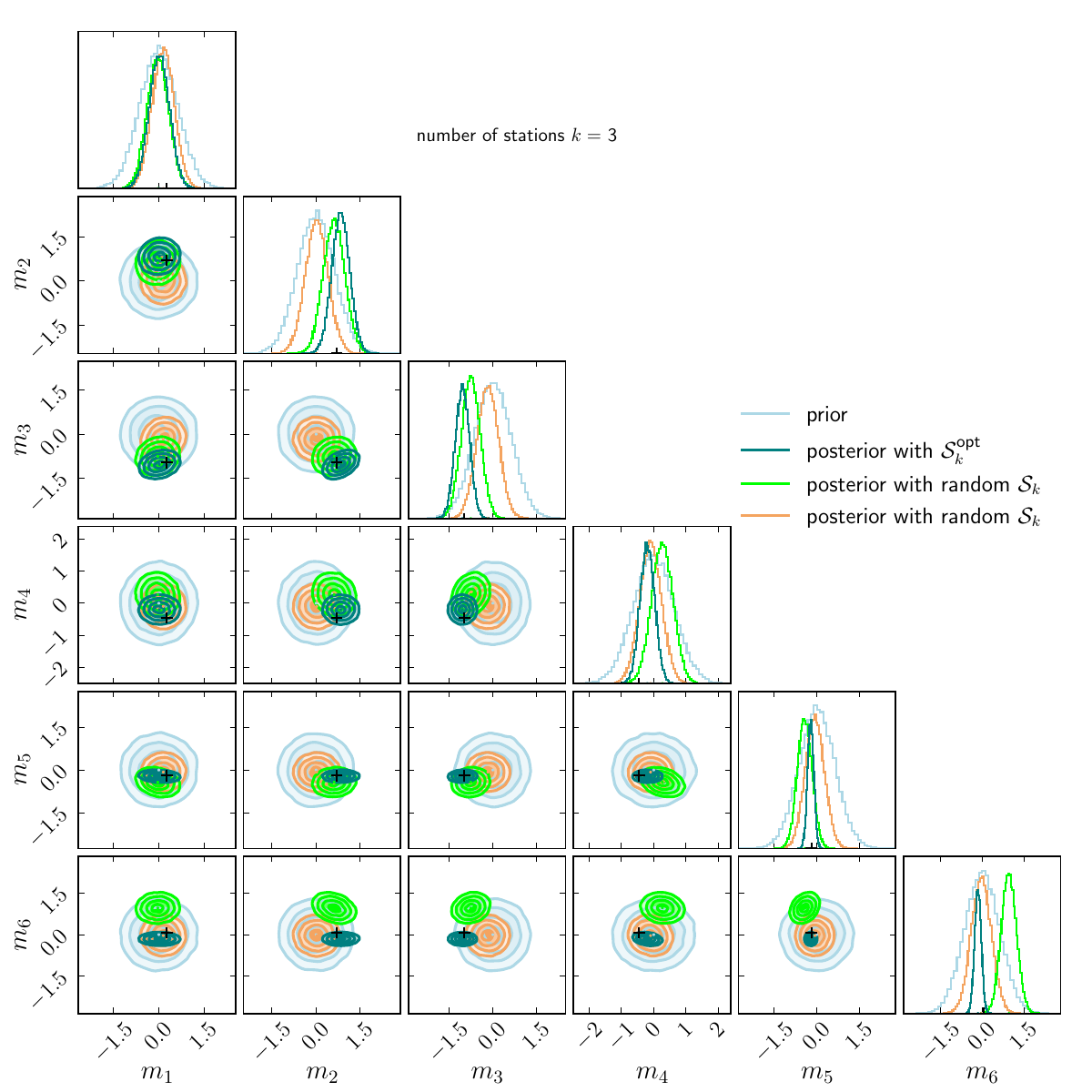}
  \includegraphics[width=0.35\textwidth]{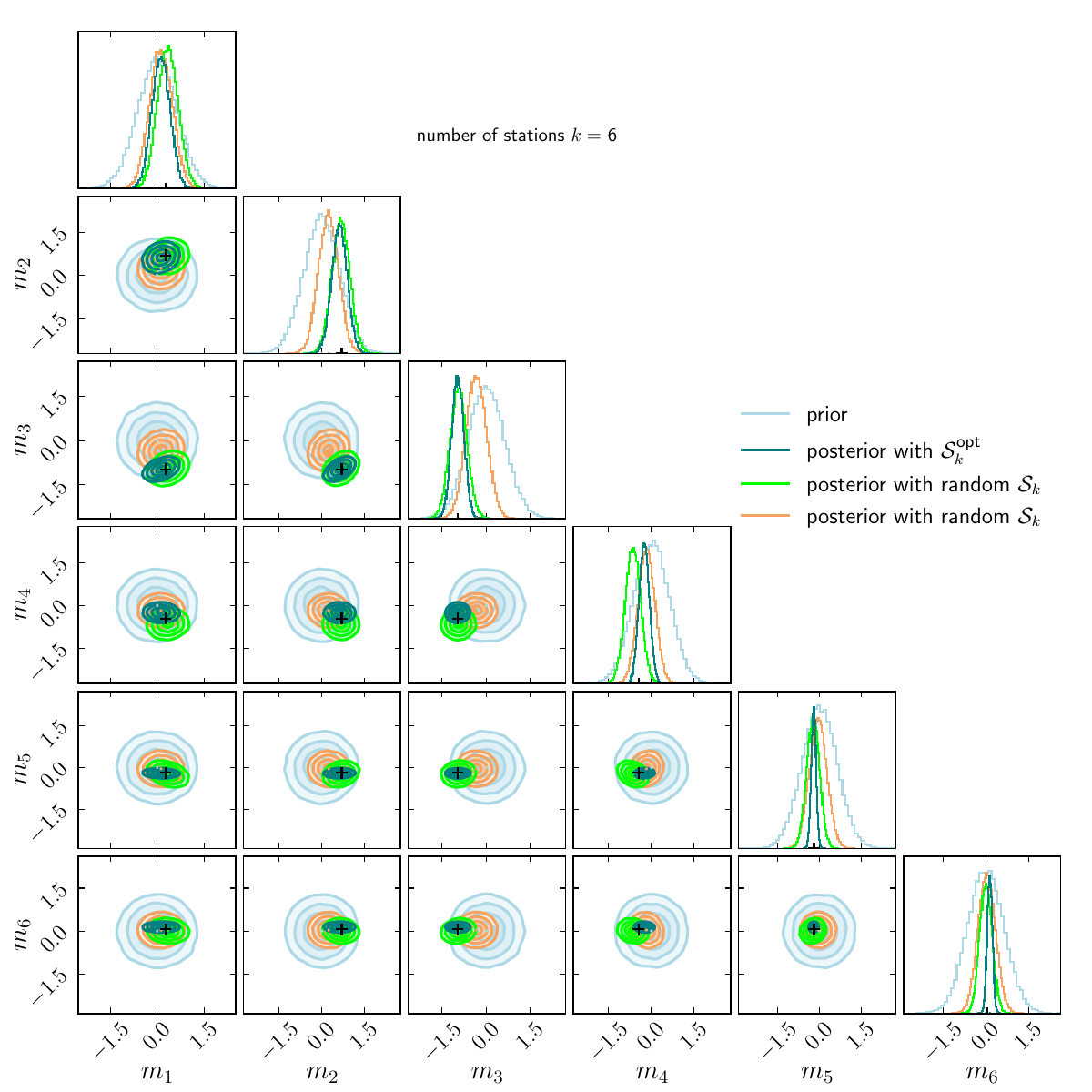}
  \\
  \includegraphics[width=0.35\textwidth]{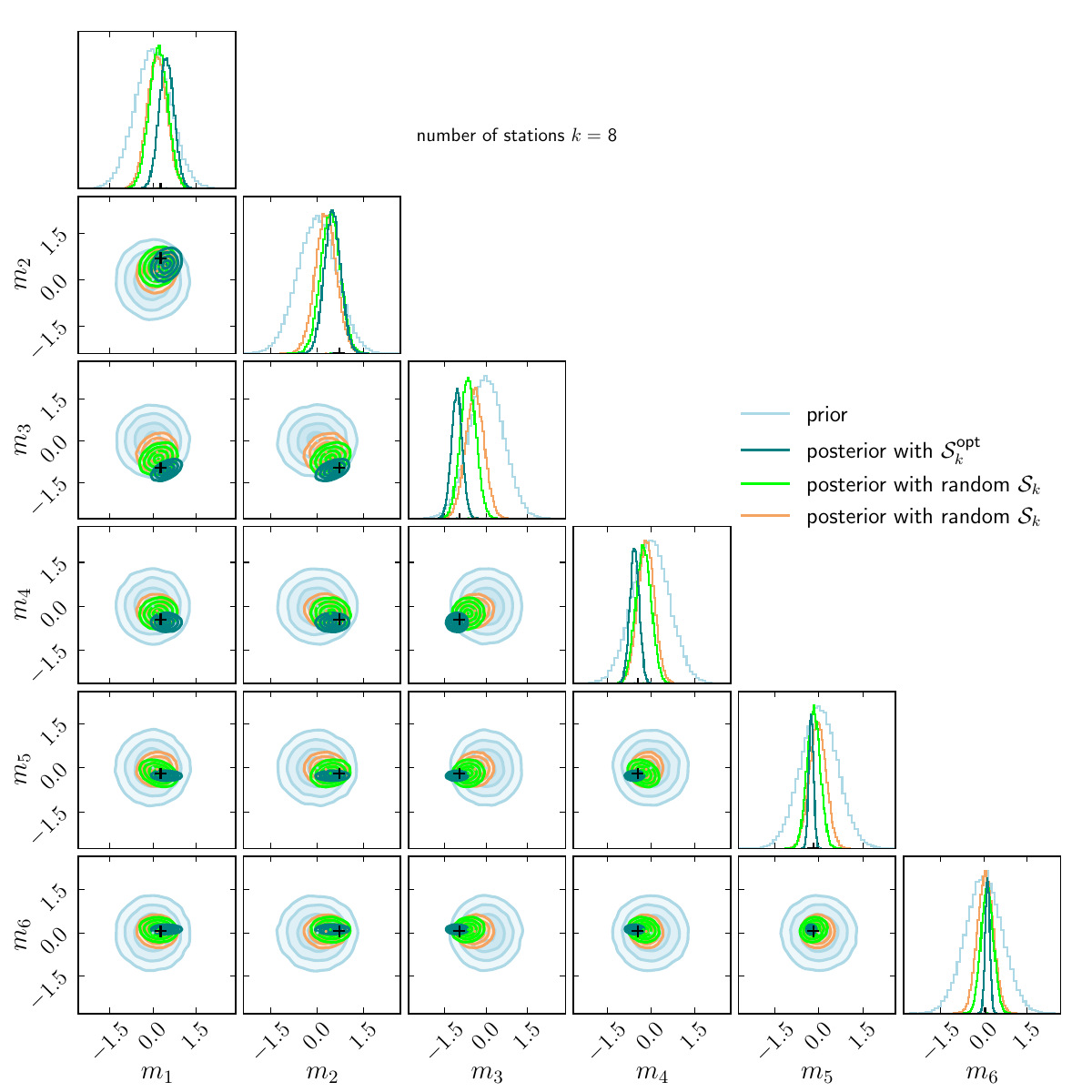}
  \includegraphics[width=0.35\textwidth]{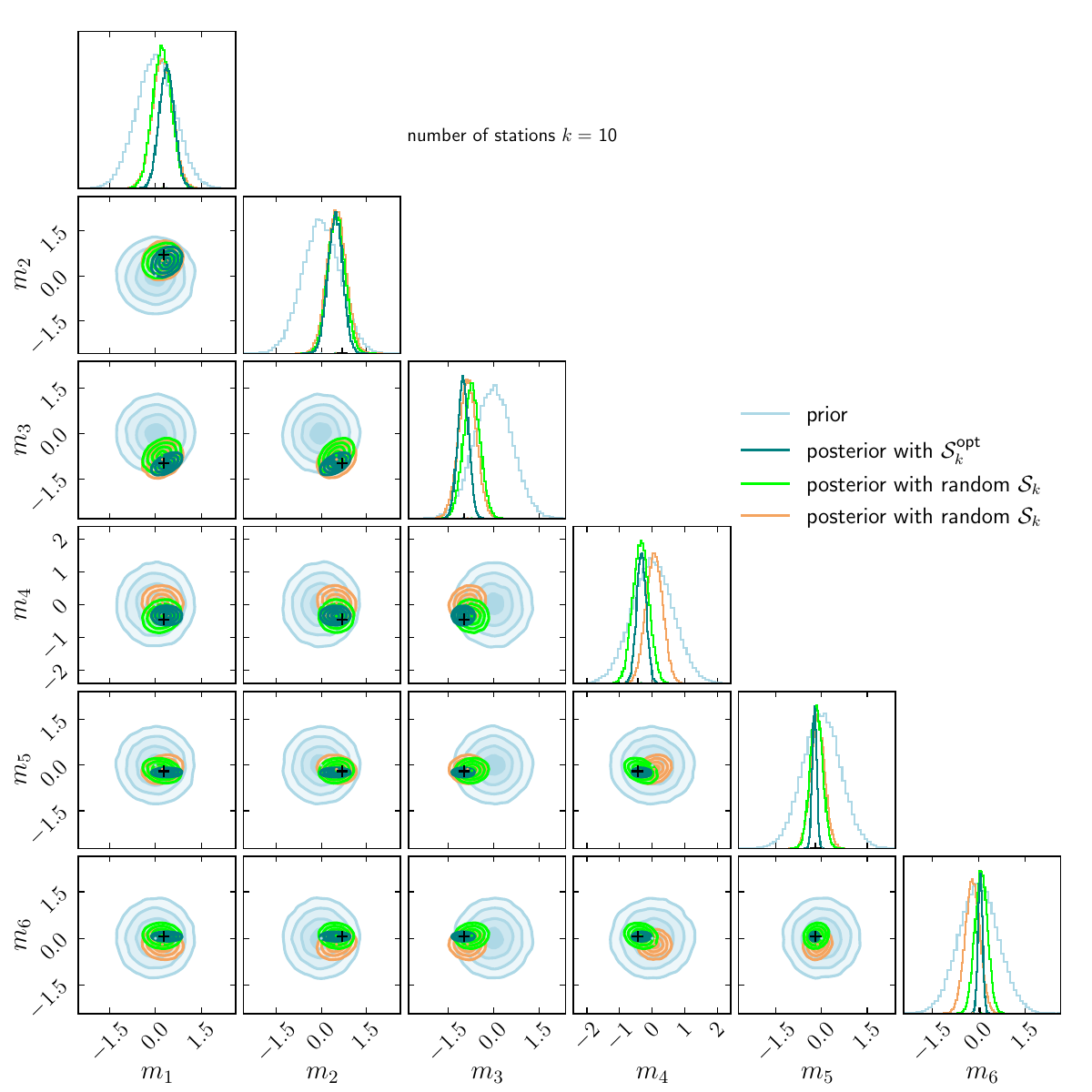}
  \caption{Comparison of the posterior distributions corresponding to different networks of stations.}
  \label{fig:compare_pdfs}
\end{figure}

It is interesting to also compare how the optimally chosen network compares, in terms of uncertainty reduction, to networks chosen randomly. Algorithm \ref{algo:seq} selects the index $i_l$ that achieves the greatest marginal increase in the value of the EIG. In other words, the search of the $l$-th best station (Algorithm \ref{algo:seq} step \ref{setp6}) is conditioned on the previous $l$-1 stations. The overall joint EIG provided by that single selected station $\hat{\bm{x}}_{\hbox{i}_l}$ is  the sum of the conditional EIGs  from the previous $l-1$ selected stations. 
\begin{eqnarray*}
  \Ig_{_{\tiny{\hbox{joint}}}}(\{\bm{x}_{\hbox{i}_l}\}) &=& \frac{1}{2} \log \left( \frac{\det \left(\Sigpost (\{\bm{x}_{\hbox{i}_l}\})\right)}{\det \left( \Sigprior\right)}\right)\\\\
   &=& \frac{1}{2} \log \left( \frac{\det \left(\Sigpost (\{\bm{x}_{\hbox{i}_l}\})\right)}{\det \left(\Sigpost (\{\bm{x}_{\hbox{i}_{l-1}}\}|\mathcal{S}^{\tiny{\hbox{opt}}}_{_{l-2}})\right)} \frac{\det \left(\Sigpost (\{\bm{x}_{\hbox{i}_{l-1}}\}|\mathcal{S}^{\tiny{\hbox{opt}}}_{_{l-2}})\right)}{\det \left(\Sigpost (\{\bm{x}_{\hbox{i}_{l-2}}\}|\mathcal{S}^{\tiny{\hbox{opt}}}_{_{l-3}})\right)} \ldots \frac{\det \left(\Sigpost(\{\bm{x}_{\hbox{i}_{1}}\}|\mathcal{S}^{\tiny{\hbox{opt}}}_{_{1}})\right)}{\det \left( \Sigprior\right)} \right)\\\\
      &=& \frac{1}{2}\sum_{j=1}^{l} \log \left( \frac{\det \left(\Sigpost (\{\bm{x}_{\hbox{i}_j}\}|\mathcal{S}^{\tiny{\hbox{opt}}}_{_{j}})\right)}{\det \left(\Sigpost (\{\bm{x}_{\hbox{i}_{j-1}}\}|\mathcal{S}^{\tiny{\hbox{opt}}}_{_{j-1}})\right)} \right)\;\; \hbox{with}\;\; \mathcal{S}^{\tiny{\hbox{opt}}}_{_0} = \emptyset\\\\
      &=& \sum_{j=1}^{l} \Ig(\{\bm{x}_{\hbox{i}_j}\})\quad \hbox{for} \quad l = 1, \ldots, k.
\end{eqnarray*}
That means that $\Ig$ is submodular in the set $\mathcal{S}^{\tiny{\hbox{opt}}}_{_{n}}$. The joint information gain of $\mathcal{S}^{\tiny{\hbox{opt}}}_{_{k}}$ is the joint information of $\mathcal{S}^{\tiny{\hbox{opt}}}_{_{k-1}}$ plus the incremental EIG of $\bm{x}_{\hbox{i}_k}$. We use this property to show the gradual increase in EIG for the greedy-optimal network $\mathcal{S}^{\tiny{\hbox{opt}}}_{_{k}}$. Figure \ref{fig:compare_nets}  represents the value of the information provided by the greedy-optimal network versus the EIG corresponding to 50 networks constructed by random selection of the corresponding stations. We observe that the cumulative information provided by the greedy-optimal network of stations is greater than the information of any network of stations with random selection of station locations. The discrepancy becomes more apparent with the number of stations. In Figure \ref{fig:compare_pdfs}, we also report the posterior distributions for the greedy-optimal network and some randomly chosen networks for cardinality $k=3, 6, 8, 10$ to show how the greedy-optimal network outperforms networks with randomly-selected station locations.

\subsection{Effect of the source depth on the network design}\label{sec:depth}
\begin{figure}
  \centering
  \includegraphics[width=0.45\textwidth]{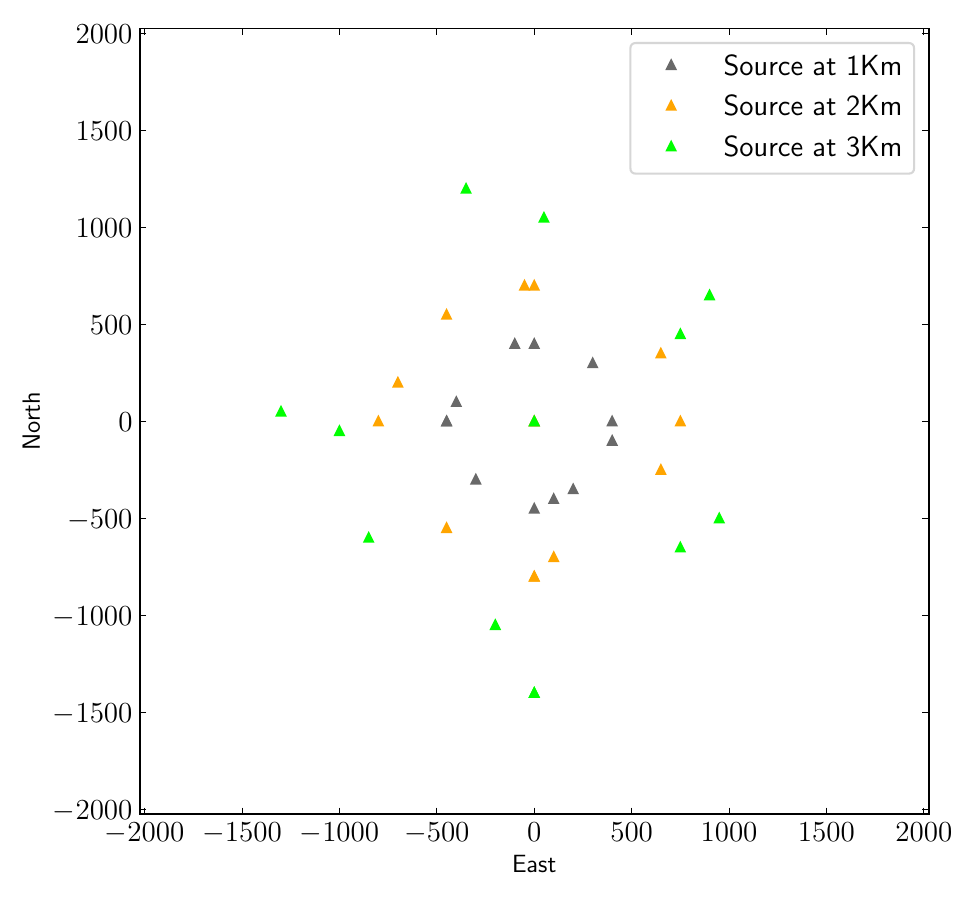}
  \caption{Distributions of $\mathcal{S}^{\tiny{\hbox{opt}}}_{12}$, the set of the first best 12 stations, for different depth of the earthquake source.}
  \label{fig:depth_effect_test1}
\end{figure}

We now want to simulate how the source depth impacts the geometry of the greedy-optimal network $\mathcal{S}^{\tiny{\hbox{opt}}}_{k}$, we perform three experiments with the earthquake-source at three different depths. The simulation parameters remain the same for the three scenarios except the value of the source depth $x^{\tiny{\hbox{src}}}_3$: domain of $8Km\times 8Km$, $161$ stations from East to West and from South to North with the same velocity-model given in Figure \ref{fig:vel_model_test1}, a trace length of 4.5 seconds and a sampling of the waveform at every 0.005 second. In Figure \ref{fig:depth_effect_test1}, each color (for the triangles) shows the station locations for greedy-optimal network with 12 sensors $\mathcal{S}^{\tiny{\hbox{opt}}}_{12}$ associated with a given depth for the earthquake-source. The triangles in gray shows the network for the earthquake-source at depth $1Km$. The khaki color triangles describe the distribution of the network obtained with source at $2Km$. The light-cyan color triangles represent the geometry of the greedy-optimal network achieved when the earthquake-source is at the depth of $3Km$. The first best station of each network associated with a value of the depth is located at the center of the domain. For each depth of the earthquake-source, the greedy-optimal network forms a ring centered around the first best station. Nevertheless, the main observation is that the  radius of an imaginary circle circumscribed to the greedy-optimal network increases with the depth of the earthquake source. More precisely, the inner and the outer radii of the ring increase with the source depth. Also the radius outer of the ring grows faster than that of the inner radius. As source depth increases, the distance from the epicenter for a ray leaving the source at a given takeoff angle increases as depth increases. The increase in radii of the optimal network reflects this increased distance.   

\section{Design of a consensus greedy-optimal network of stations}\label{sec:cons}
So far, we have assumed that both earthquake location and the velocity-models are known, while this is generally not the case in practice. In the following section we will present a number of use-cases in which some modeling assumptions are relaxed in order to better account for uncertainties in the velocity-model and  earthquake-source location. We will also present a setting in which the inversion is being performed under misspecification i.e. the data generating process is different from what the models used for inversion can describe. In all of these cases we prove that it is possible to build an optimal network called a ``consensus network'' that can take into account these uncertainties and improve the performance of the inversion process under a number of evaluation criteria. 

In section \ref{sec:gen_cons_agl} we will present the general algorithm for building a consensus network when some degree of uncertainty is present in the parameters of the problem. In section \ref{sec:con_vel} we propose to use this algorithm in the following context: a set of 2D layered velocity models is available to represent  the real, complex  3D model that would more reliably describe the subsurface. Instead of choosing only one of these models to optimize the network of stations for, we show how it is possible to build a consensus over all of the possible models and evaluate its performance. In section \ref{sec:cons_src} we repeat the same process but for a set of uncertain source locations. Finally, in section \ref{sec:model_misp} we test the performance of a consensus network under a misspecified setting i.e. a network set up to be robust to the possibility of the data-generating model to be different from the model used for inversion.

\subsection{Building process of the consensus greedy-optimal network} \label{sec:gen_cons_agl}
A natural way of addressing the problems of an uncertain velocity model (or source location) would be that of inverting jointly for the MT and the velocity (or source location) itself. However, doing so would drastically increase the number of parameters to be inferred and break the linear nature of the problem as outlined so far. The conceptual approach for the consensus network is therefore to keep the optimal experimental design  within the linear conjugate-Gaussian setting and account for the uncertainties in the velocity model through a network found by averaging the values of the EIG corresponding to a number of plausible velocity-models and/or source-locations. In other words, each iteration would be the EIG assuming that the velocity-model or the earthquake source location correspond to one of the plausible sets.

Assume that $n_{\tiny{\theta}}$ different instances (or models) of the uncertain geophysical parameter $\bm{\theta}$ are available. These could be a set of one-dimensional velocity-models $\bm{V}_r$ or earthquake-source locations $\xss$ for $r = 1, \ldots, n_{\tiny{\theta}}$.  The iterative selection of the stations is performed by maximizing the arithmetic mean value of the EIG for all velocity models, for each candidate station, obtained with the $n_{\tiny{\theta}}$ realizations of $\bm{\theta}$.

\begin{algorithm}[H]
\caption{Design of the consensus greedy-optimal network} \label{algo:consensus}
\begin{algorithmic}[1]
  \State Set $\mathcal{S} = \mathbb{N}_n$
  \State Initialize $\mathcal{S}^{\tiny{\hbox{opt}},\tiny{\hbox{ugp}}}_k = \emptyset$
  \For{\texttt{$r = 1$ to $n_{\tiny{\theta}}$}}
           \State Parameterize the data noise: evaluate $\sigma_\epsilon$, $\forall s \in \mathcal{S}$
           \State Set the prior PDF: $\mprior\left(\bm{\theta}_r\right)=0$ and $\Sigprior \left(\bm{\theta}_r\right) = \sigma_p^2 \bm{I}_d$
  \EndFor
  
  \For{\texttt{$l = 1$ to $k$}}
      \For{\texttt{$r = 1$ to $n_{\tiny{\theta}}$}}
           \State Compute  $\Ig(\mathcal{S}|\bm{\theta}_r)$ 
      \EndFor
      \State Average the EIG: $\Ibar = \frac{1}{n_{\tiny{\theta}}} \displaystyle \sum_{r =1}^{n_{\tiny{\theta}}} \Ig(\mathcal{S}|\bm{\theta}_r)$ 
      \State $\hbox{i}_r = \displaystyle \argmax_{\substack{j \in \mathcal{S}}} \Ibar(\{\bm{x}_j\})$     
      \State $\mathcal{S}^{\tiny{\hbox{opt},\tiny{\hbox{ugp}}}}_k \gets \mathcal{S}^{\tiny{\hbox{opt},\tiny{\hbox{ugp}}}}_k \cup \{\hbox{i}_r\}$
      \State $\mathcal{S} \gets \mathcal{S}\backslash \{\hbox{i}_r\}$
      \For{\texttt{$r = 1$ to $n_{\tiny{\theta}}$}}
          \State Compute the covariance matrix $\Sigpost\left(\bm{\theta}_r, \{\bm{x}_{\hbox{i}_r}\}\right)$ of the posterior PDF according to \eqref{eq:post_moments}
          \State Update the covariance matrix  of the prior PDF:  $\Sigprior\left(\bm{\theta}_r\right) \gets \Sigpost\left(\bm{\theta}_r, \{\bm{x}_{\hbox{i}_r}\}\right)$
      \EndFor
  \EndFor
  \State \textbf{return the index set} $\mathcal{S}^{\tiny{\hbox{opt},\tiny{\hbox{ugp}}}}_k$.
  \end{algorithmic}
\end{algorithm}
In Algorithm \ref{algo:consensus}, the string word standing for the uncertain geophysical parameter "$\tiny{\hbox{ugp}}$" can be replaced by  $\tiny{\hbox{vel}}$ or ${\tiny{\hbox{src}}}$.

\subsection{Consensus design for full three-dimensional velocity-models}\label{sec:con_vel}

In this section we specialize the above framework to the specific issue of approximations in the velocity models.

\subsubsection{Problem setup}
\begin{figure}
  \centering
  \includegraphics[height=7cm, width=7.5cm]{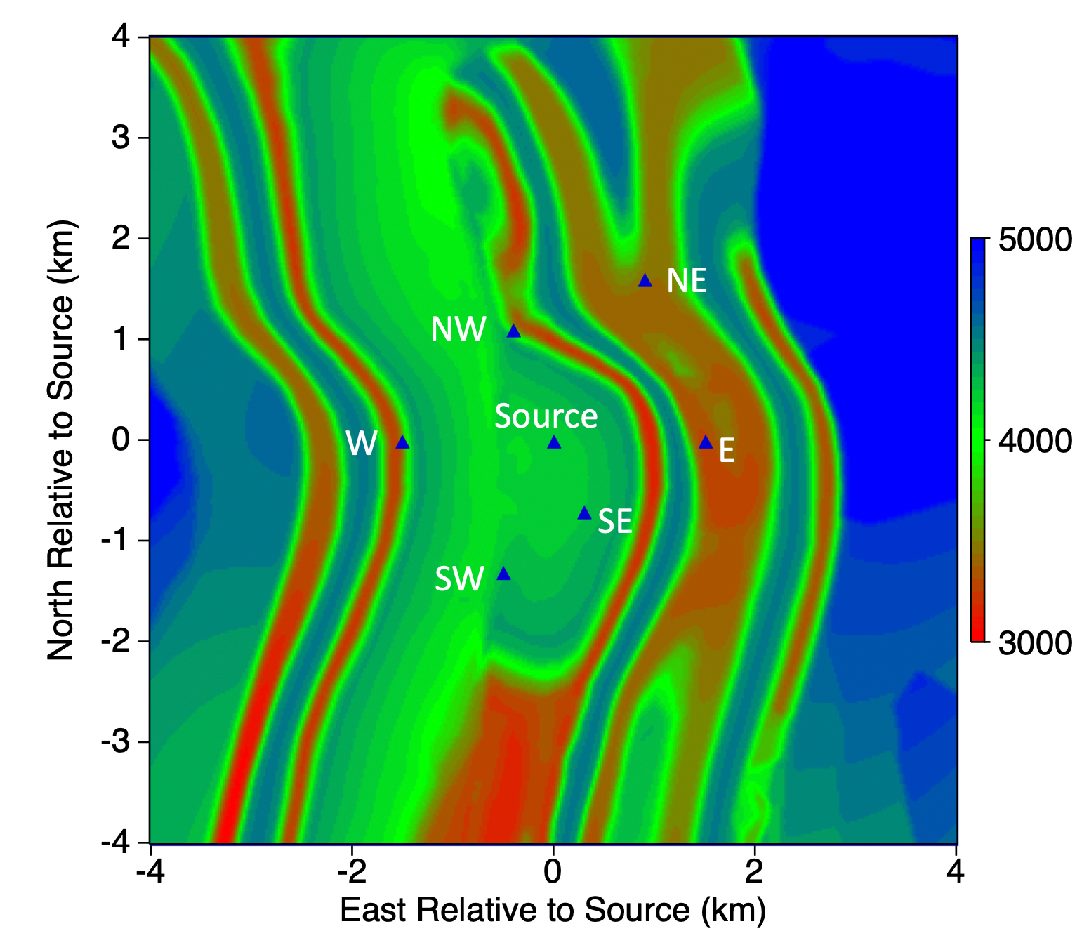}
  \includegraphics[height=7.5cm, width=5cm]{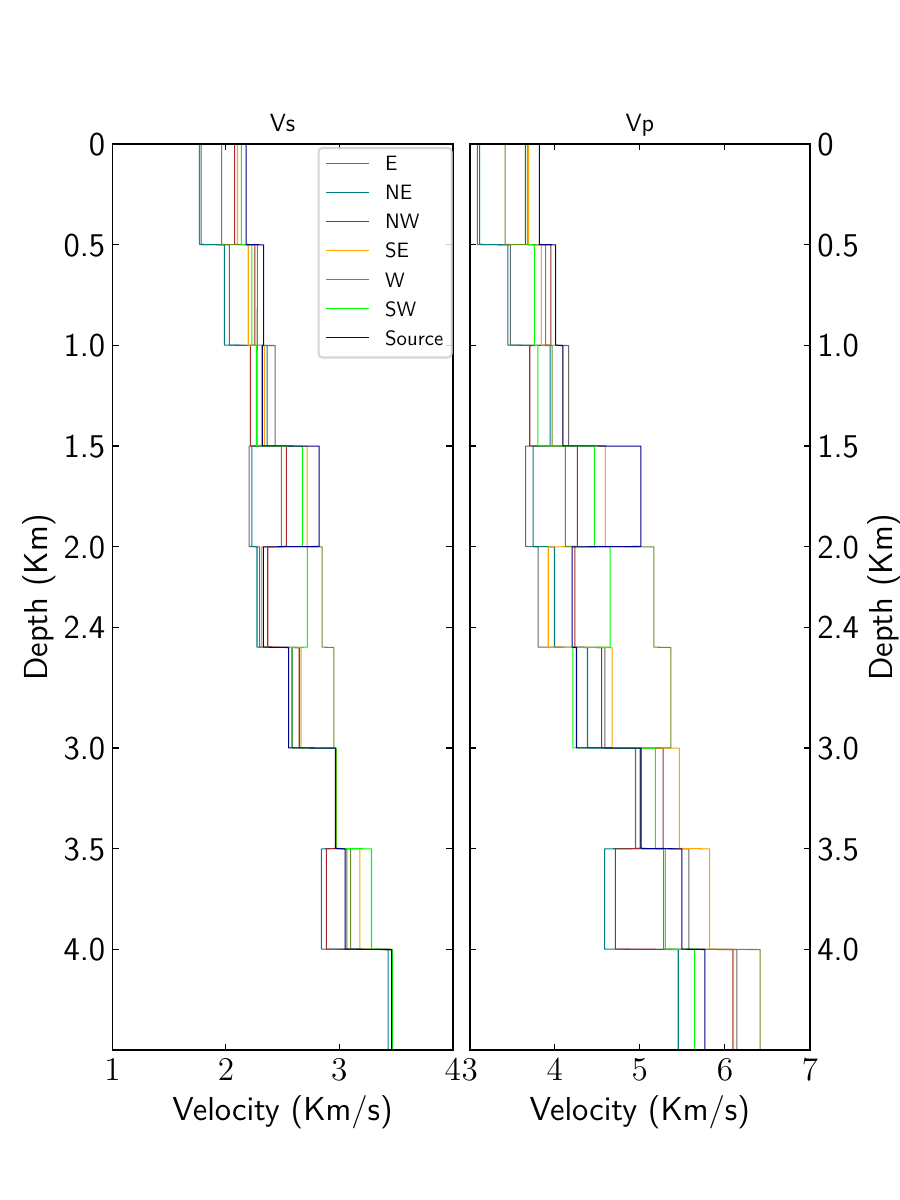}
  \caption{Left: Horizontal cross-section of the 3D P-wave velocity model at 1.1km depth for the SEG/EAGE overthrust model with locations of the 7 well-logs used to construct layered models. Right: vertical profiles representing  the seven velocity ($V_p$ and $V_s$) models constructed from the seven well-logs. }
  \label{fig:welllogs_velocity_test2}
\end{figure}

In general, an accurate three-dimensional velocity-model of a region is not available. Instead we suppose that 7 different several layered velocity-models have been built using geophysical well-logs measured at 7 specific locations. Here we consider a domain ${\D}$ = 8 Km$\times$8 Km portion of the SEG/EAGE over-thrust three-dimensional velocity-model \cite{Aminzadeh1996, House1996} for the real complex structure. The seven layered one-dimensional velocity-models are: \texttt{E}(1500, 0) located East of the origin, \texttt{NE}(900, 1600) located North-East of the origin,  \texttt{NW}(-400, 1100) located to the North-West, \texttt{SE}(300, -700) located to the South-East, \texttt{W}(-1500, 0) located to the  West of the origin,  and \texttt{SW}(-500, -1300) located  South-West of the origin, and the source \texttt{Source}(0,0) at the origin, see  left-plot of Figure \ref{fig:welllogs_velocity_test2}.

\begin{figure}
  \centering
  \includegraphics[width=.5\textwidth]{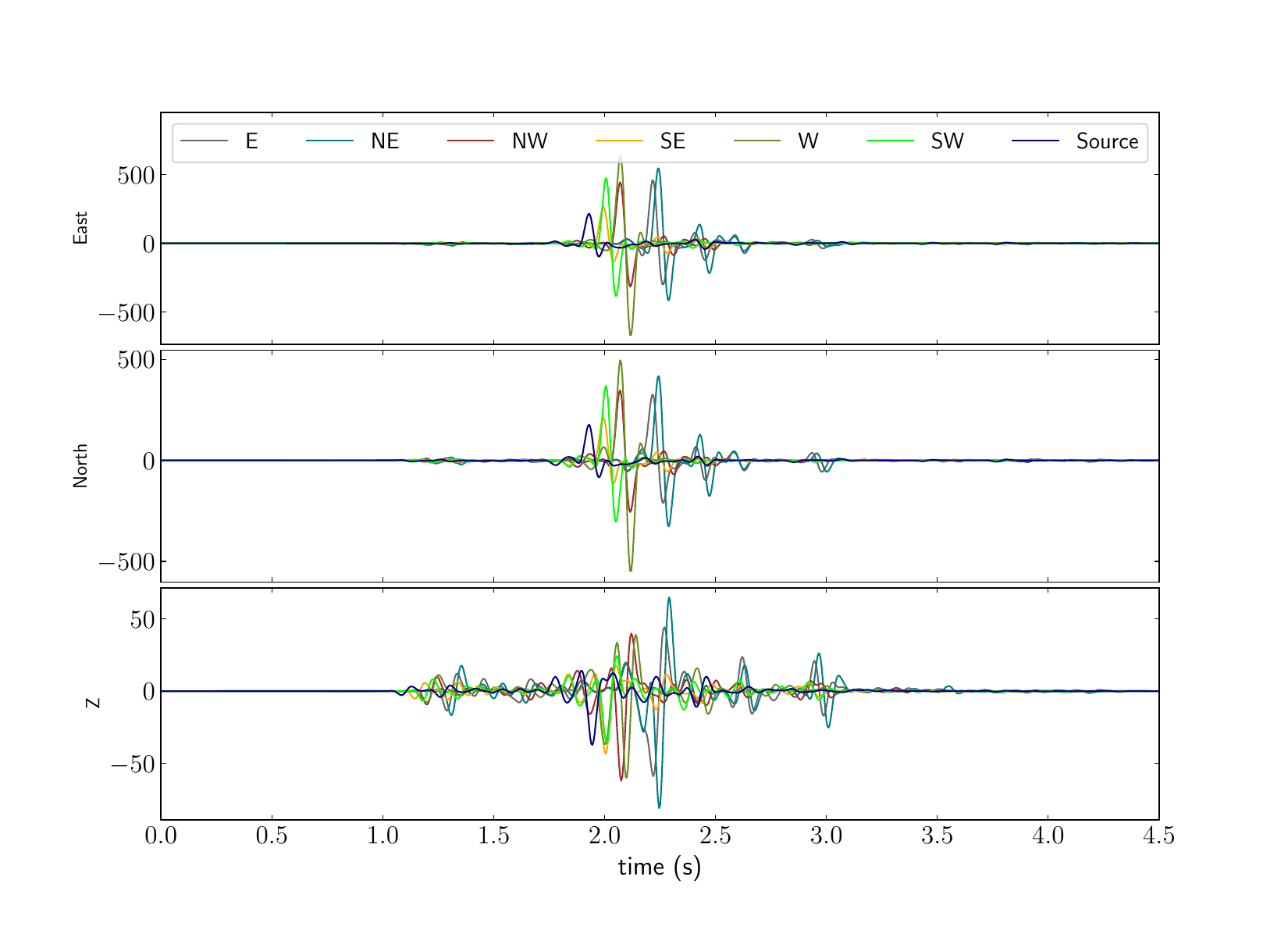}
  \caption{Three component waveforms, for the seven velocity-models, recorded at the station $\nbr$1905 for the moment tensor $\bm{m}_t$.} 
  \label{fig:waveforms_test2}
\end{figure}

For each velocity-model, the simulations are carried out as described in \ref{sec:setting_forward}. Three-component traces for the Green’s function tensor are sampled at 0.005 seconds over a time-length of 4.5 seconds yielding a total of 900 samples per trace. Figure \ref{fig:waveforms_test2} shows three-component traces at station 1905, which is located 3450 m West and 2650 m North of the source position. Traces for all seven models are shown. Note that the P-waves are strongest on the vertical component and that there is a variation in P-wave arrival time among the seven models as expected. The horizontal components are larger than the vertical component traces since the larger S-wave is seen mostly on the horizontal components of motion for a source at depth.

A possible solution for network design would be to choose one of the 7 velocity models and build an optimal network solely based on the selected model. This of course, would be detrimental if the trace data were best captured by one of the other 6 available velocity models. The consensus network addresses this problem by taking into account information from all 7 velocity models.

\subsubsection{Consensus greedy-optimal network $\mathcal{S}^{\tiny{\text{opt,vel}}}_{k}$} \label{sec:optimal_vel}
The objective of the consensus approach is to define a network of $k$ stations,  $\mathcal{S}^{\tiny{\hbox{opt,vel}}}_{k}$, that are built using a set of layered one-dimensional profiles instead of a single one. We expect this network to underperfom, i.e. translate into larger uncertainties in the posterior distribution, only in the case where the correct velocity model is among the seven picked ones for inversion. However, on average, the consensus network should perform better than the worse case scenario of picking the most inappropriate of the models. Its mean effect should also be equal to or better than the result of using optimal networks built by focusing on one of the individual layered velocity  models. We assume that the seven layered velocity models represent some average of the 3D model. The uncertain feature from Algorithm \ref{algo:consensus}, $\bm{\theta}$, is here represented by the one-dimensional velocity-model. 

\begin{figure}
  \centering
  \includegraphics[width=.5\textwidth]{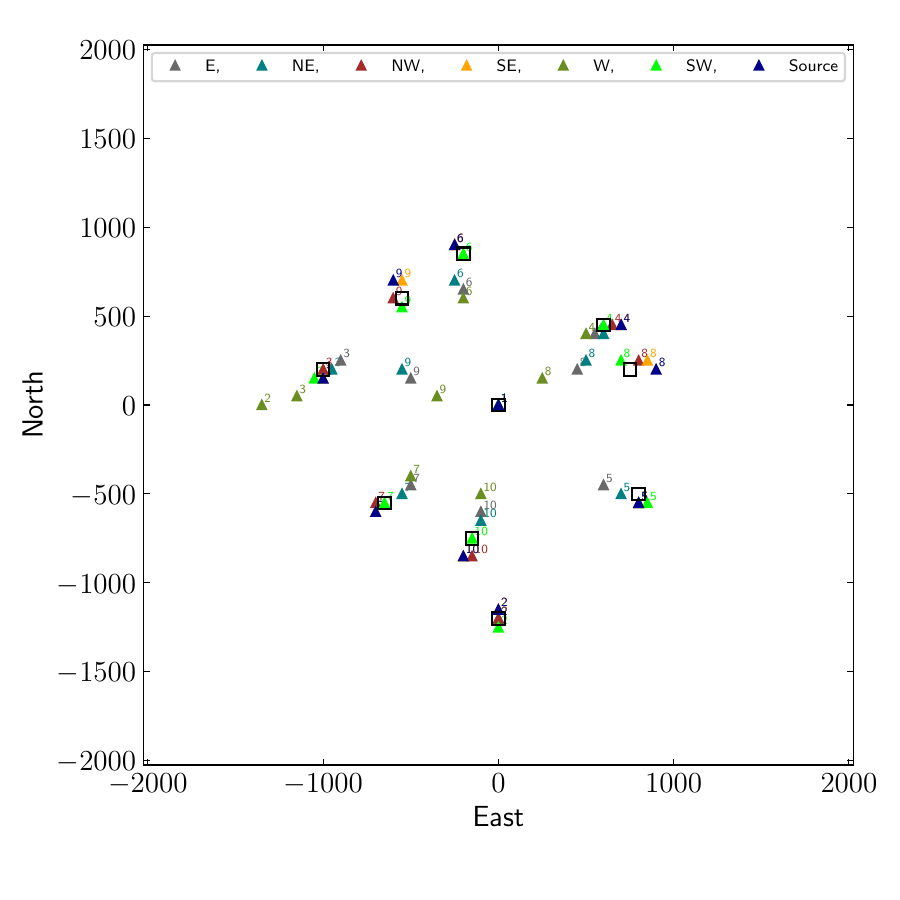}
  \caption{The consensus greedy-optimal network $\mathcal{S}^{\tiny{\hbox{opt,vel}}}_{10}$ with the 10 best stations in black unfilled squares and the $n_{\tiny{\theta}}=7$ greedy-optimal networks $\mathcal{S}^{\tiny{\hbox{opt}}}_{10}\left(\bm{V}_r\right)$. Each color is associated with a given velocity-model.}
  \label{fig:consensus_test2}
\end{figure}

Figure \ref{fig:consensus_test2} shows the consensus greedy-optimal network and the seven individual greedy-optimal networks $\mathcal{S}^{\tiny{\hbox{opt}}}_{k}\left(\bm{V}_r\right)$ for $r=1,\ldots,7$ associated with the velocity-models collected from  the  seven well-logs: $\bm{V}_1$ for the well-log \texttt{E}, $\bm{V}_2$ for the well-log \texttt{NE}, $\bm{V}_3$ for the well-log \texttt{NW}, $\bm{V}_4$ for the well-log \texttt{SE}, $\bm{V}_5$ for the well-log \texttt{W}, $\bm{V}_6$ for the well-log \texttt{SW}, and $\bm{V}_7$ for the well-log at the \texttt{Source}. The first observation is that the location of the first best station (above the source) is independent of the velocity-model. Also, the optimal network for velocity-model $\bm{V}_6$  is similar to  the consensus greedy-optimal network indicating that this velocity model in some sense represents some average of the other models.

\subsubsection{Performance analysis of the consensus greedy-optimal network $\mathcal{S}^{\tiny{\text{opt,vel}}}_{k}$}

To asses the amount of uncertainty in the posterior distribution resulting from different choices of optimal network, we report the determinant of the posterior covariance matrix as well as the Bayes risk, that is, in this case,  the trace of the posterior covariance, see \cite{Alexanderian2023}, 
\begin{eqnarray}
   \risk(\mpost) = \trace \left( \Sigpost \right).
\end{eqnarray}
\begin{figure}
  \centering
  \includegraphics[width=0.7\textwidth]{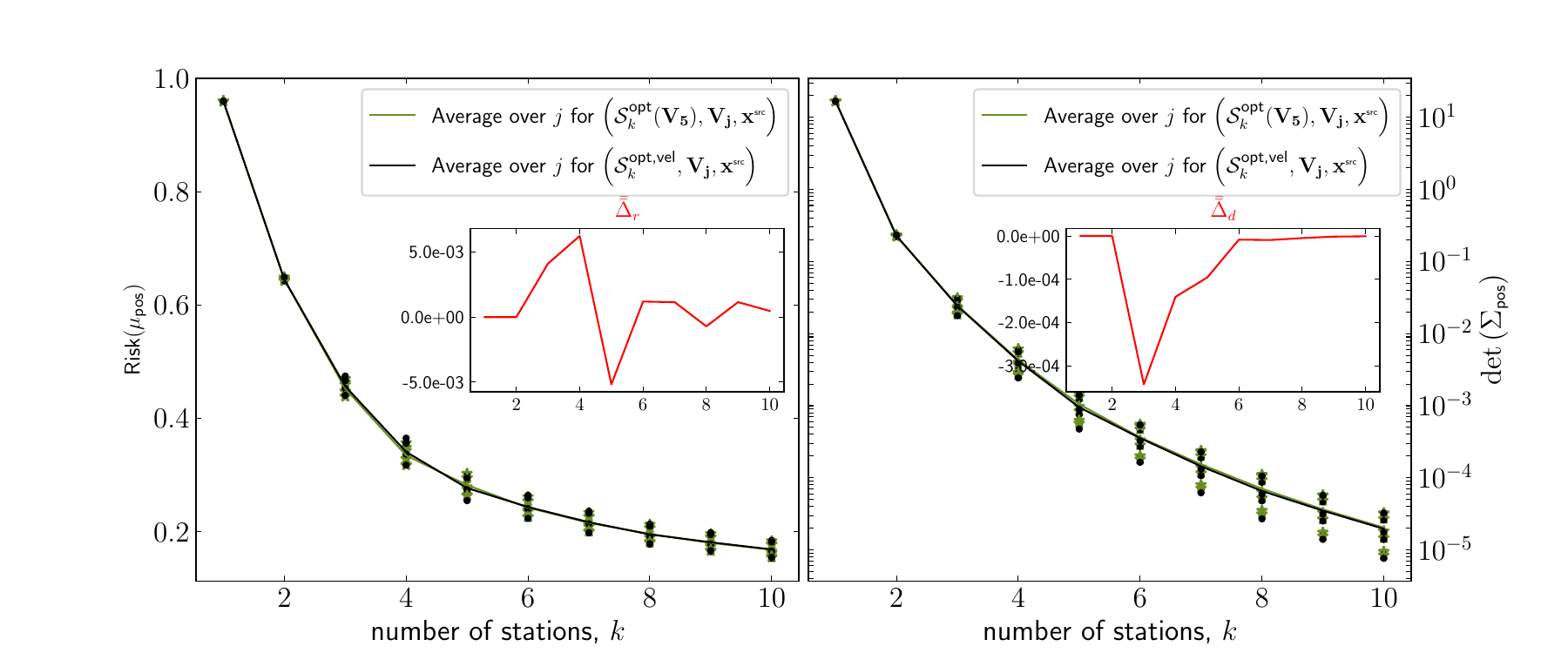}
  \caption{Comparison of the Bayes risk (left) and the determinant of the posterior covariance matrix (right) as a function of the number of stations for the consensus greedy-optimal network and the greedy-optimal network obtained with the velocity-model $\bm{V}_5$. For a fixed cardinality $k$, the green (respectively black) scattered plots is achieved by using the greedy optimal network $\mathcal{S}^{\tiny{\hbox{opt}}}_{k}\left(\bm{V}_5\right)$ combined with the velocity-model $\bm{V}_j$ and the source $\xs$  for $j=1,\cdots,7$ (respectively the consensus greedy-optimal network $\mathcal{S}^{\tiny{\hbox{opt,vel}}}_{k}$ combined with the velocity-model $\bm{V}_j$ and the source $\xs$  for $j=1,\cdots,7$) while the green (respectively black) line stands for the average over $j$ for the greedy-optimal network (respectively the consensus greedy-optimal network).}
\label{fig_vel:compare_post_v5}
\end{figure}
A posterior distribution is defined not only by the network locations, but also by the data and model used to invert the data i.e source location and velocity-model. In our numerical experiments we use the consensus network to invert data coming from each of the 7 velocity models and calculate the average trace and determinant of the posterior covariance matrices. We then repeat the exercise by instead using a network based off a single velocity model. This will mean that while in one case the network will be optimally chosen, in the others we will have optimized for the wrong model.   Figure \ref{fig_vel:compare_post_v5} shows the comparison between the consensus greedy-optimal network and the optimal network built using velocity model 5. The inner plots show the difference $\bar{\bar{\Delta}}_r$, between the average Bayes risk of the consensus greedy-optimal network and the average Bayes risk of the greedy-optimal network $\mathcal{S}^{\tiny{\hbox{opt}}}_{k}\left(\bm{V}_5\right)$, (in the left plot) and the difference of the average of the posterior determinants  $\bar{\bar{\Delta}}_d$ (in the right plot). In both cases, it is noticed that the average difference decreases in absolute value with an increase in the number of stations. This means that as the number of stations increase, the performance of the consensus greedy-optimal network, built without a specific choice of velocity model is quasi-equivalent to the performance of the greedy-optimal network that assumes an available velocity-model.

Besides the uncertainty in the posterior distribution, it is also interesting to look at a metric that accounts for the bias in the posterior's estimation of the MT.  For this, we choose the continuous ranked probability score CRPS. It is a score function that compares the cumulative distribution function CDF of the ``perfect'' posterior distribution (i.e. a delta-Dirac function centered around the  true value of the MT) to the actual  CDF of the posterior distribution $F$.  The score for each MT component $m_i$ is given by:
\begin{eqnarray*}
  \hbox{CRPS}(F,m_i) = \int_{-\infty}^{+\infty} \left(F(m_i) - \mathbbb{1}_{\{m_i \geq m_{\hbox{\tiny{true}}_i}\}}(m_i) \right)^2 d m_i,
\end{eqnarray*}
where $\mathbbb{1}$ stands for the indicator function. The CRPS measures both the bias and the level of uncertainty of each marginal posterior distribution in the way that the smaller its value the closer the posterior is to the delta Dirac centered around the  true value of the MT.
\begin{figure}
  \centering
  \includegraphics[width=0.7\textwidth]{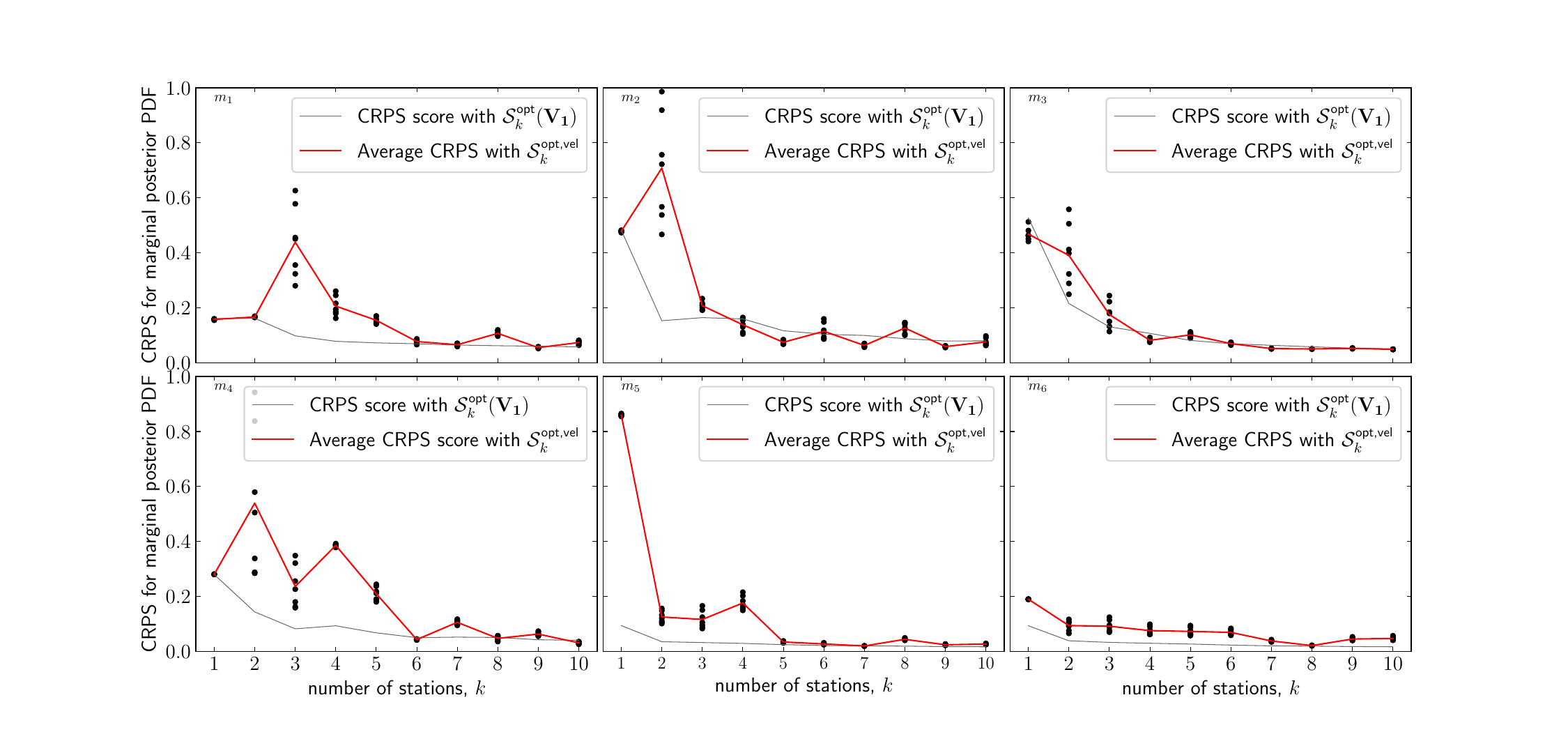}
  \caption{ CRPS scores of the marginal posterior densities for each MT component, between the greedy-optimal network obtained with the velocity-model $\bm{V}_1$ and the consensus greedy-optimal network. For a fixed cardinality $k$, the scattered distribution plot represents the CPRS value when the consensus greedy-optimal network is associated with the seven velocity-models.}
  \label{fig_vel:compare_crps_v1}
\end{figure}
We compare the consensus network and the optimal network associated to the single velocity-model $\bm{V}_1$ in terms of the CRPS scores of the marginal densities for the six components of the MT in Figure \ref{fig_vel:compare_crps_v1}. We observe that the greedy-optimal network outperforms the consensus greedy-optimal network, averaged over the seven velocity-models for a small number of stations and for most of the components of MT. However, when increasing the number of stations, the performance gap  decreases and eventually vanishes. This makes intuitive sense because, as the number of stations increases, the sensitivity of the uncertainty to station position goes down.

\subsubsection{Influence of the number of velocity models on the geometry of the consensus network  }

We now analyze how the consensus network geometry changes  depending on the number of layered velocity-models included in calculating the objective function.   We focus on the following sample sets: $\mathcal{R}_3 = \{ \bm{V}_2, \bm{V}_4, \bm{V}_5\}$, $\mathcal{R}_4 = \{ \bm{V}_3, \bm{V}_4, \bm{V}_5, \bm{V}_6\}$, $\mathcal{R}_5 = \{ \bm{V}_1, \bm{V}_2, \bm{V}_5, \bm{V}_6, \bm{V}_7\}$, $\mathcal{R}_6 = \{ \bm{V}_1, \bm{V}_2, \bm{V}_3, \bm{V}_4, \bm{V}_6, \bm{V}_7\}$, and $\mathcal{R}_7 = \{ \bm{V}_1, \bm{V}_2, \bm{V}_3, \bm{V}_4, \bm{V}_5, \bm{V}_6, \bm{V}_7\}$.
 
\begin{figure}
  \centering
  \includegraphics[width=0.3\textwidth]{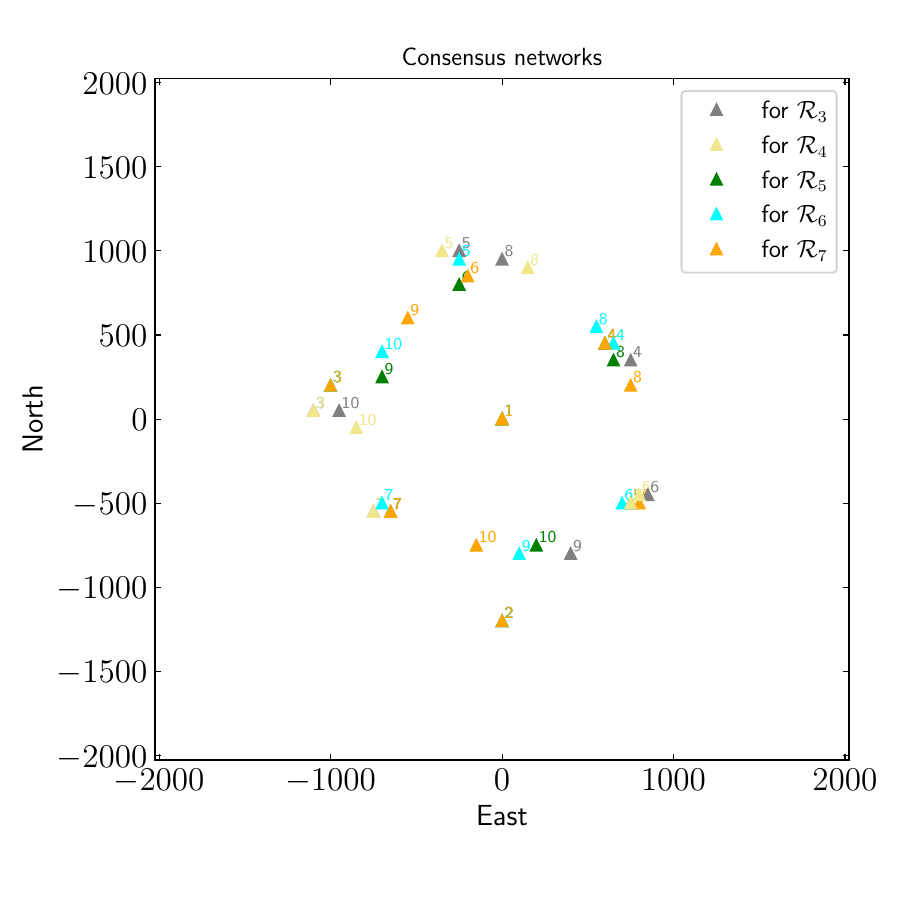}
  \caption{Consensus greedy-optimal networks according to the choice of the subset of velocity-models used to develop the optimal consensus network}
  \label{fig:consensus_net_by_set_test2}
\end{figure}

In Figure \ref{fig:consensus_net_by_set_test2} we show the geometries of the respective consensus networks. We observe that the cardinality of the set of representations does not affect the locations of the first three stations in the consensus network, while it affects the others. Overall, the general shape of the network appears to be consistent across the different choices of velocity models.

\subsection{Consensus design for distribution of earthquake-sources}\label{sec:cons_src}
In this section we repeat the exercise of the previous section \ref{sec:con_vel} where the uncertain parameter is not the velocity model, but rather the source location, in particular we consider the realistic case where there is a distribution of source locations.

\subsubsection{Physical configuration}
We fix the depth of the earthquake source and we consider a 8 Km$\times$8 Km region centered around the origin of the reference, and meshed with $161\times161$ station locations with a spacing of 50 m. We use the velocity-model considered in \cite{Chen2020} based on seismic surveys in the Kimberlina CCUS (Carbon Capture Utilization and Storage) pilot site in California. The layered velocity-model $\bm{V} = \{V_p, V_s\}$ is such that the P-wave and S-wave velocities satisfy $V_p = 1.73 V_s$.
\begin{figure}
  \centering
  \includegraphics[height=7.1cm, width=4cm]{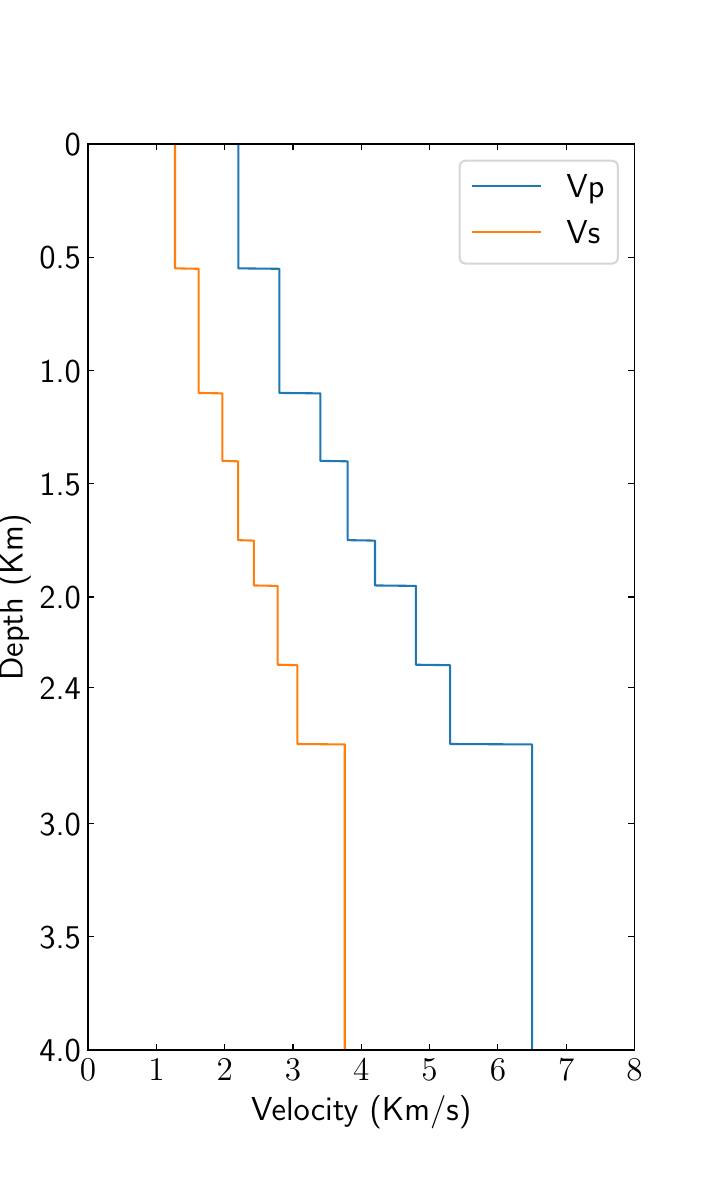}
  \includegraphics[height=6.3cm, width=7cm]{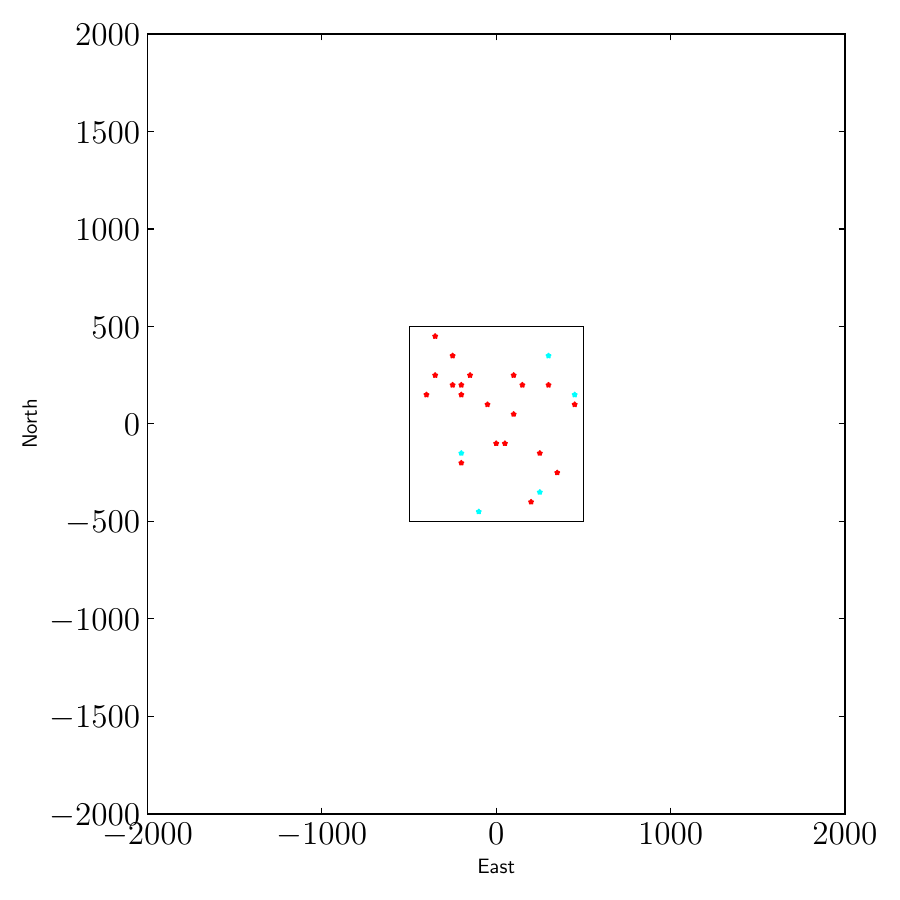}
  \caption{Velocity ($V_p$ and $V_s$) models  for the layered medium from \cite{Chen2020}, and $n_s = 25$ selected locations for earthquake-sources: the 20 red-color locations are used in the construction of the consensus greedy-optimal network.}
  \label{fig:velocity_source_test3}
\end{figure}

Figure \ref{fig:velocity_source_test3} shows the one-dimensional profiles of the P-wave and S-wave velocity-models on the left and $n_{\tiny{\theta}} = 20$ (in red color)  possible locations of the earthquakes distributed within a 1 Km $\times$ 1 Km  region. The locations are independently identically distributed samples from the uniform distribution $\mathcal{U}\left([-500 \hbox{ m}, 500\hbox{ m}]^2\right)$, and stand for possible locations of earthquake-sources. We use 20 locations  to construct the consensus greedy-optimal network while the performance analysis is addressed using all the 25 locations ($n_{\tiny{\theta}}$ plus the five locations in blue color).

\subsubsection{Consensus greedy-optimal network  $\mathcal{S}^{\tiny{\text{opt,src}}}_{k}$}
\begin{figure}
  \centering
  \includegraphics[width=1.\textwidth]{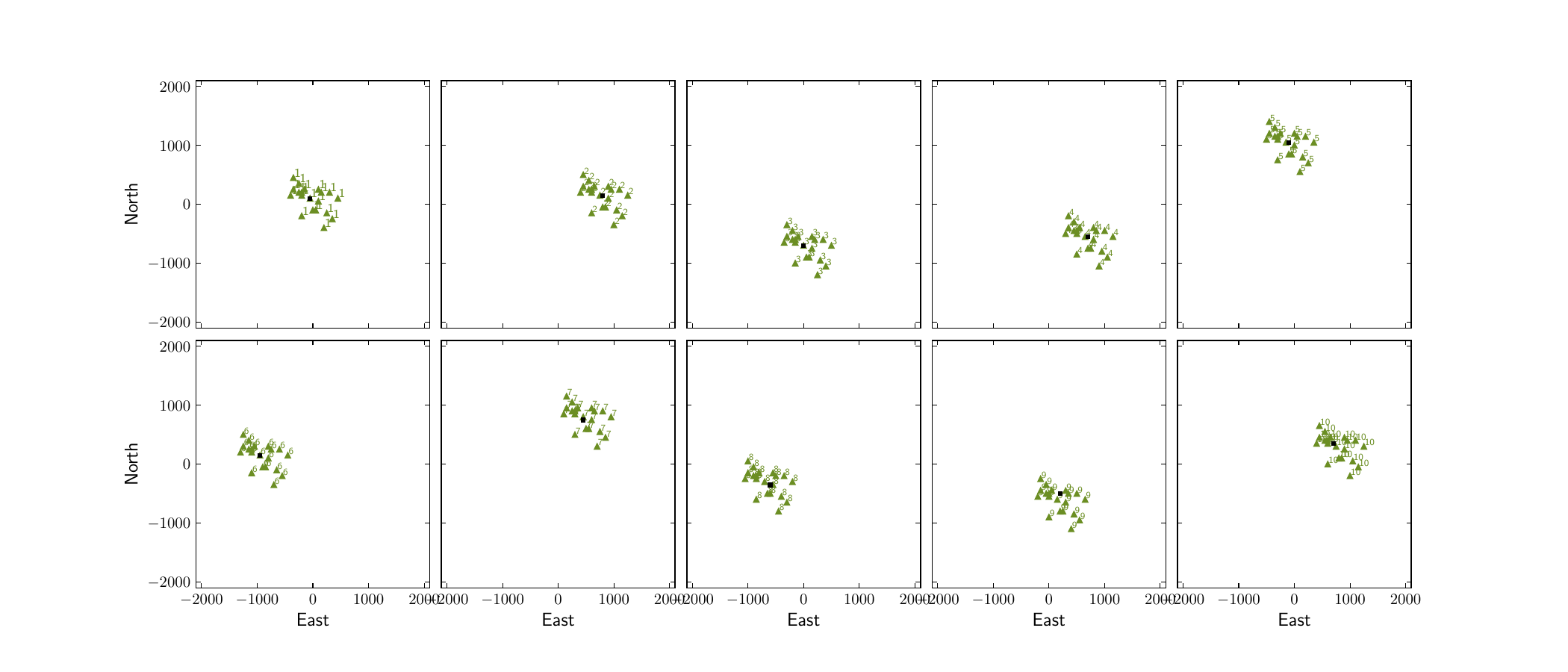}
  \caption{Selection of the 10 best stations comprising the consensus greedy-optimal network from $20$ guessed locations for the earthquake-source. Each box shows the distribution (green triangles) of one station location. The black square shows the position of station in the consensus network.}
  \label{fig:consensus_src}
\end{figure}

Figure \ref{fig:consensus_src} shows (in green) the distribution of the $k$-th best stations if we were to optimize the network for each individual source location (i.e. 20 sources, $\mathcal{S}^{\tiny{\hbox{opt}}}_{10}(\xss)$). On the same plot (in black) we report the associated $k$-th best station for the consensus greedy-optimal network  $\mathcal{S}^{\tiny{\hbox{opt,src}}}_{10}$ obtained by averaging the EIG over $20$ assumed event locations. We observe that the $k$-th station of the consensus greedy-optimal network is always within the imaginary cluster described by the 20 $k$-$th$ stations but is not necessarily at the center of it. This  barycentric configuration indicates a difference in the value of the contribution of possible source locations; the ones close to the unknown true location contribute better to the design of the consensus network.

\subsubsection{Performance analysis of the consensus greedy-optimal network $\mathcal{S}^{\tiny{\text{opt,src}}}_{k}$}
There is a total of 20 assumed locations for the earthquake-sources and each of them has an associated greedy-optimal network $\mathcal{S}^{\tiny{\hbox{opt}}}_{k}(\xss)$ built using Algorithm \ref{algo:seq}. The consensus greedy-optimal network is constructed with only 20 assumed locations from the 25 shown in the right panel in Figure \ref{fig:velocity_source_test3}. 

\paragraph{Test 1: an analysis based on the network of stations.} In this test we intend to compare the performance of the consensus network built for a distribution of source locations compared to that of the network optimized for a single source only.
For that, in addition to the covariance of the posterior distribution we will consider the Bayes risk of the posterior mean and the CRPS scores for each marginal posterior distribution of MT. 
\begin{figure}
  \centering
  \includegraphics[width=.7\textwidth]{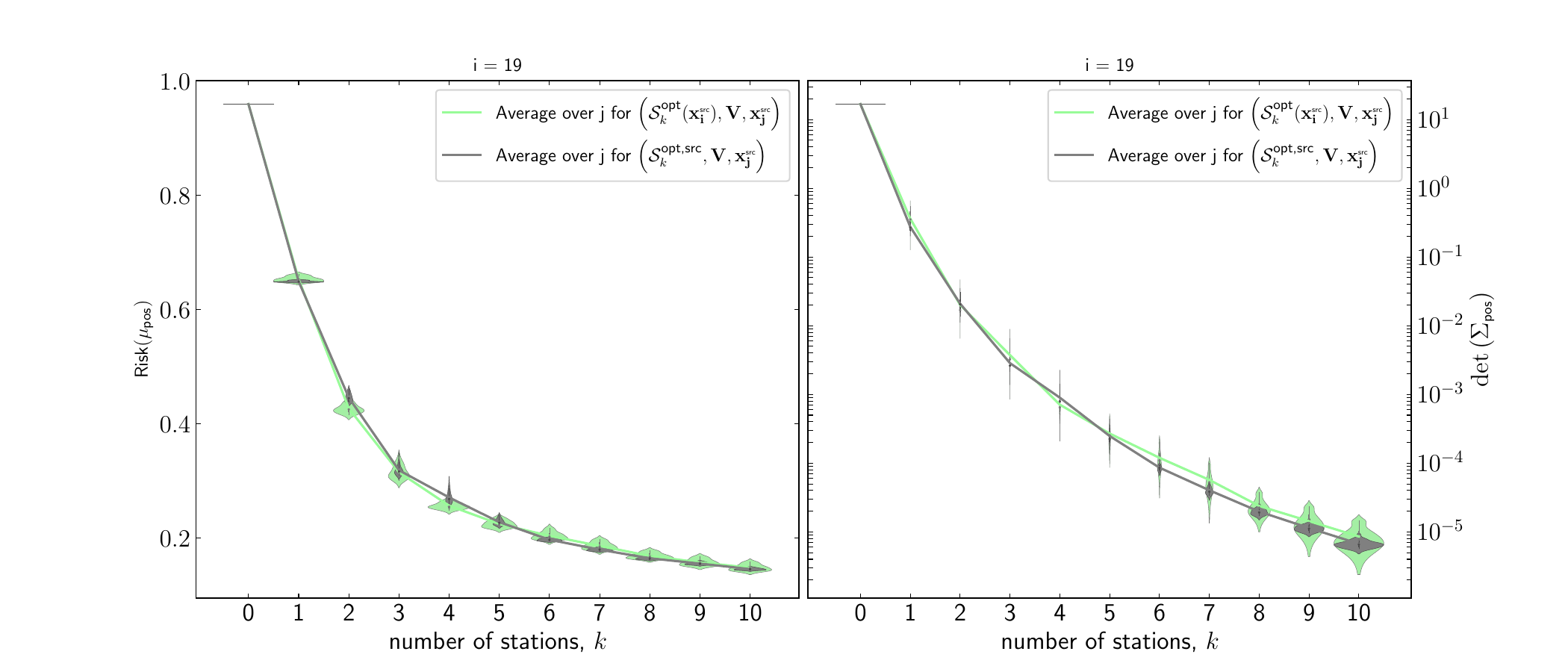}
  \caption{Performance of the consensus network against the individual greedy-optimal network $\nbr19$: Evolution of the Bayes risk distribution (left) and the posterior determinant (right) with cardinality for the consensus greedy-optimal network and the greedy-optimal network associated with the location candidate 19. }
  \label{fig_src:perf_cons_netw1}
\end{figure}
\begin{figure}
  \centering
  \includegraphics[width=.7\textwidth]{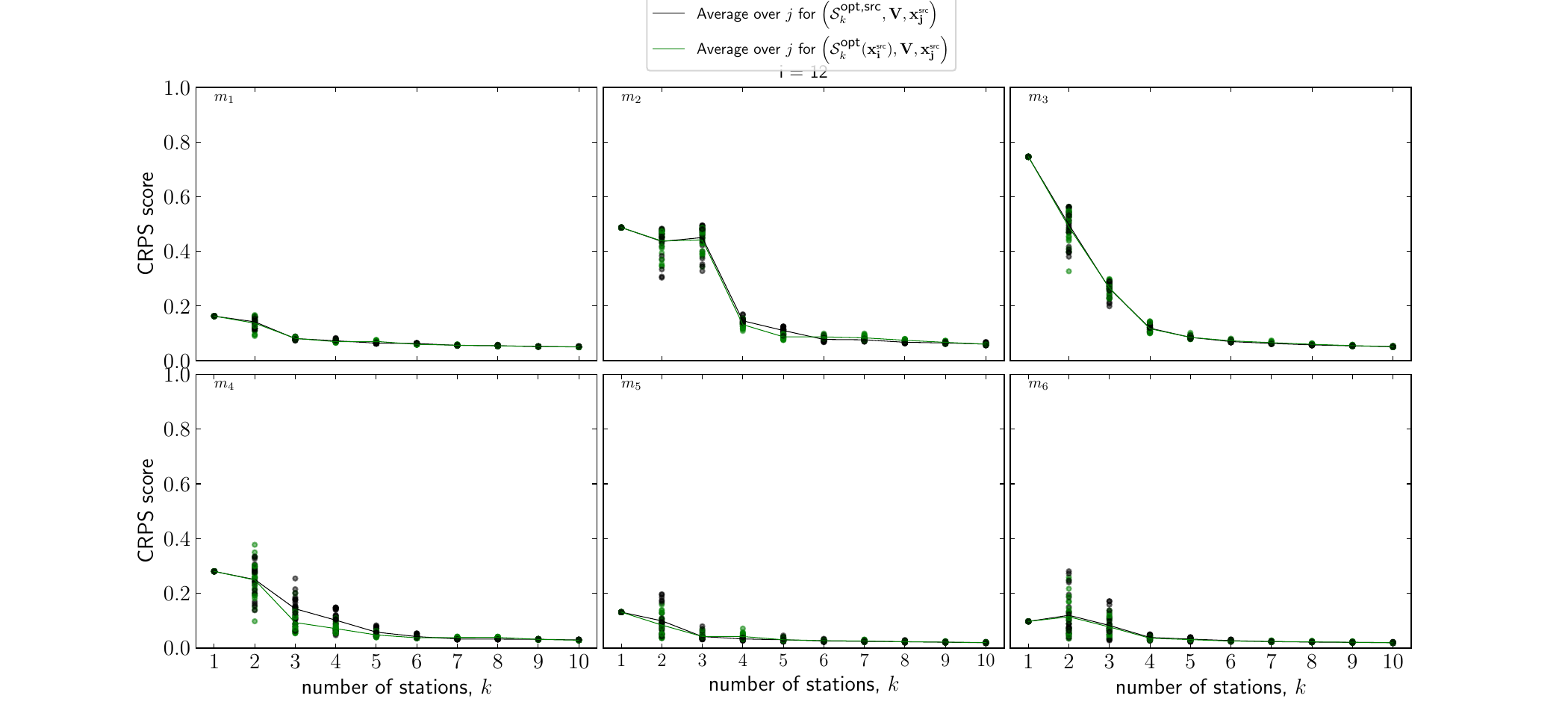}
  \caption{Scattered distribution and mean of CRPS score achieved by the consensus greedy-optimal network (in black) and by the individual greedy-optimal network optimized for event $\nbr12$ (in green). Each panel provides results for one component $m_i$ of the MT.}
  \label{fig_src:cprs_cons_netw2}
\end{figure}
 The left-plot (respectively the right-plot) in Figure \ref{fig_src:perf_cons_netw1} shows the distribution, over all the assumed locations, of the Bayes risk (respectively the determinant of the posterior covariance $\det \left( \Sigpost \right)$) in terms of the number of stations $k$ for the consensus greedy-optimal network and for the greedy-optimal network optimized for event $\nbr 19$. 
 
Figure \ref{fig_src:cprs_cons_netw2} compares the average CRPS score of the consensus greedy-optimal network and the greedy-optimal network over the distribution of 25 locations, and displays a similar performance of the two networks.         
 Practically, it is important to assess also the consensus greedy-optimal network against a greedy-optimal network associated with a source location that was not considered when deploying the consensus algorithm. Since the location $\nbr 19$, associated with the greedy-optimal network, is used in building process of the consensus greedy-optimal network, we observe in  Appendix \ref{app:test1}  the evolution of the posterior quality for the consensus greedy-optimal network against greedy-optimal networks (Figure \ref{fig_src:perf_cons_netw1_app} for the candidate location $\nbr 23$  and  Figure \ref{fig_src:crps_cons_netw2_app} for the candidate location $\nbr 21$)  that were not included in the design of the consensus greedy-optimal network. 

The main observation here is that the consensus greedy-optimal network outperforms the greedy-optimal networks in reducing the uncertainty level in the MT estimate, on average for a distribution of earthquake source locations. This becomes more clear with  an increase in the number of stations.

\paragraph{Test 2: an analysis based on the source location.}
Each chosen earthquake location within an assumed distribution of locations provides a greedy-optimal network while the consensus greedy-optimal network captures the aggregation of the information from all 20 locations. The task here consists in comparing the consensus greedy-optimal network to the 25 greedy-optimal networks for each source location. For a given earthquake location, we examine the average performance of the distribution of the 25 greedy-optimal networks against the performance of the consensus greedy-optimal network. In practice, we compute each measured value using the Green matrix associated with the consensus greedy-optimal network and with that for a chosen location. The same location is used together with all the 25 individual greedy-optimal networks, in turn, to construct the distribution of the measured value of the posterior PDF, for a fixed number of stations. 

\begin{figure}
  \centering
  \includegraphics[width=.7\textwidth]{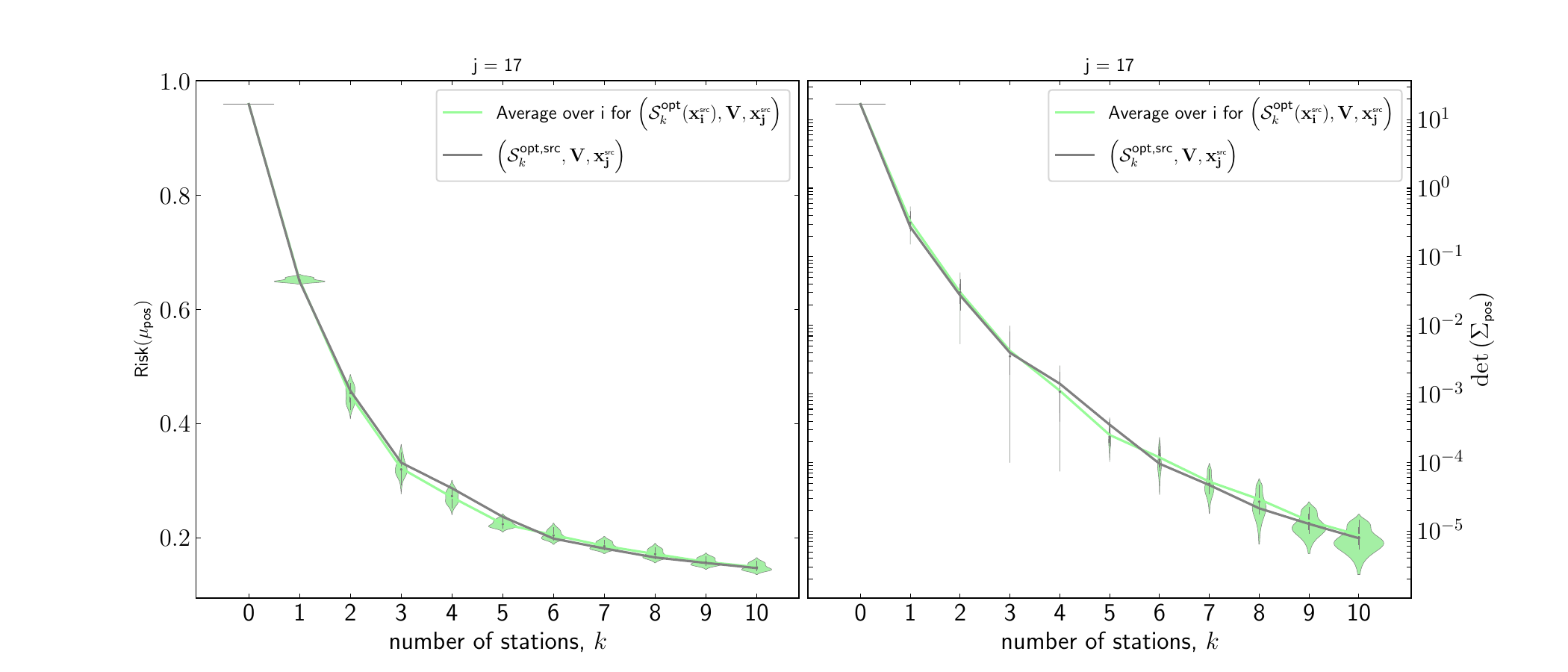}
  \caption{Performance  comparison, when looking the source location $\nbr17$, between the consensus greedy-optimal network and the average over the 25 greedy-optimal networks.}
  \label{fig_src:perf_cons_src1}
\end{figure}
\begin{figure}
  \centering
  \includegraphics[width=.7\textwidth]{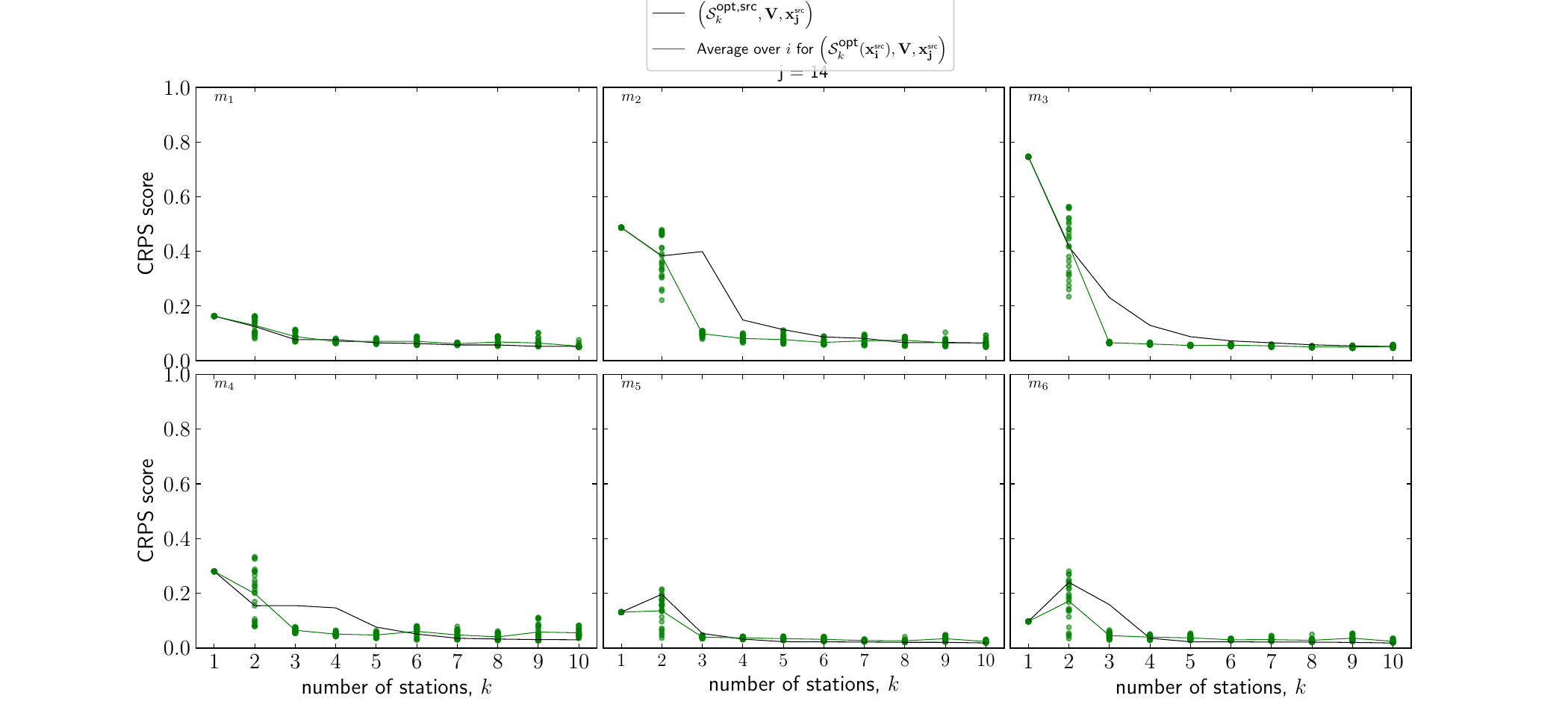}
  \caption{CRPS score of the consensus greedy-optimal network and the average CRPS score over the 25 greedy-optimal networks,  when looking the source location $\nbr14$. Each panel provides results for one component $m_i$ of the MT.}
  \label{fig_src:cprs_cons_src2}
\end{figure}

Figures \ref{fig_src:perf_cons_src1}  shows the evolution of the distributions  of the Bayes risk and the determinant of the posterior covariance  when the number of stations varies. The green-line, in both figures, is the average of the distribution for a given cardinality while the gray-line depicts the evolution of those quantities of measure for the consensus greedy-optimal network in terms of the number of stations. A more detailed behavior for each marginal posterior distribution is shown in Figure \ref{fig_src:cprs_cons_src2}, where it is observed that for a low cardinality, the consensus greedy-optimal network falls behind the average greedy-optimal network for some components  of MT but improves for higher cardinality to level the performance of the average greedy-optimal network.
Here also we present two types of source location: one that participates as a possible location to be used in the building process of the consensus greedy-optimal network and the second type of source location that is not involved, see Appendix \ref{app:test2} (Figure \ref{fig_src:perf_cons_src1_app} with source location $\nbr 24$ and Figure \ref{fig_src:cprs_cons_src2_app} with source location $\nbr 23$). In both cases, the performance of the consensus greedy-optimal network is clear.

\subsection{Performance comparison under model misspecification} \label{sec:model_misp}
Generally the analysis of an inverse problem is based on the assumption that the data generating process can be characterized exactly by the parametric model used to describe the phenomenon. In other words,  observed data  are assumed to come exactly from the forward predictive model. In reality, however, that is most often not the case, that is, the model used to describe the data does not capture the true data generating process in its entirety. We call this setting that of model misspecification. In this section, we analyze the impact of misspecified data-generating model in the performance of the consensus network in reference to the performance of the greedy-optimal networks. 
\begin{proposition}[Bayes risk]
The Bayes risk under misspecification through the forward model's Green matrix $\bm{G}$ distinct to the data-generating process's Green matrix $\tilde{\bm{G}}$ is given by
\begin{eqnarray}
   \risk(\mpost) &=& \trace(\Sigpost) + \trace\left(\Sigpost (\tilde{\bm{H}} - \bm{H})(\Sigprior +\mprior \cdot \mprior^\top)(\tilde{\bm{H}} - \bm{H})^\top \Sigpost\right) \nonumber \\
   \label{eq:bayesrisk} & & \hspace{1.5cm} - \trace\left(\Sigpost \left((\tilde{\bm{H}} - \bm{H}) + (\tilde{\bm{H}} - \bm{H})^\top \right) \Sigpost\right),
\end{eqnarray}
where $\bm{H} = \bm{G}^\top\bm{\Sigma}^{-1}_{\epsilon}\bm{G}$, and $\tilde{\bm{H}} = \bm{G}^\top\bm{\Sigma}^{-1}_{\epsilon}\tilde{\bm{G}}$.
\end{proposition}
A detailed derivation of \eqref{eq:bayesrisk} is provided in Appendix \ref{app:miss}. The motivation is to assess the applicability of the consensus approach to more realistic scenarios of MT inversion. We do so by analyzing the misspecification robustness of the consensus greedy-optimal network. As the posterior covariance is invariant to misspecified data, we typically examine the evolution of the Bayes risk \eqref{eq:bayesrisk} for the consensus greedy-optimal network and the greedy-optimal network.

\subsubsection{Misspecification analysis for the network $\mathcal{S}^{\tiny{\text{opt,vel}}}_{k}$}
For an uncertain velocity-model, we introduce misspecification by altering the velocity-model congruity in the characterization of the Green matrices, which means that the data $\bm{Y}$ generation occurs with a velocity-model different from the one in the formulation of the forward predictive model. That is the Green matrix $\tilde{\bm{G}} \myeq \tilde{\bm{G}}\left(\xs, {\bm{V}}_i, \mathcal{S}_k \right)$ while the Green matrix for the forward predictive model is  $\bm{G} \myeq \bm{G}\left(\xs, \bm{V}_j, \mathcal{S}_k\right)$ for $i \neq j$.  
\begin{figure}
  \centering
  \includegraphics[width=.3\textwidth]{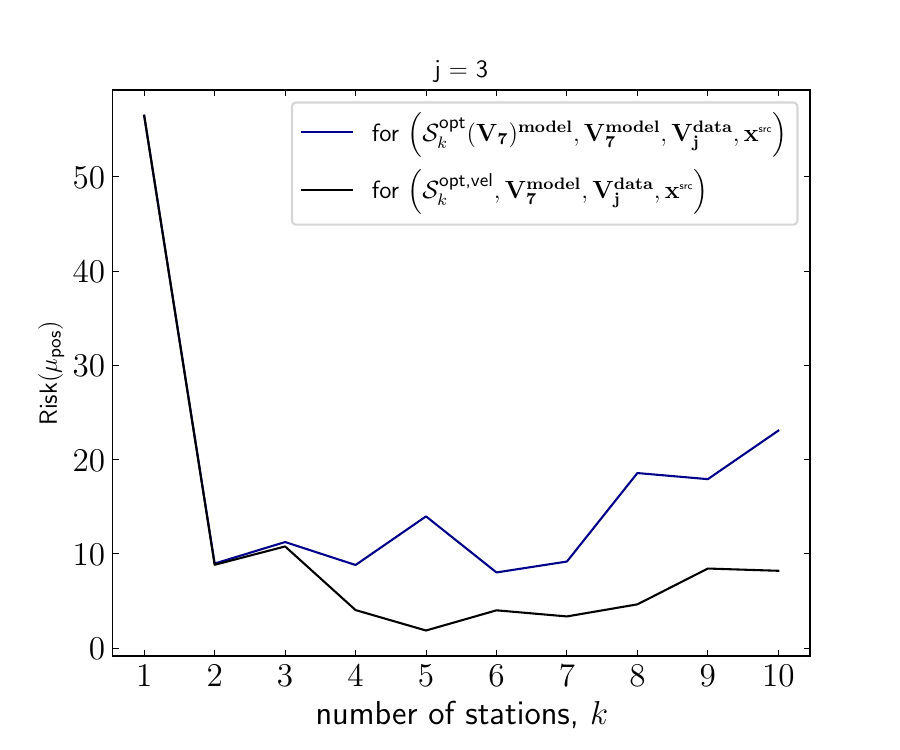}
  \caption{Bayes risk comparison between the consensus greedy-optimal network and the greedy-optimal network associated with velocity-model $\bm{V}_3$ in misspecified setting.}
  \label{fig_miss_vel:risk_7_vel_3}
\end{figure}

\begin{figure}
  \centering
  \includegraphics[width=.3\textwidth]{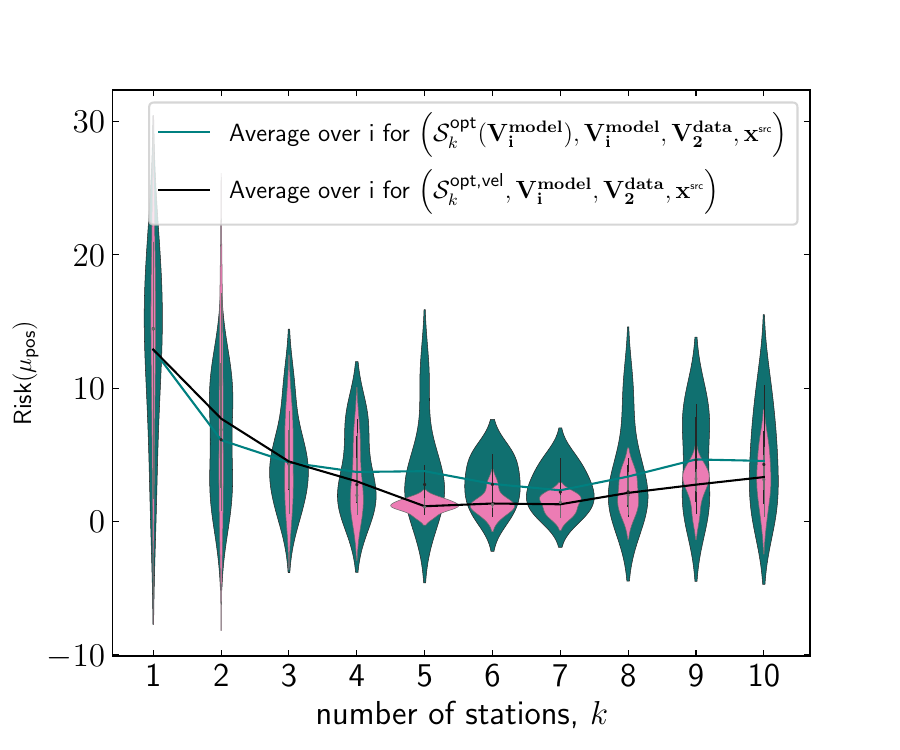}
  \includegraphics[width=.3\textwidth]{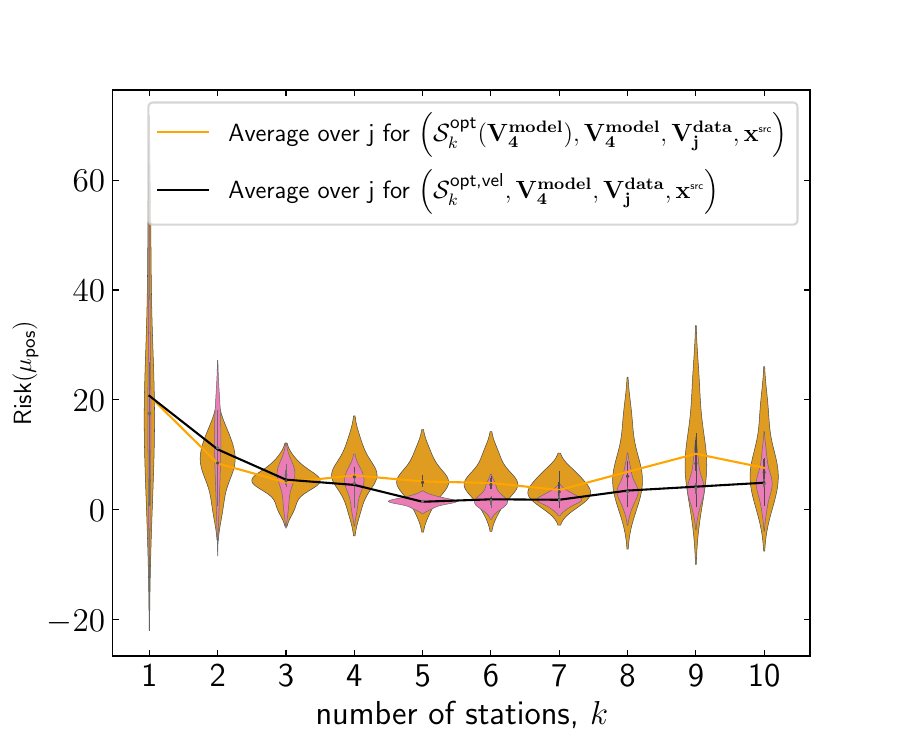}
  \caption{Comparison of distributions of Bayes risk comparison  the consensus greedy-optimal network and the greedy-optimal network associated with velocity-model $\bm{V}_3$ in misspecified setting}
  \label{fig_miss_vel:risk_2_risk4}
\end{figure}

We start by examining a single scenario where the data are assumed to be generated from the velocity-model $\bm{V}_3$ while the Green matrix in the forward predictive model considers the velocity-model $\bm{V}_7$. The value of the Bayes risk \eqref{eq:bayesrisk} achieved by the consensus greedy-optimal network and by the greedy-optimal network  associated with the velocity-model $\bm{V}_7$ are shown in Figure \ref{fig_miss_vel:risk_7_vel_3}. We note the superiority of the consenus greedy-optimal network in reducing the loss of inverting MT under misspecification. That observation is also, at a more general level, seen when we consider the distribution of the Bayes risk over the velocity-models  in the forward predictive model (left-plot Figure \ref{fig_miss_vel:risk_2_risk4} with misspecified data collected from the velocity-model  $\bm{V}_2$) and over the velocity-models in the data collection process (right-plot of Figure \ref{fig_miss_vel:risk_2_risk4} for a forward predictive model built with the velocity-model $\bm{V}_4$). 

\begin{figure}
  \centering
  \includegraphics[width=.3\textwidth]{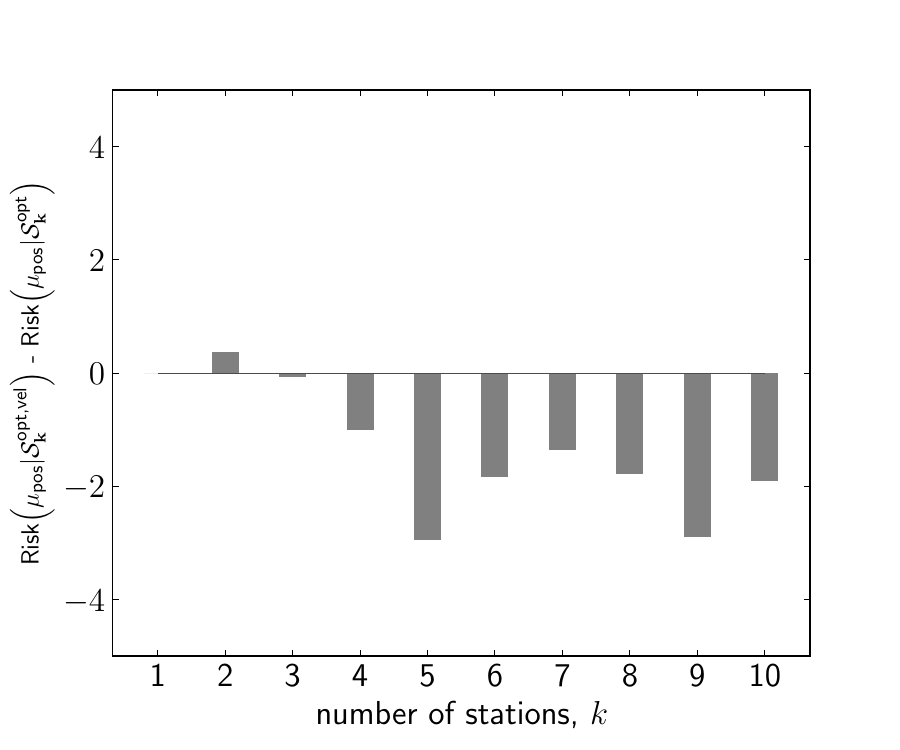}
  \includegraphics[width=.3\textwidth]{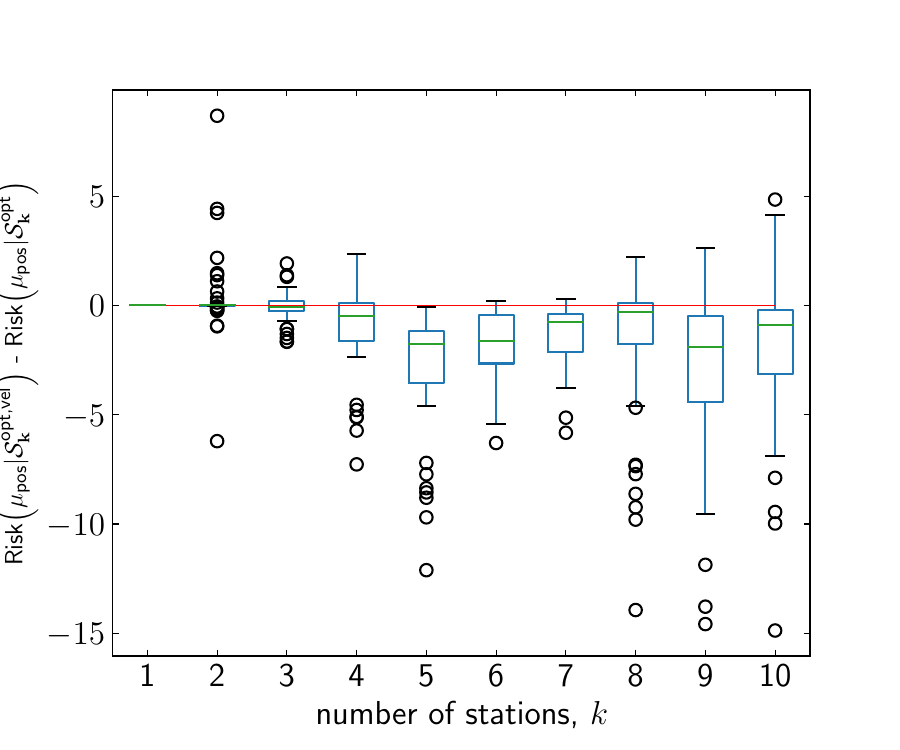}
  \caption{Bar-plot (left) and Box-plot (right) of the distribution of the difference of Bayes risk over all possible misspecified settings with the velocity-model.}
  \label{fig_miss_vel:risk_all_bar_box}
\end{figure}

To get a more general picture of the performance comparison, we look at the distribution over all possible misspecified settings given by two distincts velocity-models in the forward predictive model and the data-generating model, of the difference of the Bayes risk between the consensus greedy-optimal network and the set of all greedy-optimal networks. The bar-plot representation (left-plot of Figure \ref{fig_miss_vel:risk_all_bar_box}) of the difference distributions indicates a clear advantage of the consensus greedy-optimal network when considering a number of stations higher than three. Representation in box-plot (right-plot of Figure \ref{fig_miss_vel:risk_all_bar_box}) shows the details of the comparison for fixed cardinality. At the end, the consensus approach provides a network of stations, without accessing  the full structure of the velocity-model, that is as robust as the greedy-optimal network based on the assumption of available layered velocity-model, and that is more performance in reducing the loss of inverting MT in misspecified settings. 

\subsubsection{Misspecification analysis for the network $\mathcal{S}^{\tiny{\text{opt,src}}}_{k}$}
\begin{figure}
  \centering
  \includegraphics[width=.3\textwidth]{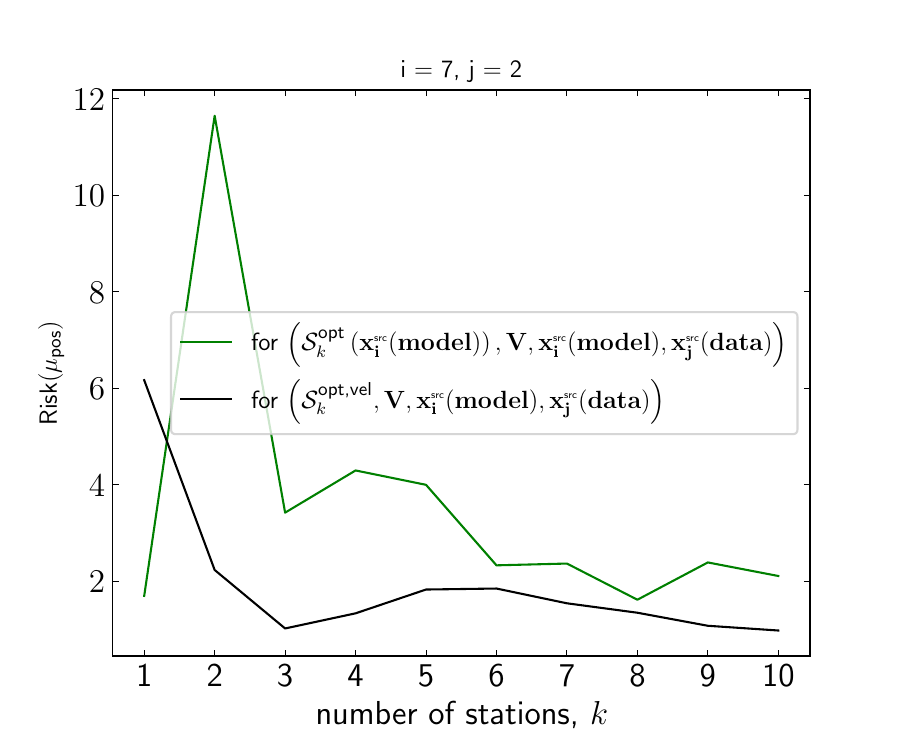}
  \includegraphics[width=.3\textwidth]{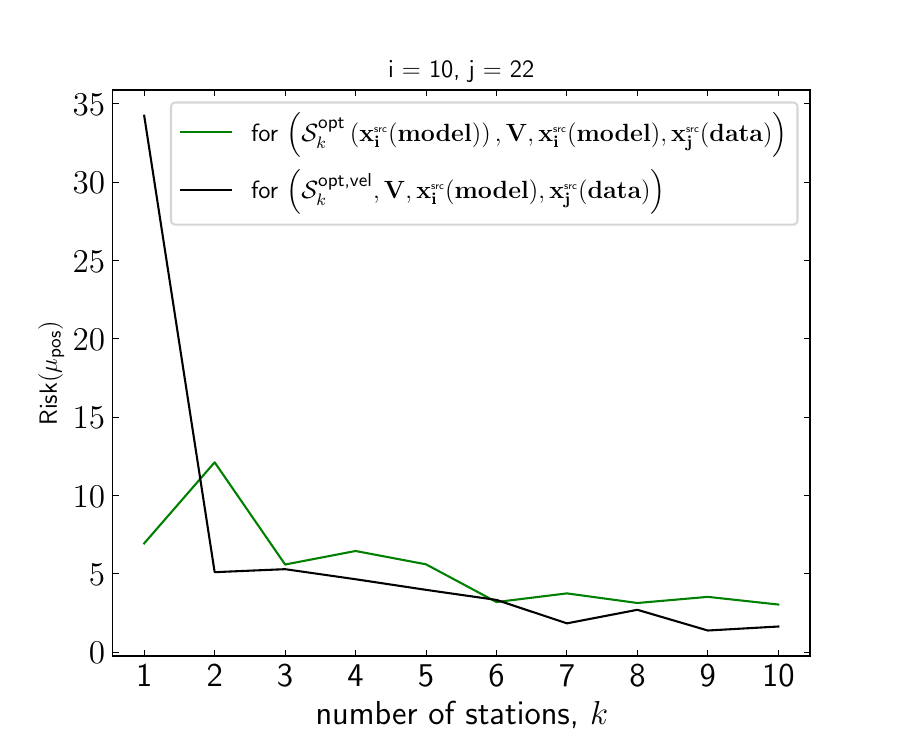}
  \caption{Bayes risk comparison, under misspecification, between the consensus greedy-optimal network and greedy-optimal networks. The misspecification is formulated as follows: the forward model is associated with the location candidate $i$ while the data generating process considers the location candidate $j$.}
  \label{fig_miss_src:risk_7_10}
\end{figure}

In this section, we analyze the robustness of the consensus greedy-optimal network by distinguishing   the Green matrix $\bm{G}$ in the forward predictive model from the Green matrix $\tilde{\bm{G}}$ in the data acquisition  through the earthquake-source location used for the inference: $\bm{G} \myeq \bm{G}\left(\xs_i, {\bm{V}}, \mathcal{S}_k \right)$ and $\bm{G} \myeq \bm{G}\left(\xs_j, {\bm{V}}, \mathcal{S}_k \right)$, $i \neq j$ for a given network $\mathcal{S}_k$, and a velocity-model  $\bm{V}$. To investigate the robustness of the consensus approach for uncertain earthquake-source location, we start by inspecting the variation of the Bayes risk using the consensus greedy-optimal network and the greedy-optimal network, both combined with the source location $\xs_7$ and with misspecified data from the location $\xs_2$ (left-plot Figure \ref{fig_miss_src:risk_7_10}). The right-plot of Figure \ref{fig_miss_src:risk_7_10} displays the analogy of the consensus greedy-optimal network and the greedy-optimal network  via the Bayes risk variation in terms of the number of stations when the data misspecification is through a source location that was not considered in the construction of the consensus greedy-optimal network. In both cases, we remark that the consensus-greedy optimal network surpasses the greedy-optimal network in compressing the level of uncertainty in the MT estimate.

\begin{figure}
  \centering
  \includegraphics[width=.3\textwidth]{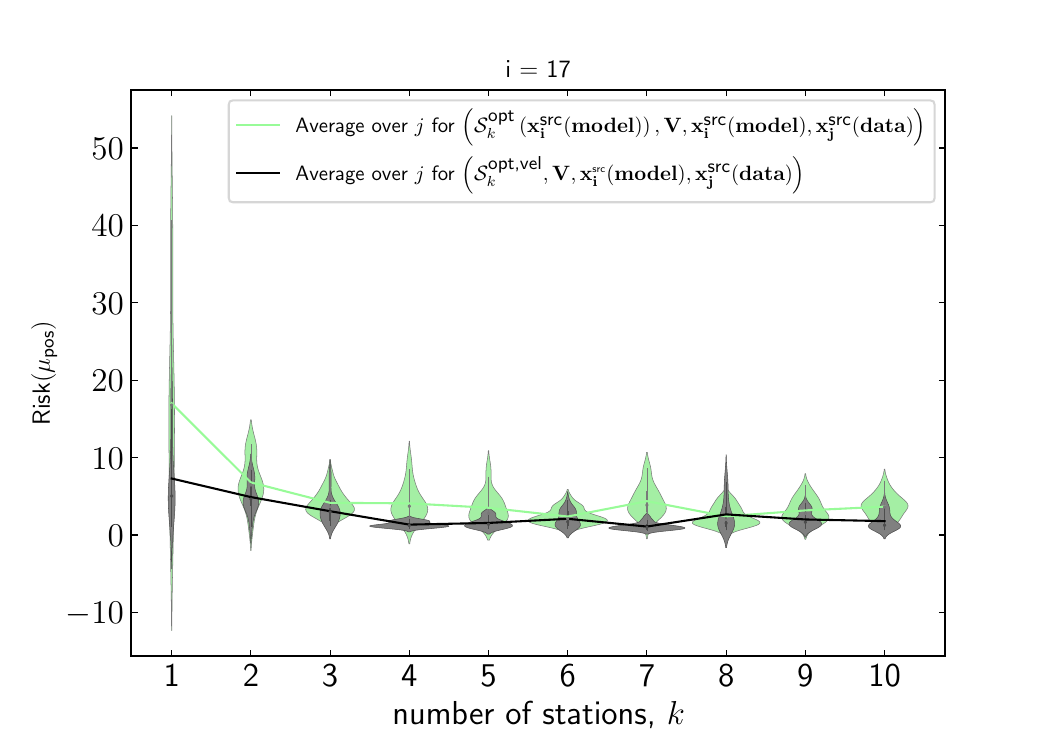}
  \includegraphics[width=.3\textwidth]{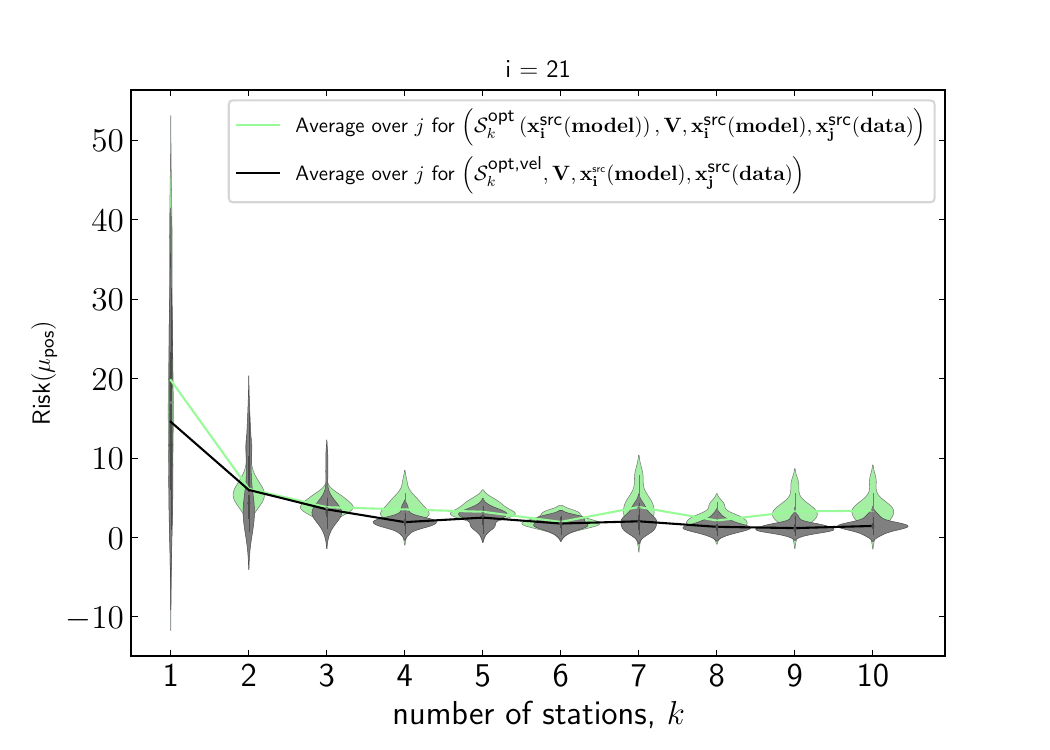}
  \caption{Distributions of Bayes risk, over a sample of Green matrix $\tilde{\bm{G}}$, for  the consensus greedy-optimal network and the greedy-optimal network associated with earthquake-source location $\xs_{17}$ (left-plot) and earthquake-source location $\xs_{21}$ (right-plot).}
  \label{fig_miss_src:risk_all_data_17_21}
\end{figure}

\begin{figure}
  \centering
  \includegraphics[width=.3\textwidth]{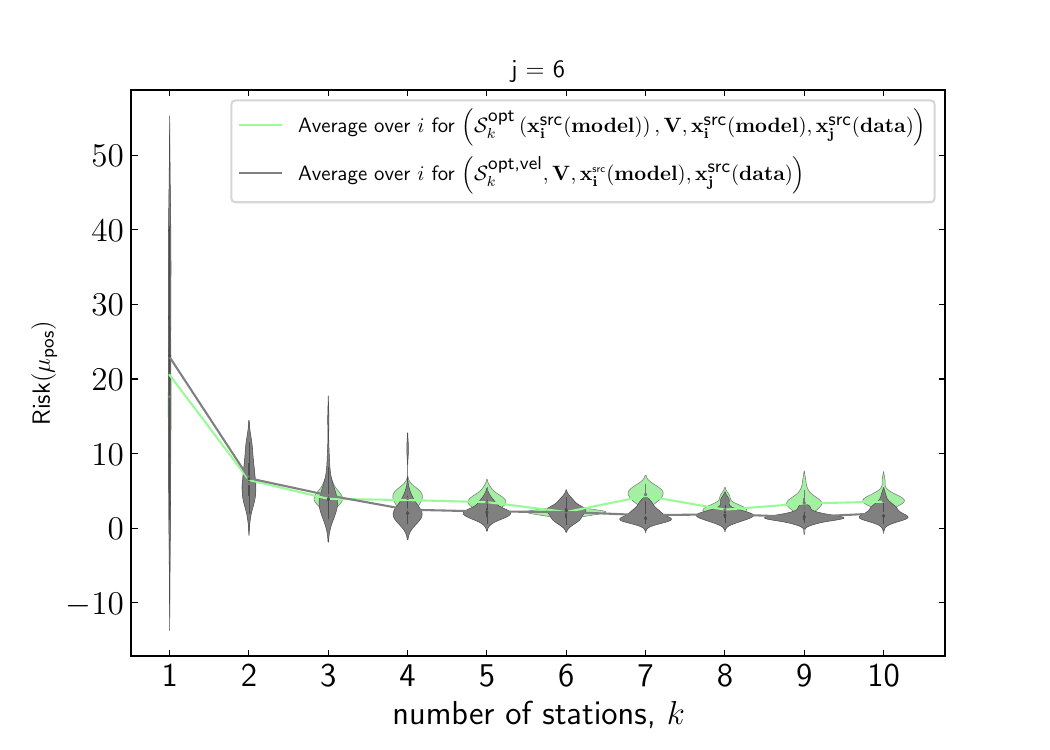}
  \includegraphics[width=.3\textwidth]{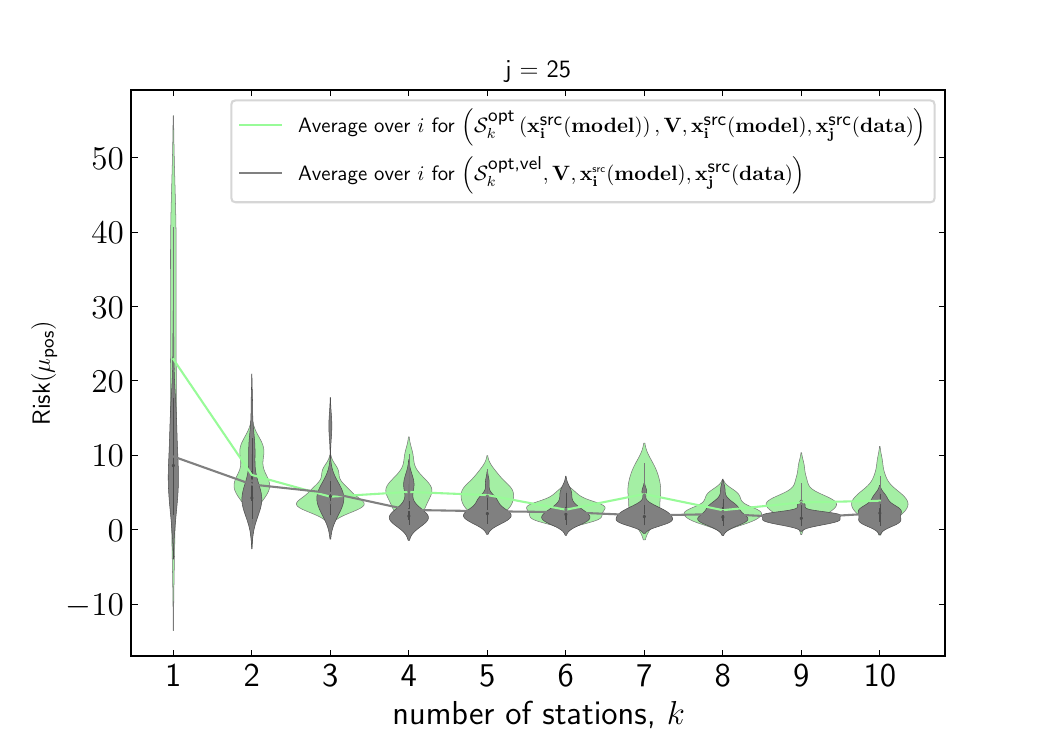}
  \caption{Distributions of Bayes risk, over a sample of Green matrix $\bm{G}$, for  the consensus greedy-optimal network and the greedy-optimal network associated with earthquake-source location $\xs_{6}$ (left-plot) and earthquake-source location $\xs_{25}$ (right-plot)}
  \label{fig_miss_src:risk_all_model_6_25}
\end{figure}

Additonal numerical results confirm the last statement to a greater extent  when comparing the distribution of the Bayes risks over a sample of earthquake-source locations misspecifying the data generating process combined with a given source location for the Green matrix in the forward predictive model (Figures \ref{fig_miss_src:risk_all_data_17_21}), and when comparing the distribution of the Bayes risk over a sample of earthquake-source location for the Green matrix in the forward model  mixed with a given earthquake-source location that misspecifies the data acquisition process (Figure \ref{fig_miss_src:risk_all_model_6_25}). 

\begin{figure}
  \centering
  \includegraphics[width=.3\textwidth]{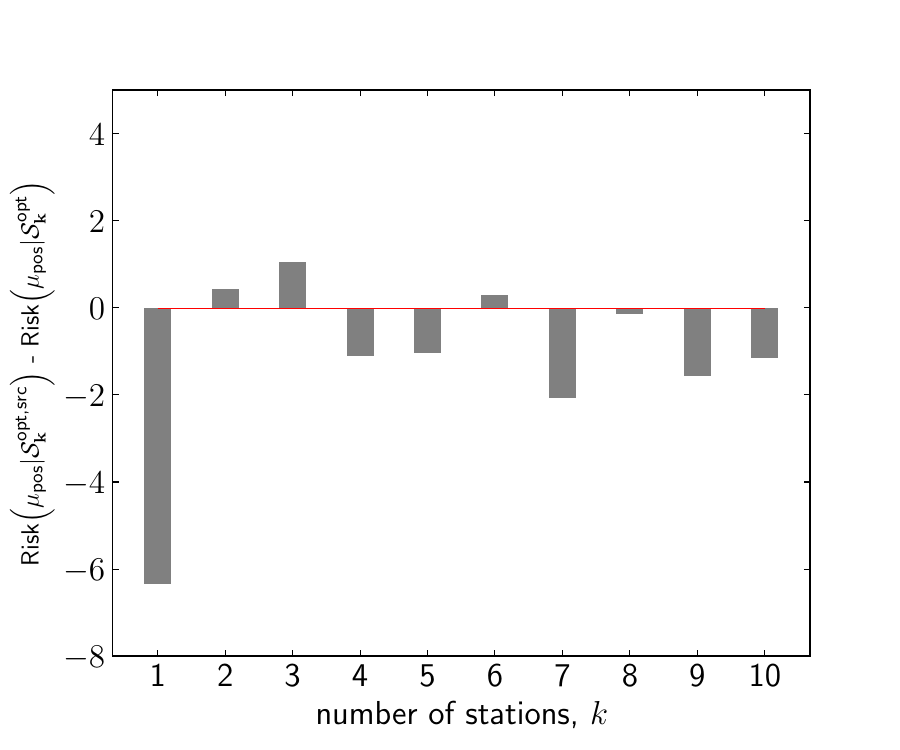}
  \includegraphics[width=.3\textwidth]{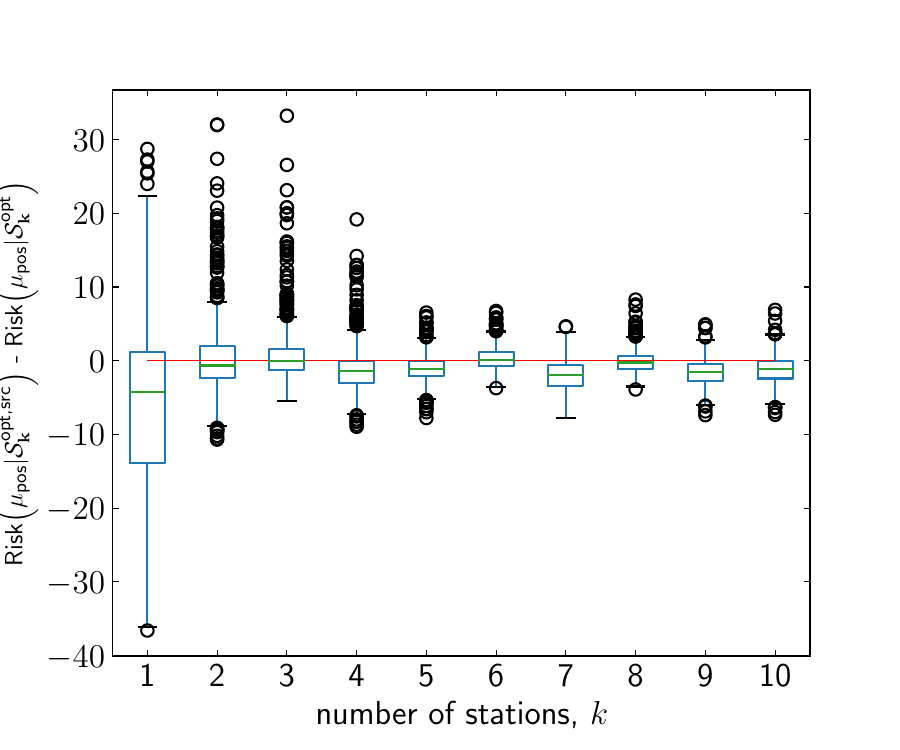}
  \caption{Bar-plot (letf) and Box-plot (right) of the distribution of the difference of Bayes risk over all possible misspecified settings with the earthquake-source location.}
  \label{fig_miss_src:risk_all_bar_box}
\end{figure}

To encapsulate the robustness of the consensus greedy-optimal network and assess its overall performance in reference to the greedy-optimal networks, we examine the Bayes risk difference between the consensus greedy-optimal network and the greedy-optimal networks. In Figure \ref{fig_miss_src:risk_all_bar_box}, we show the distribution of the Bayes risk difference: the bar-plot representation at left shows a moderate advantage of the consensus approach and this without knowing the exact earthquake-source location in comparison to the greedy-optimal network construction. 

\section{Conclusions}
We have presented and discussed a new methodology for designing station networks that reduce the uncertainty in the Bayesian learning of the seismic moment tensor. The key aspect is the formulation of the optimal sensor placement problem as a Bayesian experimental design problem with a linear conjugate-Gaussian  structure for the moment tensor inversion. One important step that alleviates the computational complexity is the adoption of a heuristic greedy approach for solving the combinatorial optimization problem resulting from the Bayesian experimental design formulation. We begun by assuming the velocity-model and the earthquake-source location are known and we performed numerical experiments to assess the level of uncertainty in the posterior distribution, and to analyze the properties of the devised greedy-optimal network. In case the velocity-model or the earthquake-source location is not accessible, numerical results confirmed the good performance of a consensus approach, which consists in addressing the maximization of the information gain averaged over the set of representations of features.  Essentially, we show that when we don't know the location of the earthquake-source or the velocity-model of the medium, there is no network of stations that can perform better than the consensus greedy-optimal network, on average. The robustness of the consensus greedy-optimal network in misspecified settings together with the simplicity of the design procedure is significant for the efficient learning of the seismic moment tensor in realistic scenarios.

One future consideration could be the analysis of that methodology with the windowing of the waveforms for the purpose of reducing further the computational complexity. Another consideration would be the appropriate bandwidth of the data, which likely is associated with the uncertainty in the velocity model. However, the important challenge for future work would be to preserve the simplicity and the ease of application of the methodology associated with the linear conjugate-Gaussian structure while assuming that the velocity-model and the earthquake-source location are both parts of the parameter-of-interest. 

\section*{Acknowledgments}
The research reported in this publication was supported by funding from the CPG (College of Petroleum Engineering \& Geosciences) at
King Fahd University of Petroleum and Minerals, under Global Partnership Program with MIT (Massachusetts Institute of Technology).

\bibliographystyle{acm}
\bibliography{reference}

\newpage
\appendix

\section{Complement plots for Test 1}\label{app:test1}

\begin{figure}[H]
  \centering
  \includegraphics[width=.6\textwidth]{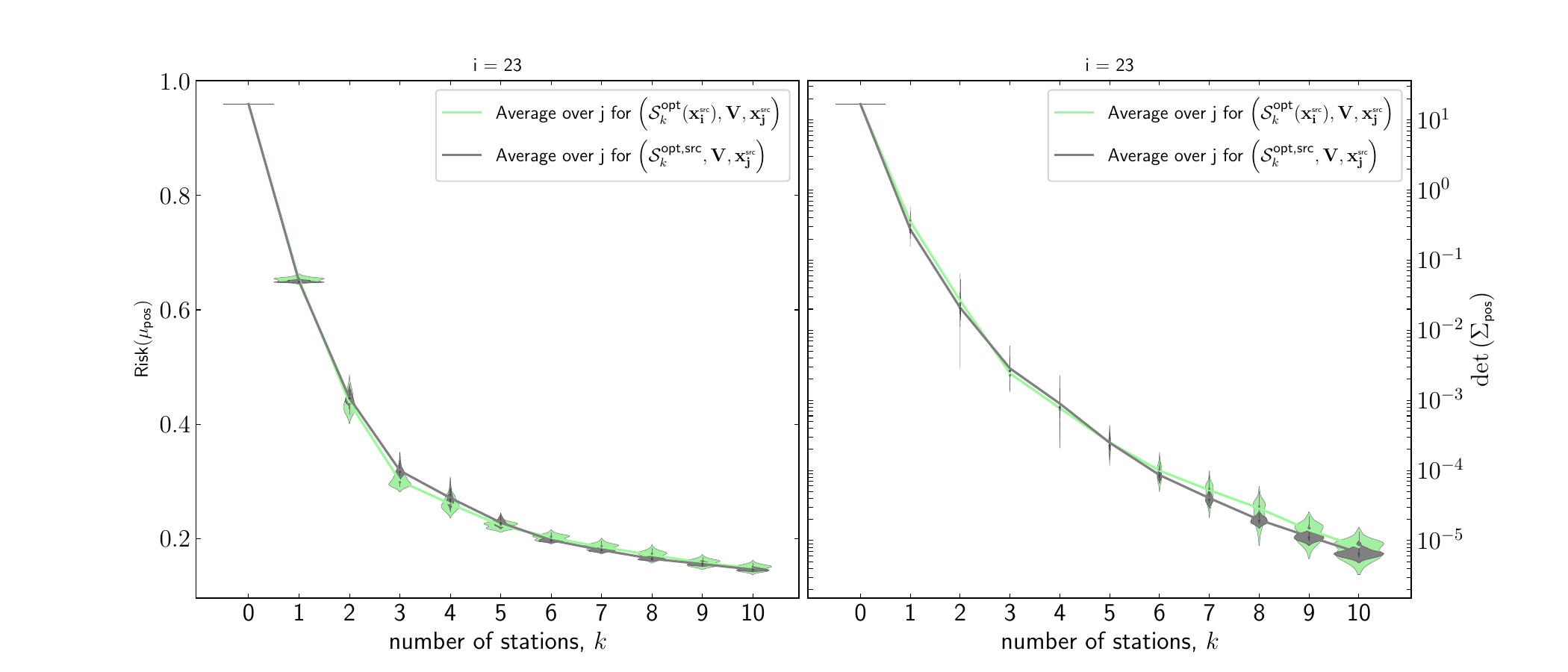}
  \caption{Performance of the consensus network against the individual greedy-optimal network $\nbr23$}
  \label{fig_src:perf_cons_netw1_app}
\end{figure}

\begin{figure}[H]
  \centering
  \includegraphics[width=.6\textwidth]{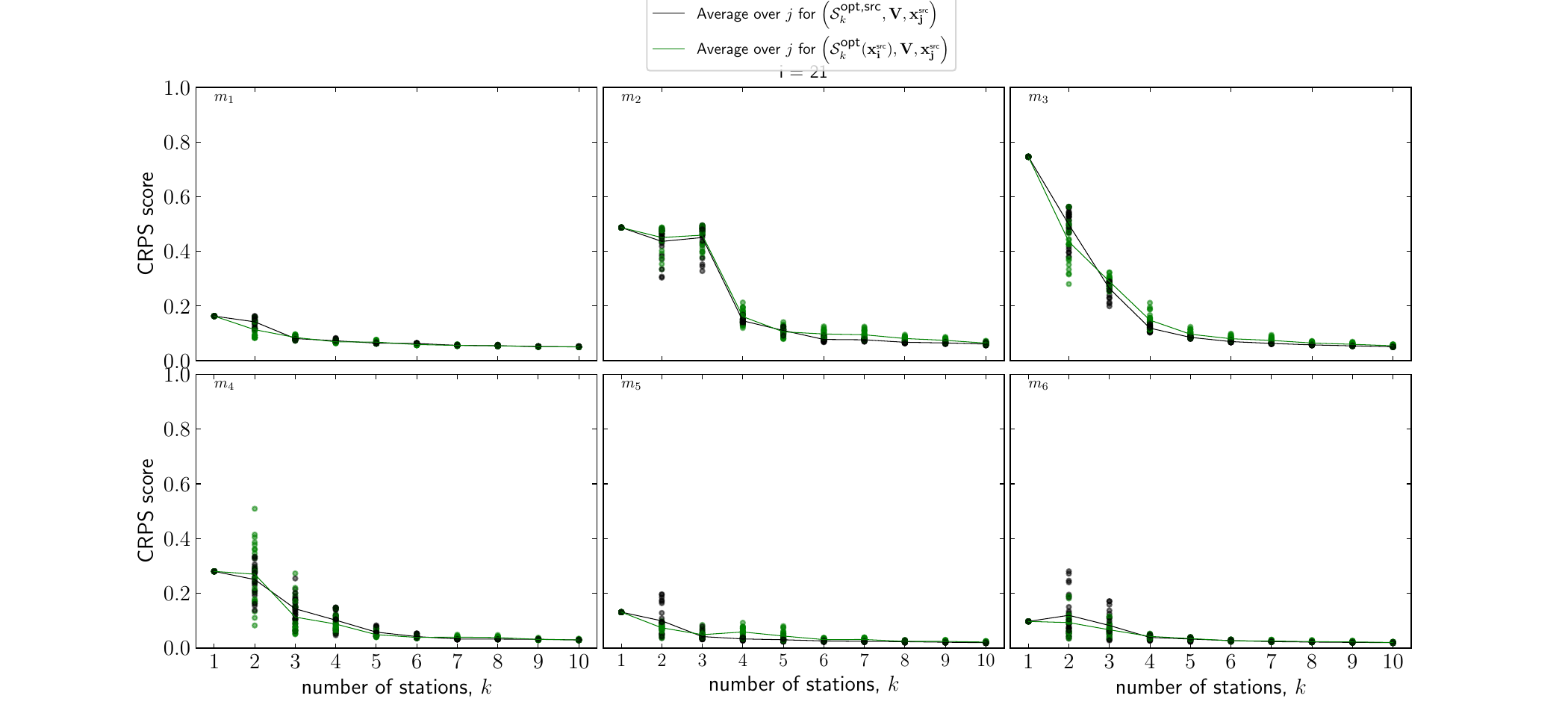}
  \caption{Performance of the consensus network against the individual greedy-optimal network $\nbr21$}
  \label{fig_src:crps_cons_netw2_app}
\end{figure}

\section{Complement plots for Test 2}\label{app:test2}
\begin{figure}[H]
  \centering
  \includegraphics[width=.7\textwidth]{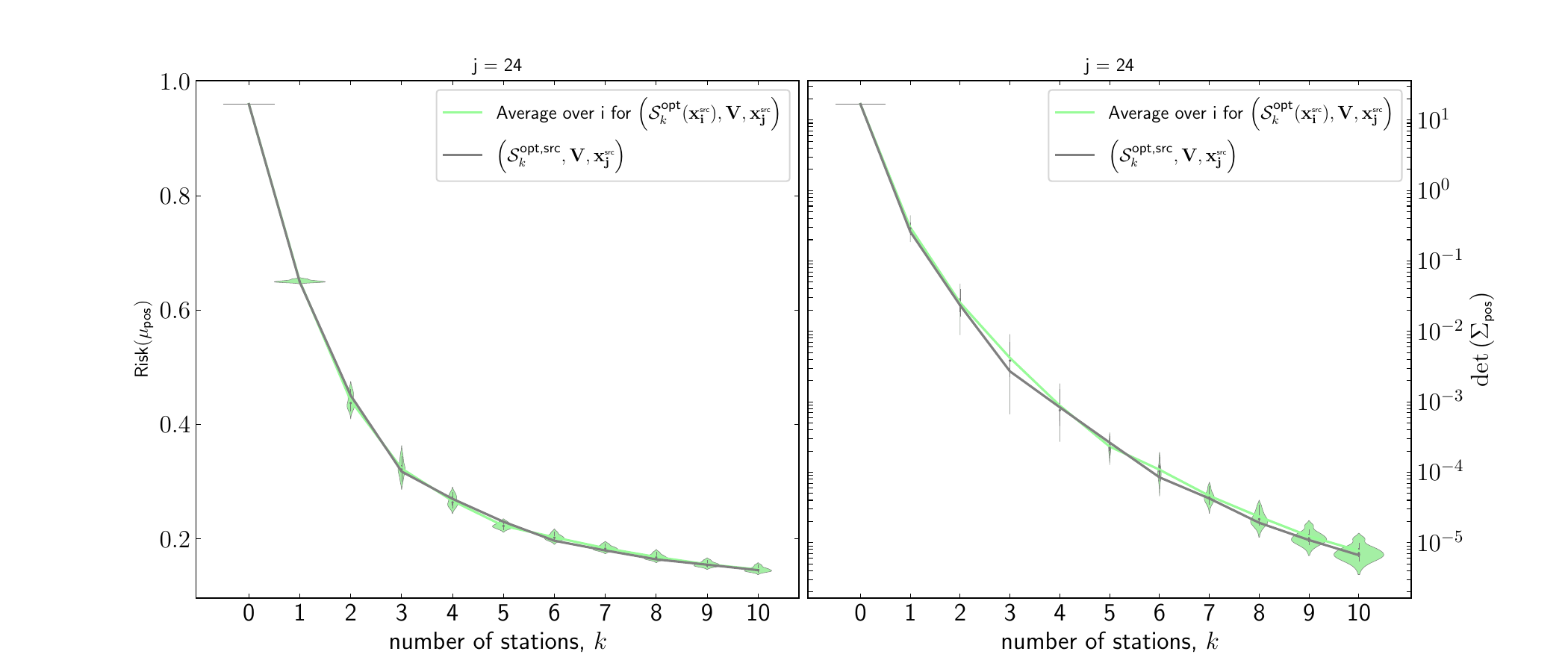}
  \caption{Performance of the consensus network when looking the source}
  \label{fig_src:perf_cons_src1_app}
\end{figure}
\begin{figure}[H]
  \centering
  \includegraphics[width=.7\textwidth]{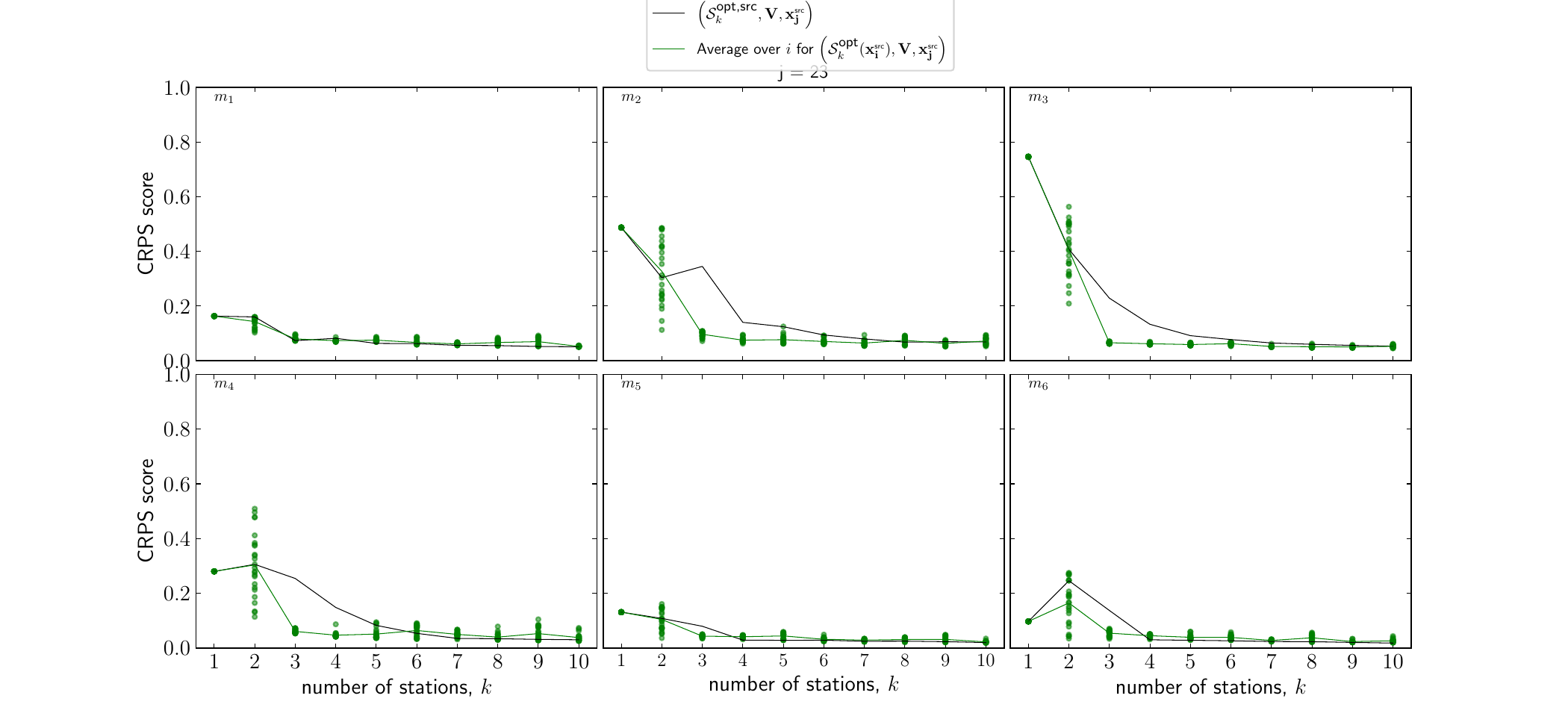}
  \caption{Performance of the consensus network when looking the source}
  \label{fig_src:cprs_cons_src2_app}
\end{figure}

\section{Derivation of the Bayes risk in misspecification}\label{app:miss}
Under model misspecification, the data is collected from the network of stations $\mathcal{S}_{k,\tiny{\text{data}}}$ that is different from the one we use for the inference, $\bm{Y} = \tilde{\bm{G}} \bm{m}^\top + \bm{\varepsilon}$. That affects the posterior mean but the posterior covariance matrix remains the same \eqref{eq:post_momentsY}. By definition, the Bayes risk is
\begin{eqnarray}
  \risk = \mathbb{E}_{\bm{m}} \left[ \left\|\mpost - \mathbb{E}_{\bm{Y}|\bm{m}}\left[\mpost \right]  \right\|^2\right] +  
          \mathbb{E}_{\bm{m}} \left[\left\| \mathbb{E}_{\bm{Y}|\bm{m}}\left[ \mpost  \right]- \bm{m} \right\|^2 \right].
\end{eqnarray}
Using the expression of the posterior mean, we have
\begin{eqnarray*}
  \mathbb{E}_{\bm{Y}|\bm{m}}\left[\mpost \right] &=& \mathbb{E}_{\bm{Y}|\bm{m}}\left[ \Sigpost \left(\Sigprior^{-1} \mprior + \bm{G}^\top\bm{\Sigma}^{-1}_{\epsilon}\bm{Y}\right)\right] \quad \hbox{with} \quad \bm{Y} = \tilde{\bm{G}} \bm{m}^\top\\
  &=&  \Sigpost \left(\Sigprior^{-1} \mprior + \bm{G}^\top\bm{\Sigma}^{-1}_{\epsilon}\tilde{\bm{G}} \bm{m}^\top\right).
\end{eqnarray*}
We introduce the multivariate random variable $\bm{Z}$
\begin{eqnarray*}
   \bm{Z} &=& \mpost  - \mathbb{E}_{\bm{Y}|\bm{m}}\left[\mpost \right]\\
          &=& \Sigpost \left(\Sigprior^{-1} \mprior + \bm{G}^\top\bm{\Sigma}^{-1}_{\epsilon}\bm{Y}\right) -  \Sigpost \left(\Sigprior^{-1} \mprior + \bm{G}^\top\bm{\Sigma}^{-1}_{\epsilon}\tilde{\bm{G}} \bm{m}^\top\right) \\
          &=&  \Sigpost \bm{G}^\top\bm{\Sigma}^{-1}_{\epsilon}  \left(\bm{Y} -  \tilde{\bm{G}} \bm{m}^\top\right).
\end{eqnarray*}
The expectation of $\bm{Z}$ with respect to the data $\bm{Y}$ is 
\begin{eqnarray*}
  \mathbb{E}_{\bm{Y}|\bm{m}}\left[\bm{Z} \right] 
  &=& \mathbb{E}_{\bm{Y}|\bm{m}}\left[ \Sigpost \bm{G}^\top\bm{\Sigma}^{-1}_{\epsilon}  \left(\bm{Y} -  \tilde{\bm{G}} \bm{m}^\top\right) \right] \\
  &=&   \Sigpost \bm{G}^\top\bm{\Sigma}^{-1}_{\epsilon}  \left(\tilde{\bm{G}} \bm{m}^\top -  \tilde{\bm{G}} \bm{m}^\top\right) \\
  &=& 0.
\end{eqnarray*}
Since $\bm{Y} \sim \mathcal{N}\left(\tilde{G} \bm{m}^\top, \bm{\Sigma}_{\epsilon} \right)$, the random variable $\bm{Z}$ is zero-mean Gaussian wit variance
\begin{eqnarray*}
  \mathbb{V}_{\bm{Y}|\bm{m}}\left[\bm{Z} \right] 
  &=&  \mathbb{V}_{\bm{Y}|\bm{m}}\left[ \Sigpost \bm{G}^\top\bm{\Sigma}^{-1}_{\epsilon}  \left(\bm{Y} -  \tilde{G} \bm{m}^\top\right)  \right]\\
  &=&  \mathbb{V}_{\bm{Y}|\bm{m}}\left[ \Sigpost \bm{G}^\top\bm{\Sigma}^{-1}_{\epsilon} \bm{Y}  \right]\\
  &=&  \trace\left(  \Sigpost \bm{G}^\top\bm{\Sigma}^{-1}_{\epsilon} \mathbb{V}_{\bm{Y}|\bm{m}}\left[\bm{Y}  \right]  \left( \Sigpost \bm{G}^\top\bm{\Sigma}^{-1}_{\epsilon}\right)^\top \right) \\
  &=&  \trace\left(  \Sigpost \bm{G}^\top\bm{\Sigma}^{-1}_{\epsilon}\bm{\Sigma}_{\epsilon}   \left( \Sigpost \bm{G}^\top\bm{\Sigma}^{-1}_{\epsilon}\right)^\top \right) \\
 &=&  \trace\left(  \Sigpost \bm{G}^\top  \left( \Sigpost \bm{G}^\top\bm{\Sigma}^{-1}_{\epsilon}\right)^\top \right) \\
 &=&  \trace\left(  \Sigpost \bm{G}^\top  \bm{\Sigma}^{-1}_{\epsilon}  \bm{G} \Sigpost \right) \quad \text{since } \Sigpost \text { and }{\Sigma}_{\epsilon} \text{ are symmetric}.
\end{eqnarray*} 
By definition the variance of $\bm{Z}$ is 
\begin{eqnarray*}
   \mathbb{V}_{\bm{Y}|\bm{m}}\left[\bm{Z} \right] 
   &=&  \mathbb{E}_{\bm{Y}|\bm{m}}\left[ \left\|\bm{Z}  - \mathbb{E}_{\bm{Y}|\bm{m}}\left[\bm{Z} \right] \right\|^2 \right] \\
   &=&  \mathbb{E}_{\bm{Y}|\bm{m}}\left[ \left\|\bm{Z}  \right\|^2 \right] \quad \text{because} \quad  \mathbb{E}_{\bm{Y}|\bm{m}}\left[\bm{Z} \right] = 0\\
   &=&  \mathbb{E}_{\bm{Y}|\bm{m}}\left[ \left\| \mpost  - \mathbb{E}_{\bm{Y}|\bm{m}}\left[\mpost \right]  \right\|^2 \right],
\end{eqnarray*}
which implies that
\begin{eqnarray}
\label{eq:risk_p1}
  \mathbb{E}_{\bm{Y}|\bm{m}}\left[ \left\| \mpost  - \mathbb{E}_{\bm{Y}|\bm{m}}\left[\mpost \right]  \right\|^2 \right] =  \trace\left(  \Sigpost \bm{G}^\top  \bm{\Sigma}^{-1}_{\epsilon}  \bm{G} \Sigpost^\top \right).
\end{eqnarray}

Let us introduce another random variable $\bm{W} = \mathbb{E}_{\bm{Y}|\bm{m}}\left[ \mpost \right] - \bm{m}$, its expectation is
\begin{eqnarray*}
  \mathbb{E}_{\bm{m}}\left[ \bm{W}  \right]  
  &=& \mathbb{E}_{\bm{m}}\left[ \Sigpost \left(\Sigprior^{-1} \mprior + \bm{G}^\top\bm{\Sigma}^{-1}_{\epsilon}\bm{Y}\right) - \bm{m}\right]  \quad \hbox{with} \quad \bm{Y} = \tilde{G} \bm{m}^\top\\
  &=&  \Sigpost \Sigprior^{-1} \mprior + \mathbb{E}_{\bm{m}}\left[ \Sigpost \bm{G}^\top\bm{\Sigma}^{-1}_{\epsilon}\tilde{\bm{G}}\bm{m}\right] - \mathbb{E}_{\bm{m}}\left[ \bm{m}\right] \\
   &=&  \Sigpost \Sigprior^{-1} \mprior +  \Sigpost \bm{G}^\top\bm{\Sigma}^{-1}_{\epsilon}\tilde{\bm{G}}\mprior - \mprior\\
   &=&  \Sigpost \left( \Sigprior^{-1} +   \bm{G}^\top\bm{\Sigma}^{-1}_{\epsilon}\tilde{\bm{G}} - \Sigpost^{-1} \right) \mprior\\
   &=&  \Sigpost \left( \Sigprior^{-1} +   \bm{G}^\top\bm{\Sigma}^{-1}_{\epsilon}\tilde{\bm{G}} -\bm{G}^\top\bm{\Sigma}^{-1}_{\epsilon}\bm{G} - \Sigprior^{-1} \right) \mprior  \quad \text{because} \quad  \Sigpost^{-1} =  \bm{G}^\top\bm{\Sigma}^{-1}_{\epsilon}\bm{G} + \Sigprior^{-1}  \\
   &=&  \Sigpost \left( \bm{G}^\top\bm{\Sigma}^{-1}_{\epsilon}\tilde{\bm{G}} -\bm{G}^\top\bm{\Sigma}^{-1}_{\epsilon}\bm{G}  \right) \mprior  \\
   &=&  \Sigpost \left( \tilde{\bm{H}} -\bm{H}  \right) \mprior,
\end{eqnarray*} 
and its variance is
\begin{eqnarray*}
 \mathbb{V}_{\bm{m}}\left[ \bm{W}  \right]  
  &=& \mathbb{V}_{\bm{m}}\left[ \mathbb{E}_{\bm{Y}|\bm{m}}\left[ \mpost \right] - \bm{m} \right] \\ 
  &=& \mathbb{V}_{\bm{m}}\left[ \Sigpost \left(\Sigprior^{-1} \mprior + \bm{G}^\top\bm{\Sigma}^{-1}_{\epsilon}\bm{Y}\right) - \bm{m}\right]  \quad \hbox{with} \quad \bm{Y} = \tilde{G} \bm{m}^\top\\
  &=& \mathbb{V}_{\bm{m}}\left[ \Sigpost \Sigprior^{-1} \mprior + \Sigpost \bm{G}^\top\bm{\Sigma}^{-1}_{\epsilon}\tilde{\bm{G}}\bm{m} - \bm{m}\right] \\
  &=& \mathbb{V}_{\bm{m}}\left[ \Sigpost \bm{G}^\top\bm{\Sigma}^{-1}_{\epsilon}\tilde{\bm{G}}\bm{m} - \bm{m}\right] \\
  &=& \mathbb{V}_{\bm{m}}\left[ \left(\Sigpost \tilde{\bm{H}} - \bm{I}_6\right)\bm{m}\right] \\
  &=& \trace \left( \left(\Sigpost \tilde{\bm{H}} - \bm{I}_6\right) \Sigprior \left(\Sigpost \tilde{\bm{H}} - \bm{I}_6\right)^\top \right).
\end{eqnarray*}
We know that
\begin{eqnarray*}
   \Sigpost \tilde{\bm{H}} - \bm{I}_6 &=& \Sigpost \left(\tilde{\bm{H}} + \bm{H} -\bm{H} + \Sigprior^{-1}- \Sigprior^{-1} \right) - \bm{I}_6\\
     &=& \Sigpost \left(\tilde{\bm{H}} -\bm{H} - \Sigprior^{-1}\right) + \Sigpost \left(\bm{H} + \Sigprior^{-1} \right) - \bm{I}_6\\
     &=& \Sigpost \left(\tilde{\bm{H}} -\bm{H} - \Sigprior^{-1}\right) + \Sigpost \Sigpost^{-1}  - \bm{I}_6\\
     &=& \Sigpost \left(\tilde{\bm{H}} -\bm{H} - \Sigprior^{-1}\right). 
\end{eqnarray*}
Consequently, it comes that
\begin{eqnarray}
\mathbb{V}_{\bm{m}}\left[ \bm{W}  \right]
   &=& \trace \left( \left( \Sigpost \left(\tilde{\bm{H}} -\bm{H} - \Sigprior^{-1}\right)\right) \Sigprior \left(\Sigpost \left(\tilde{\bm{H}} -\bm{H} - \Sigprior^{-1}\right)\right)^\top\right)\nonumber\\
   &=& \trace \left(\left(\Sigpost \left(\tilde{\bm{H}} -\bm{H}\right) - \Sigpost \Sigprior^{-1}\right)\Sigprior \left(\Sigpost \left(\tilde{\bm{H}} -\bm{H}\right) - \Sigpost\Sigprior^{-1}\right)^\top\right)\nonumber\\
   &=& \trace \left(\left(\Sigpost \left(\tilde{\bm{H}} -\bm{H}\right) - \Sigpost \Sigprior^{-1}\right)\Sigprior \left(\Sigpost \left(\tilde{\bm{H}} -\bm{H}\right) - \Sigpost\Sigprior^{-1}\right)^\top\right)\nonumber\\
   &=& \trace \left(\Sigpost \left(\tilde{\bm{H}} -\bm{H}\right) \Sigprior \left(\tilde{\bm{H}} -\bm{H}\right)^\top\Sigpost \right)
     - \trace \left( \Sigpost \left( \left(\tilde{\bm{H}} -\bm{H}\right) - \left(\tilde{\bm{H}} -\bm{H}\right)^\top\right) \Sigpost \right)\nonumber \\
    & &  + \trace\left(\Sigpost\Sigprior^{-1}\Sigpost\right). \label{eq:risk_p2}
\end{eqnarray}
By definition of $\bm{Z}$ and $\bm{W}$, the Bayes risk can be recast as
\begin{eqnarray}
\label{eq:risk_p3}
  \risk =  \mathbb{E}_{\bm{Y}|\bm{m}}\left[ \left\|\bm{Z} \right\|^2 \right] + \mathbb{E}_{\bm{m}}\left[ \left\|\bm{W} \right\|^2 \right],
\end{eqnarray}
where $\mathbb{E}_{\bm{Y}|\bm{m}}\left[ \left\|\bm{Z} \right\|^2 \right] =  \trace\left(  \Sigpost \bm{H} \Sigpost \right)$ and $\mathbb{E}_{\bm{m}}\left[ \left\|\bm{W} \right\|^2 \right] = \trace(\mathbb{V}\left[\bm{W}\right]) +  \left\|\mathbb{E}_{\bm{m}}\left[ \bm{W}  \right]\right\|^2$ according the total variance law. Gathering \eqref{eq:risk_p1} and \eqref{eq:risk_p2} in the equality \eqref{eq:risk_p3}, it yields
\begin{eqnarray*}
  \risk &=& \trace\left(  \Sigpost \bm{H} \Sigpost \right) + \trace\left(\Sigpost\Sigprior^{-1}\Sigpost\right) +  
  \trace \left(\Sigpost \left(\tilde{\bm{H}} -\bm{H}\right) \Sigprior \left(\tilde{\bm{H}} -\bm{H}\right)^\top\Sigpost \right) \\
     & & - \trace \left( \Sigpost \left( \left(\tilde{\bm{H}} -\bm{H}\right) - \left(\tilde{\bm{H}} -\bm{H}\right)^\top\right) \Sigpost \right) + \left\|\Sigpost \left( \tilde{\bm{H}} -\bm{H}  \right) \mprior\right\|^2 \\
    &=& \trace\left(  \Sigpost \left( \bm{H} + \Sigprior^{-1} \right) \Sigpost \right)  +  
  \trace \left(\Sigpost \left(\tilde{\bm{H}} -\bm{H}\right) \Sigprior \left(\tilde{\bm{H}} -\bm{H}\right)^\top\Sigpost \right) \\
     & & - \trace \left( \Sigpost \left( \left(\tilde{\bm{H}} -\bm{H}\right) - \left(\tilde{\bm{H}} -\bm{H}\right)^\top\right) \Sigpost \right) + \trace\left(\Sigpost \left( \tilde{\bm{H}} -\bm{H}  \right) \mprior \mprior^\top \left( \tilde{\bm{H}} -\bm{H}  \right)^\top \Sigpost^\top \right) \\
     &=& \trace\left(  \Sigpost \right)  +  
  \trace \left(\Sigpost \left(\tilde{\bm{H}} -\bm{H}\right)^\top \left(\Sigprior + \mprior \mprior^\top \right) \left(\tilde{\bm{H}} -\bm{H}\right)\Sigpost \right) \\
     & & - \trace \left( \Sigpost \left( \left(\tilde{\bm{H}} -\bm{H}\right) - \left(\tilde{\bm{H}} -\bm{H}\right)^\top\right) \Sigpost \right).
\end{eqnarray*}
The derivation is therefore complete.

\end{document}